\documentclass[twocolumn]{aastex631}

\usepackage{booktabs}
\usepackage{amsmath}
\usepackage{graphicx}
\usepackage{algorithm}
\usepackage{algorithmic}
\usepackage{textgreek}
\usepackage{silence}
\newcommand{\om}{\Omega_{m}}
\newcommand{\thetae}{\theta_{\rm E}}
\newcommand{\xl}{x_{\rm l}}
\newcommand{\yl}{y_{\rm l}}
\newcommand{\zl}{z_{\rm l}}
\newcommand{\zs}{z_{\rm s}}
\newcommand{\sigmav}{\sigma_{\rm v}}

\begin{document}

\title{Strong Lensing Cosmology with Population-level Calibrated Neural Ratio Estimation}

\author[0000-0002-5386-7076]{Sreevani Jarugula}
\affiliation{Scientific Computing Division, Fermi National Accelerator Laboratory, Batavia, IL 60510}

\author[0000-0001-6706-8972]{Brian D. Nord}\thanks{Now Independent Researcher}
\affiliation{Scientific Computing Division, Fermi National Accelerator Laboratory, Batavia, IL 60510}
\affiliation{Department of Astronomy and Astrophysics, University of Chicago, Chicago, IL 60637}
\affiliation{KICP, University of Chicago, Chicago, IL 60637}

\author[0000-0003-1281-7192]{Aleksandra \'Ciprijanovi\'c}
\affiliation{Scientific Computing Division, Fermi National Accelerator Laboratory, Batavia, IL 60510}
\affiliation{Department of Astronomy and Astrophysics, University of Chicago, Chicago, IL 60637}
\affiliation{NSF-Simons AI Institute for the Sky (SkAI), 172 E. Chestnut St., Chicago, IL 60611}

\author[0000-0003-1237-4301]{Shubhendu Trivedi}\thanks{Now at Google DeepMind}
\affiliation{Scientific Computing Division, Fermi National Accelerator Laboratory, Batavia, IL 60510}

\begin{abstract}
Strong gravitational lensing contains key information about cosmic acceleration. Modern and next-generation galaxy imaging surveys are expected to provide high-quality data on $\mathcal{O}(10^5)$ galaxy-galaxy lensing systems.
The plethora and complexity of the data are likely to present computational challenges for parameter inference methods for fitting high-dimensional likelihoods, which are often analytically intractable.
Neural Ratio Estimation (NRE) efficiently computes individual likelihood ratios that can be combined into population-level posteriors.
We use simulations to study the capacity of NRE to jointly predict the dark energy equation-of-state parameter $w$ and the total matter density $\om{}$ from lensing images and companion spectroscopic information.
We also introduce a post hoc posterior coverage calibration procedure that mitigates the model overconfidence that is typically found in neural density estimation applications.
Our experiments show that the errors on both parameters decrease with increasing inference population sizes.
In particular, for 100 lenses in a standard $\Lambda$CDM Universe, our calibrated NRE model achieves median fractional uncertainty of $22.8\%$ in $w$ and $2.9\%$ in $\om{}$.
This proof of concept demonstrates a potentially scalable approach for efficient cosmological parameter inference with large populations of galaxy-scale lenses observed in future surveys.
\end{abstract}

\section{Introduction} 
\label{sec:intro}

In the standard cosmological model ($\Lambda$CDM), the evolution of a flat Universe is driven by two major components. 
First, the total matter density $\om{}$, mostly comprising Dark Matter (DM), which drives the hierarchically forming and gravitationally bound large-scale structures -- from subhaloes to galaxies to the cosmic web.
Second, space-time is accelerating in its expansion due to Dark Energy (DE), which is a scalar field characterized by an equation of state with pressure-density ratio $w=p/\rho$. 
In a Universe with a flat geometry the DE density is determined exactly by the total matter density, where $\Omega_{\rm DE} = 1- \om{}$ and in $\Lambda$CDM, the equation of state is fixed at $w=-1$.
The fundamental natures of both DE and DM remain unknown, and modern experiments use combinations of complementary probes -- e.g., the Cosmic Microwave Background (CMB), the cosmic distance ladder, galaxy clusters and clustering, and gravitational lensing -- to constrain cosmological parameters.

Many combinations of probes have found consistency with the $\Lambda$CDM~\citep{planck20}.
However, investigations of systematics effects and updated joint probe analyses have uncovered significant disagreement in cosmology parameters~\citep{leizerovich23}.
In particular, the Hubble Tension originates in estimates of $H_{0}$ that are $>4\sigma$ higher from the cosmic distance ladder than from the CMB measurements~\citep{Freedman_2021, 2022ApJ...934L...7R, 10.1093/mnras/sty418, DiValentino_2021, huterer2023hubble, Khalife_2024, Cai_2026}.
Secondarily, the $S_8$ Tension arises from estimates of $S_{8}\equiv\sqrt{\om{}/0.3}\sigma_8$ that are $\sim 3\sigma$ lower from galaxy surveys than from CMB experiments~\citep{PhysRevD.107.123538, PhysRevD.108.023520, PANTOS2026102286}.
These discrepancies may indicate new physics in the form of systematic effects in individual probes or an updated cosmological model~\citep[e.g.,][]{10.1093/mnras/stad3107, 2023ARNPS..73..153K, Zu2026}.
These discrepancies invite the collection of larger, more information-dense data sets and the development of analysis techniques that maximize the extraction of information from existing and new cosmological probes.

Complementary to other probes, strong gravitational lensing can constrain DE, DM, and the expansion rate through distances and the large-scale structure.
Different lens configurations provide access to different parameters.
Time-delay cosmography~\citep{1964MNRAS.128..307R} relies on quasars or supernovae to constrain the Hubble constant and is independent of all the other probes ~\citep{10.1111/j.1365-2966.2004.08160.x, 10.1093/mnras/stab484, 2021A&A...656A.153S, Treu2022,  Treu2024, Li_2024}. 
There exist dedicated long-term programs like H0LiCOW to make high-precision measurements of $H_0$~\citep{10.1093/mnras/stx483}, which have led to corroborating evidence of the Hubble Tension~\citep{10.1093/mnras/stz3094}.
Alternatively, through the ratio of angular diameter distances, static galaxy-galaxy lensing provides $H_0$-independent joint constraints on $w$ and $\om{}$ when the lens stellar velocity dispersion is known~\citep{2001PThPh.105..887F, 2006EAS....20..161K,  LI2023101234}. 
State-of-the-art image-based model-fitting of individual lenses relies on Monte Carlo sampling of analytic, tractable (i.e., Gaussian) likelihoods to produce posteriors. After deriving lens parameters, further population-level analytic modeling of $\mathcal{O}(10)-\mathcal{O}(10^2)$ lenses gives 
constraints on $w$-$\om{}$~\citep{10.1093/mnras/stz1902, cao15, 10.1111/j.1365-2966.2010.16725.x, litian24}, but still requiring assumptions of tractable Gaussian parameter distributions.

Several modern experiments --- the Vera Rubin Observatory's Legacy Survey of Space and Time ~\citep[LSST;][]{ivezic08, 2025RSPTA.38340117S}, the Euclid Wide Survey~\citep{euclid22, Lines2025}, the Roman Space Telescope High Latitude Wide Area Survey~\citep{spergel15}, and the China Space Station telescope~\citep[CSST;][]{10.1093/mnras/stae1865}  --- are expected to obtain data on $\mathcal{O}(10^5)$ strong lenses~\citep{holloway23}.
Spectroscopic projects, such as the 4MOST Strong Lensing Spectroscopic Legacy Survey ~\citep[4SLSLS;][]{collett23} and the Dark Energy Spectroscopic Instrument~\citep[DESI;][]{2021AAS...23755404A, 10.1093/mnras/staf089}, can supply redshift and velocity-dispersion measurements needed for lens models to constrain $w$ and $\om{}$. 
Nevertheless, modeling galaxy-scale lenses is highly computationally expensive.
Therefore, more efficient lens modeling techniques and more flexible parameter inference approaches may be necessary to exhaust the capacity of very large next-generation data sets.

Artificial Intelligence (AI) models are characterized by their numerous internal parameters, which makes them be useful for predictions from large and complex data sets. 
In particular, Simulation-based Inference (SBI) is a growing suite of AI-based methods 
for parameter inference when high-fidelity data simulators are available and likelihoods are high-dimensional and intractable~\citep{cranmer20, 2020JOSS....5.2505T, zeghal2022neuralposteriorestimationdifferentiable, 2024arXiv240409636G, deistler25, 2025PhRvD.112j3526F}.
When output densities must be expressive (i.e., non-Gaussian), Neural Likelihood Estimation~\citep[NLE;][]{papamakarios18, 2025arXiv250708734B, 2026PhRvD.113l4064E}, Neural Posterior Estimation~\citep[NPE;][]{Papamakarios2016,Greenberg2019}, Neural Score Estimation~\citep{2022arXiv221004872S, 2023A&A...672A..51R, 2025arXiv251104792B}, Surjective Sequential Neural Likelihood \citep[SSNL][]{2023arXiv230801054D} employ flexible density estimators, like normalizing and masked autoregressive flows~\citep{2016arXiv160604934K, papamakarios18, 2017arXiv170507057P} and diffusion~\citep{linhart2026diffusionposteriorsamplingsimulationbased, rosso25}.
NPE has been used to estimate parameters of individual objects across numerous phenomena in astrophysics and cosmology, especially for gravitational wave (GW) signals~\citep{2020PhRvD.102j4057G, PhysRevLett.127.241103, 2024arXiv240302443K} and strong lens images~\citep{2021mlps.conf...95L, 2023ApJ...943....4L, 2024arXiv241016347S, Poh_2025, 2025arXiv251117732D, 2025AJ....170...44E}.

However, estimating cosmological parameters shared by an entire population is the goal under consideration. 
In general, it is not formally permissible to combine posteriors from individual objects into a single global posterior, so conventional NPE is not appropriate.
Hierarchical Neural Posterior Estimation (HNPE) and related methods have been developed, but not widely tested~\citep{3540261.3541290, 2025arXiv250604558F, tran2017hierarchical}.
Additionally, a novel approach in GW astronomy learns the population posterior from individually learned sub-population posteriors~\citep{2024PhRvD.109f4056L}.
Combining NLE outputs for population-level parameter estimates does not appear to have been implemented in the literature.
Alternatively, Neural likelihood Ratio Estimation~\citep[NRE;][]{2018arXiv180512244B, hermans20, 2021ANIPS..34..129M} uses relatively simple classification neural networks instead of flows or diffusion models and has been expanded and refined through multiple studies in physics and astronomy~\citep{2022JCAP...09..004C, 10.1093/mnras/stac3014, 2022arXiv221006170M,  2023arXiv231110571C, 2024MNRAS.530.4107A, Rizvi2024}.
Ratios from individual lenses can be combined and then joined with a prior to achieve posteriors of lens population distributions~\citep{Brehmer_2019}. 

Inferences require strict assessment for fidelity and trustworthiness.
In this regard, SBI has been the subject of intensive study and the target of new diagnostics and metrics.
Simulation-based calibration (SBC) describes a growing suite of diagnostics of posterior quality~\citep{talts18, 2021arXiv210104653L} -- e.g., rank statistic histograms, posterior coverage, and TARP~\citep{lemos2023samplingbasedaccuracytestingposterior}.
Many studies show that SBI models tend to be overconfident, generating posteriors that are too narrow~\citep{hermans21}. Balanced NRE (BNRE) is designed to mitigate overconfidence by increasing uncertainties in parameter regimes~\citep{delaunoy22}; BNRE has not yet been applied explicitly in population-level inference.
Post hoc calibration of ratios with histograms aims to correct inferences~\citep{Brehmer_2019}, but statistical guarantees haven't been demonstrated explicitly.
\citet{2024arXiv241112068F} show that neural estimators can, in principle,
match the statistical guarantees of established methods such as approximate
Bayesian computation, but only for well-specified models. In practice, however,
no density estimator or post hoc calibration method yet guarantees a correctly
calibrated posterior for real data. The tendency toward overconfidence
therefore remains the central obstacle to trustworthy SBI.

In this work, we study the capacity of NRE for deriving trustworthy constraints on $w$ and $\om{}$ from populations of galaxy-galaxy lenses. 
We present a proof of concept that focuses on 1) building a population-level posterior from individual likelihood ratios and 2) mitigating the longstanding tendency of SBI models to be overconfident. 
Therefore, we reduce the computational burden -- i.e., allow for a simpler classification network -- by assuming an idealized context for the data.
The NRE model uses two inputs: the input simulated lens images with simplified lens models and a single band and spectroscopically derived parameters.
After combining the likelihood ratios of individual lenses, we apply a novel post hoc calibration method to mitigate posterior overconfidence.
Finally, we analyze the confidence and accuracy of the model with and without calibration.

In Sec.~\ref{sec:stronglensing}, we introduce strong gravitational lensing theory, including the degeneracy between the two cosmological parameters of interest. 
Then, in Sec.~\ref{sec:data}, we describe our study's data sets for model-building and testing --- all of which are derived from simulations.
Sec.~\ref{sec:nre} contains a brief derivation of the NRE approach for population-level posterior estimation, the NRE model architecture and training; it also introduces our post hoc calibration procedure. 
Sec.~\ref{sec:results} describes the main results of the experiments. 
In Sec.~\ref{sec:discussion}, we discuss the stress tests and current limitations of the NRE-based approach presented here.
In Sec.~\ref{sec:conclusion}, we recapitulate the key findings of this study, accompanied by an outlook on potential future analyses.
Finally, in the Appendix, we provide additional details on the NRE model architecture and stress tests of the model.

\section{Strong Gravitational Lensing} 
\label{sec:stronglensing}

Strong gravitational lensing occurs when a massive object, such as a galaxy (lens), deflects the light from a background object (source) to produce magnified and distorted images of the source in the image plane. 
For a point-like source, such as a QSO or a supernova, lensing can create multiple images of the source. 
The basic formalism below introduces galaxy-galaxy lensing, where the source is a galaxy (an extended object), and lensing causes arc-like structures in the image plane. 
More detailed discussions can be found in~\citet{2010ARA&A..48...87T, 2024SSRv..220...12S, Shajib2024}.

\subsection{Formalism}
\label{sec:formalism}

The lens equation relates the source position $\beta$ to the lensed source position $\theta$ through the deflection angle $\alpha(\theta)$ as 
\begin{equation}
    \beta\ = \theta\ - \alpha(\theta).
\end{equation}
\noindent The deflection field is the gradient of the lensing potential $\psi$:
\begin{equation}
    \alpha(\theta)\ = \nabla \psi(\theta)\ = \frac{1}{\pi}\int d\theta^{\prime}\ \frac{\theta\ - \theta^{\prime}}{|\theta\ - \theta^{\prime}|^{2}}\kappa(\theta^{\prime}),
\end{equation}
where the convergence $\kappa(\theta) = \Sigma(\theta)/\Sigma_{cr}$, and $\Sigma(\theta)$ is the deflector mass surface density projected onto the lens plane.
The critical mass surface density $\Sigma_{cr}$ is  
\begin{equation}
    \Sigma_{cr}\ = \frac{1}{4\pi\ G}\frac{D_{\rm s}}{D_{\rm ls}D_{\rm l}}, 
\end{equation}
\noindent where $G$ is the gravitational constant; and $D_{\rm l}$, $D_{\rm s}$, and $D_{\rm ls}$ are the angular diameter distances between the observer and the lens, the observer and the source, and the lens and the source, respectively. 
In $w$CDM cosmology (where $w$ can vary from $-1$), $D$ depends on the the source and lens redshifts ($\zs{}$ and $\zl{}$, respectively) and on the cosmological parameters ($H_{0}$, $\om{}$, and $w$):
\begin{equation}
    \label{eqn:angular_distance}
    D(z, H_{0}, \om{}, w)\ = \frac{1}{1+z}\frac{c}{H_{0}}\int_{0}^{z}\frac{dz^{\prime}}{h(z^{\prime},\om{}, w)},
\end{equation} 
\noindent where the Hubble parameter is
\begin{equation} 
    h^{2}(z,\om{}, w)\ = \om{}(1+z)^{3} + (1-\om{})(1+z)^{3(1+w)}.
\end{equation}
\noindent For a circular deflector (i.e., lens), the Einstein radius $\thetae{}$ is the radius within which the average surface mass density is equal to the critical density so that the average convergence $\left\langle \kappa \right\rangle = 1$. 

In the case of non-circular deflectors, the mass enclosed within the Einstein radius is~\citep{kormann94} 
\begin{equation}
    \thetae{}\ = \sqrt{\frac{4GM(\thetae{})}{c^{2}}\frac{D_{\rm ls}}{D_{\rm l}D_{\rm s}}}.
\end{equation}

The majority of the lenses are early type (elliptical) galaxies, whose mass densities are well described by an SIE profile~\citep{oguri10}, where the Einstein radius becomes
\begin{equation}
    \thetae{}\ = 4\pi \left(\frac{\sigmav{}}{c} \right)^{2} \frac{D_{\rm ls}}{D_{\rm s}},
    \label{eqn:einstein_radius}
\end{equation}
\noindent where $\sigmav{}$ is the stellar velocity dispersion, and $c$ is the speed of light~\citep{kormann94}.
The convergence of a lens with an SIE profile is  
\begin{equation}
    \kappa(x_{\phi_{\rm l}}, y_{\phi_{\rm l}})\ = \frac{1}{2}\left(\frac{\thetae{}}{\sqrt{q_{\rm l}x_{\phi_{\rm l}}^{2} + y_{\phi_{\rm l}}^{2}/q_{\rm l}}} \right),
\end{equation}
\noindent where ($x_{\phi_{\rm l}}, y_{\phi_{\rm l}}$) are the positions aligned to the major and minor lens axes, $q_{\rm l}$ is the major/minor axis ratio, and $\phi_{\rm l}$ is the angle of rotation between ($x_{\phi_{\rm l}},y_{\phi_{\rm l}}$) and the centre of the lens mass coordinates ($\xl{}$, $\yl{}$) in the image plane.
The lens ellipticity is related to the major and minor axis as
\begin{equation}
    \begin{split}
    l_{\rm e1}\ = \frac{1-q_{\rm l}}{1+q_{\rm l}}cos(2\phi_{\rm l}),\\
    l_{\rm e2}\ = \frac{1-q_{\rm l}}{1+q_{\rm l}}sin(2\phi_{\rm l}).
    \end{split}
\end{equation}
\noindent We do not include external shear in our analysis. 

We model the source light distribution with a S\'ersic profile~\citep{sersic68}:
\begin{equation}
    \label{eqn:source_sersic}
    I(r) = I_{e}exp\left(-b_{n} \left( \left(\frac{r}{R} \right)^{1/n} - 1 \right) \right),
\end{equation}
\noindent where $I_{e}$ is the intensity at the effective radius $R$, and 
$r = \sqrt{q_{\rm s}x_{\phi_{\rm s}} + y_{\phi_{\rm s}}/q_{\rm s}}$; the positions ($x_{\phi_{\rm s}}$, $y_{\phi_{\rm s}}$) are aligned to the semimajor axis of the S\'ersic light profile, and $q_{\rm s}$ is the source major-to-minor axis ratio. 
In the range $0 < n < 10$, the coefficient $b$ can be approximated as~\citep{ciotti99}
\begin{equation}
    b_{n}\ \approx\ 2n\ - \frac{1}{3} + \frac{4}{405n} + \frac{46}{25515n^{n}}.
\end{equation}
\noindent Finally the source ellipticity is 
\begin{equation}
    \begin{split}
    s_{\rm e1}\ = \frac{1-q_{\rm s}}{1+q_{\rm s}}cos(2\phi_{\rm s}),\\
    s_{\rm e2}\ = \frac{1-q_{\rm s}}{1+q_{\rm s}}sin(2\phi_{\rm s}).
    \end{split}
\end{equation}

\subsection{Degeneracy in cosmology parameters}
\label{sec:degeneracy}

For SIE lenses in a flat $w$CDM cosmology, the Einstein radius is fully specified by the velocity dispersion, the lens and source redshifts, and the cosmological parameters (see Eqn.'s~\ref{eqn:angular_distance} and~\ref{eqn:einstein_radius}).
Even when the velocity dispersion and redshifts are known, varying combinations of $w$ and $\om{}$ give the same Einstein radius: the cosmological parameters are degenerate with respect to a single lens.
Therefore, an individual lens is insufficient to jointly infer $w$ and $\om{}$.
However, two or more lenses, with different redshifts can increase the leverage for constraining capability of lens images.

\begin{figure}[!htbp]
 \centering
    \includegraphics[width=1.0\linewidth, trim={0cm 0cm 1cm 0.8cm}]{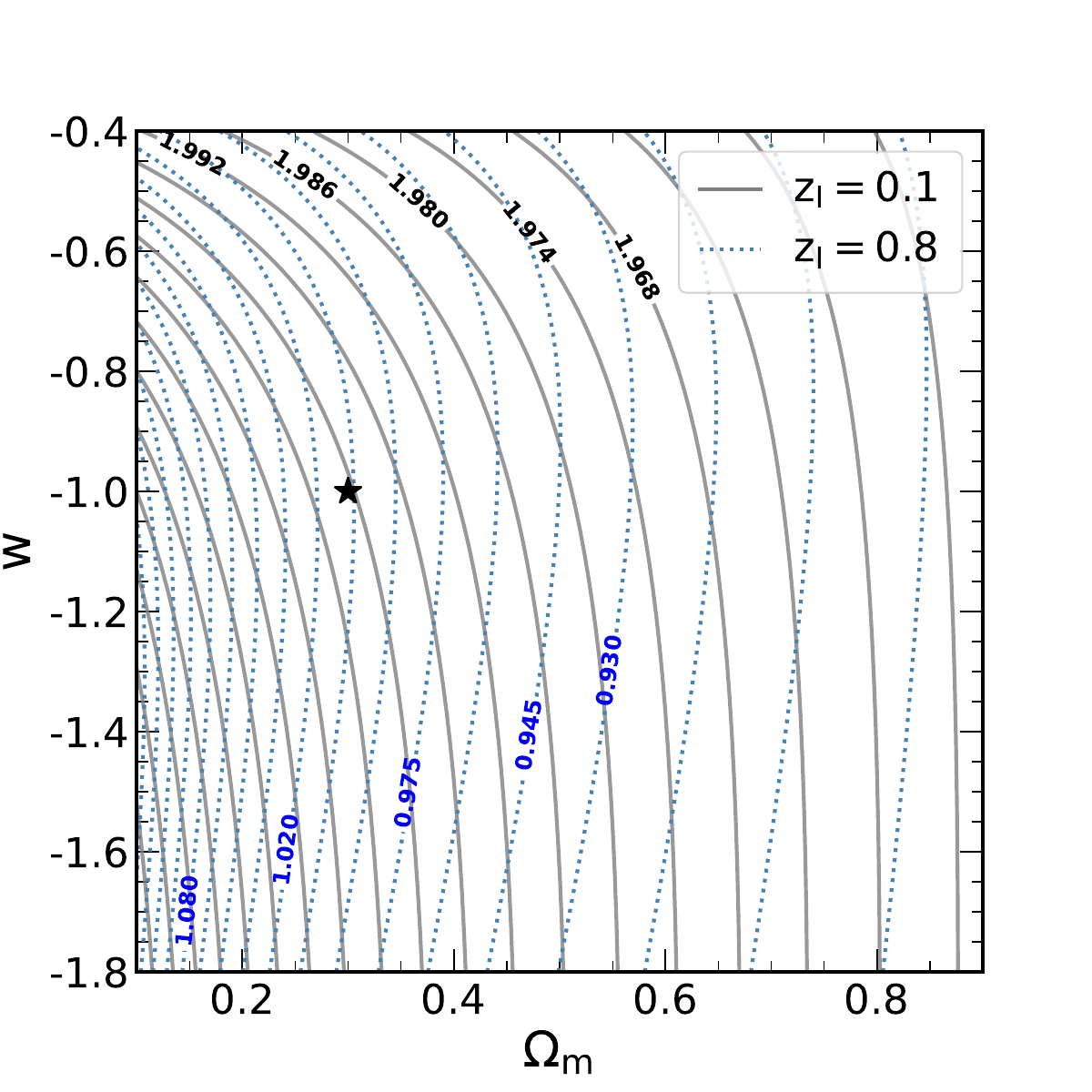}
    \caption{Einstein radius $\thetae{}$ (arcsec; Eqn.~\ref{eqn:einstein_radius}) as a function of $w$ and $\om{}$ for lenses at low redshift ($\zl{}=0.1$; gray) and high redshift ($\zl{}=0.8$; blue).
    The configurations use the same source redshift ($\zs{}=2.0$) and stellar velocity dispersion ($\sigmav{}$=275\ $\mathrm{kms}^{-1}$).
    The canonical $\Lambda$CDM cosmology is marked with a star.
    }
    \label{fig:w_om_degeneracy}
\end{figure}

We illustrate this behavior in Fig.~\ref{fig:w_om_degeneracy}, where two sets of contours show constant values of the Einstein radius with lens redshifts $\zl{} = 0.1$ and $\zl{} = 0.8$, respectively.
Both configurations have the same source redshift $\zs{} = 2.0$ and lens velocity dispersion $\sigmav{} = 275\,\mathrm{km\,s^{-1}}$. 
The true cosmology is indicated by a star.
The curve for the Einstein radius for the single lens at redshift $\zl{} = 0.1$ intersects the true cosmology, but does not constrain it. 
The curve for the radius at the other redshift $\zl{} = 0.8$ intersects the low-redshift lens curve and the true cosmology, breaking the $w$-$\om{}$ degeneracy.
In practice, modeling lens images adds uncertainty to the Einstein radius estimates, which would add a finite width to each contour curve. Two lenses constrain $w$ and $\om{}$ only to the region where their broadened bands overlap, and adding more lenses progressively shrinks this region, tightening the constraints and averaging down the measurement scatter.

\section{Data}
\label{sec:data}

The lens model in this study uses the following parameters to produce lens images:
\begin{itemize}
    \item cosmology parameters: $w$, $\om{}$;     
    \item lens and source redshift: $\zl{}, \zs{}$;
    \item lens stellar velocity dispersion: $\sigmav{}$;    
    \item lens mass profile: $\xl{}, \yl{}, l_{\rm e1}, l_{\rm e2}$;
    \item source light profile: $m_{\rm s}, R, n, s_{\rm e1}, s_{\rm e2}$.
\end{itemize}
\noindent The apparent magnitude $m_{\rm s}$ is typically obtained from observations so that it is used in place of the intensity $I_{\rm e}$ (Eqn.~\ref{eqn:source_sersic}). 
We summarize some astrophysics parameters as $\zeta \equiv (\zl{}, \zs{}, \sigmav{})$, while the remaining nuisance astrophysics parameters are summarized with $\nu$.

For $w$, we use a uniform prior $\mathcal{U}(-1.8, -0.35)$. 
The upper bound is motivated by the physical requirement $w < - 1/3$, while the lower bound is chosen to provide a generous margin around the current observational consensus of $w=-1$. 
For $\om{}$, we assume a uniform prior $\mathcal{U}(0.1,0.9)$.
This range ensures the total matter density remains strictly positive and well-defined, while allowing the model to include a wide range of cosmologies near $0.3$, which is favored by current observations. 
We fix  $H_{0} = 70\,\mathrm{km\,s^{-1}\,Mpc^{-1}}$. 

The lens position ($\xl{}, \yl{}$) is drawn from a normal distribution centered at zero because the lens is assumed to be near the center of the image.
The source position is fixed at the image center (0,0).
The priors for $\zeta$ encompass the peak of the parameter distributions from the realistic DES lens simulations~\citep{collett15}. 
These parameters typically have significant observational errors; however, to keep the model training simple, $\zeta$ are used as exact inputs for the model.
All the other parameters have uniform priors that bound current observations.
The astrophysical parameters are sampled independently of the cosmological parameters and independently of each other.
The prior ranges in the training data are listed in Table~\ref{table:params}. 

To produce images representative of survey observing conditions, we adopt instrumental and observational characteristics from DES Data Release 1~\citep{abbott18}. 
We use a pixel scale of 0.263 arcsec pixel$^{-1}$, a CCD gain of 6.083 $e^{-}$/count, and a read noise of 7.0 $e^{-}$. 
We select the $g$-band filter to maximize contrast for the lensing arcs in the images: lens galaxies are typically red ellipticals that are fainter at bluer wavelengths, and source galaxies are often bluer, star-forming galaxies that are brighter in the $g$ band. 
The sky brightness, seeing, and effective number of exposures are based on empirical survey values; the magnitude zero point is set to 30.0.
We assume that the lens light is perfectly subtracted from the images. 
In practice, lens light subtraction is commonly performed prior to lens modeling, and some automated approaches have been tested ~\citep{hezaveh17}. 
All images have a size of $32 \times 32$ pixels.
We simulate lens images with \texttt{deeplenstronomy}~\citep{morgan21}, which is built on \texttt{lenstronomy}~\citep{birre15, birrer18}. 

\begin{table*}
\noindent\begin{minipage}[b]{0.99\linewidth}
  \caption{Prior parameter distributions for generating training and validation data. 
  }
  \label{table:params}
  \centering
  \begin{tabular}{lll}
 \hline   Parameter & Name & Priors \\ \toprule  \hline
 \multicolumn{3}{c}{Cosmology}\\
 \midrule
 $w$                & Dark Energy equation of state & $\mathcal{U}(-1.8, -0.35)$ \\
 $\om{}$            & Total Matter density          & $\mathcal{U}(0.1, 0.9)$ \\
 $H_{0}$ [km/s/Mpc] & Hubble constant               & $70.0$ \\
 \midrule
 \multicolumn{3}{c}{Lens }\\
 \midrule
  $\sigmav{}$ [km/s] & velocity dispersion & $\mathcal{U}(225, 300)$ \\
  $\xl{}$ [arcsec]  & lens position x  & $\mathcal{N}(0.0, 0.8)$ \\ 
  $\yl{}$ [arcsec] & lens position y  & $\mathcal{N}(0.0, 0.8)$ \\ 
  $\zl{}$  & lens redshift & $\mathcal{U}(0.3, 0.8)$ \\
  $l_{\rm e1}$  &  lens ellipticity   & $\mathcal{U}(-0.1,0.1)$ \\ 
  $l_{\rm e2}$  &  lens ellipticity  & $\mathcal{U}(-0.1,0.1)$ \\ 
 \midrule
 \multicolumn{3}{c}{Source Light }\\
 \midrule
  $m_{\rm s}$  & magnitude & $\mathcal{U}(19, 24)$ \\ 
  $R$ [arcsec] & half-light radius & $\mathcal{U}(0.1,3.0)$ \\ 
  $n$   & Sersic index  & $\mathcal{U}(0.5,8.0)$ \\ 
  $s_{\rm e1}$  & source ellipticity  & $\mathcal{U}(-0.1,0.1)$ \\ 
  $s_{\rm e2}$  & source ellipticity  & $\mathcal{U}(-0.1,0.1)$ \\
  $\zs{}$ & source redshift & $\mathcal{U}(1.5, 2.5)$ \\
 \hline
\end{tabular}
\end{minipage}
\end{table*}

\begin{figure*}[!ht]
 \centering
    \includegraphics[width=0.8\linewidth, trim={0cm 3cm 0 3cm},clip]{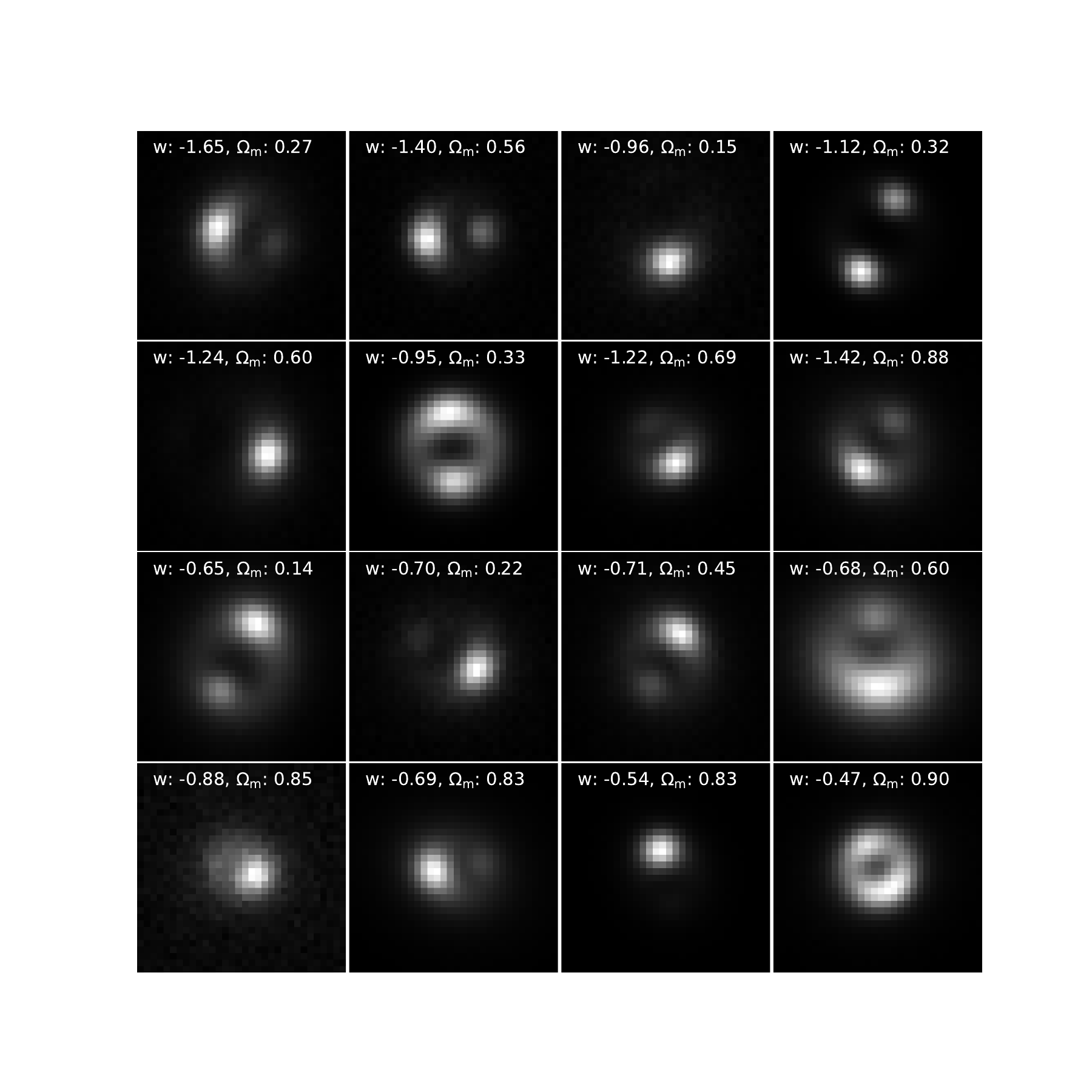}
    \caption{Randomly selected image examples from the training dataset. The images correspond to $g$-band observations with DES-like noise. 
    Only the lensed source light is included, mimicking ideal lens-light subtraction.}
    \label{fig:train_data}
\end{figure*}

We use four data sets to build the NRE model: training, validation, test, and calibration.
The training and validation datasets consist of 1,990,000 and 398,000 images, respectively. We sample the cosmology and astrophysics parameters according to the priors in Table~\ref{table:params}.
Examples of training images are shown in Fig.~\ref{fig:train_data}.

\begin{figure*}[!ht]
 \centering
    \includegraphics[width=0.5\linewidth, trim={0cm 0cm 0 0cm},clip]{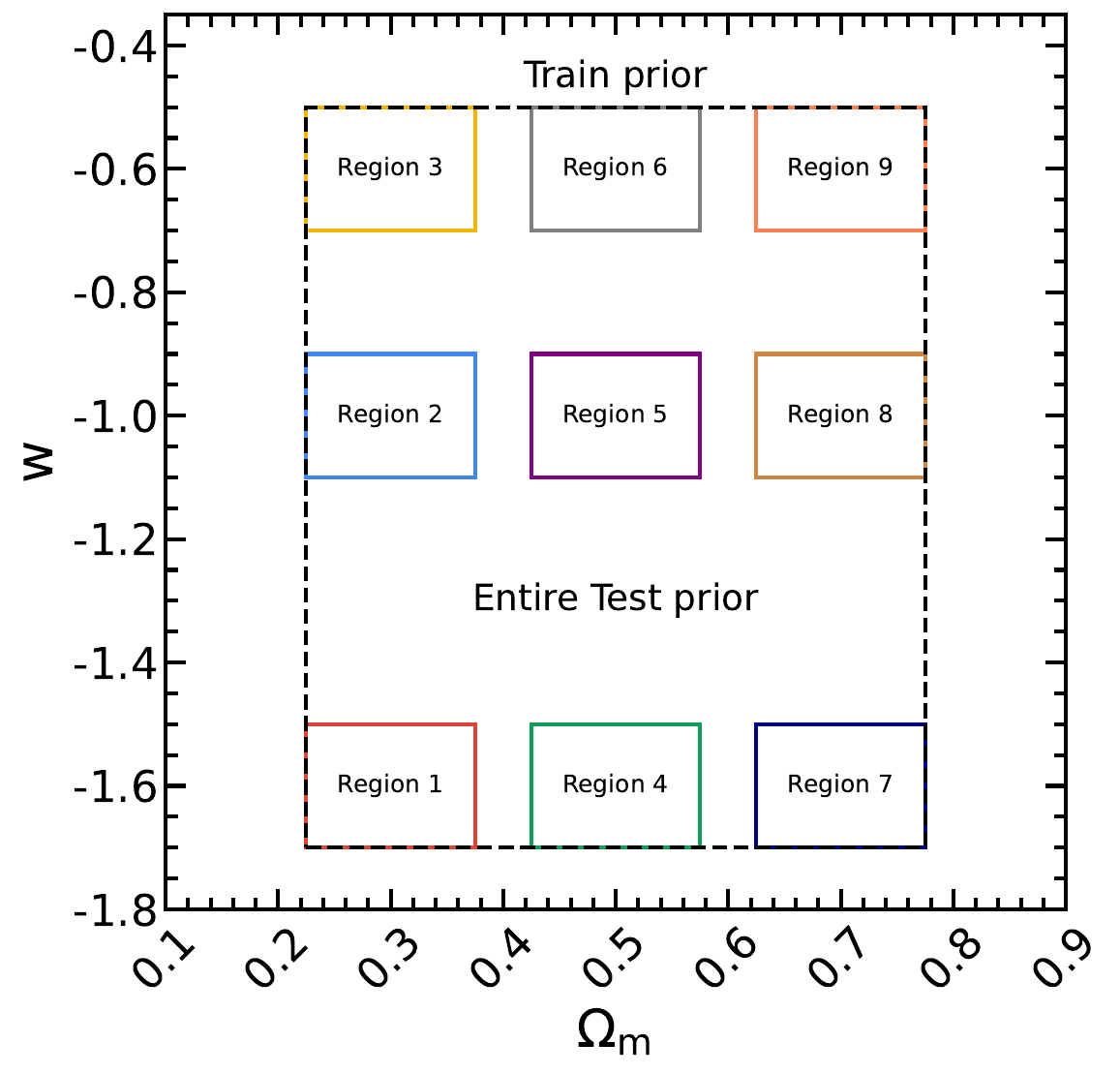}
    \caption{Priors used to generate the datasets in the $w$-$\om{}$ plane. The training prior comprises the full area displayed: $w \in [-1.8, -0.35]$ and $\om{} \in [0.1, 0.9]$.
    The priors for the Entire Test Region (dashed) and the nine Sub-Regions (colored boxes) are listed in Table~\ref{tab:emd_regions}.
    \label{fig:w_om_regions}
    }
\end{figure*}

We construct 10 independently sampled test datasets -- the Entire Region and nine Sub-Regions within the Entire Region.
The astrophysics parameter test priors are the same as those used for training.
For the cosmology parameter priors, the test ranges are chosen to be well within the training ranges to prevent edge-effect biases in model prediction.
The Entire Test Set encompasses nine smaller Sub-Regions. 
Within each test region (Entire Test Region or Sub-Region), we sample 100 pairs of cosmological parameters ($w$, $\om{}$) according to the priors in Table~\ref{tab:emd_regions} and Fig.~\ref{fig:w_om_regions} and for each cosmology pair, we generate 100 lenses according to the priors in Table~\ref{table:params}. 

For the calibration dataset, we sample 50 pairs of cosmological parameters 
and for each cosmology, we generate 100 lenses. 
Further details of the calibration dataset and procedure are provided in Sec.~\ref{sec:nre_calibration}.

All images are preprocessed before being passed to the neural network. Each image is first normalized by dividing by the sum of its pixel values over the spatial dimensions, ensuring unit total intensity. The images are then multiplied by the total number of pixels, rescaling them such that the average pixel value is unity while preserving the relative spatial intensity distribution.

The training and test datasets are publicly available on Zenodo\footnote{\url{https://zenodo.org/records/20804324}}. The code used in this analysis is available in our GitHub repository\footnote{\url{https://github.com/deepskies/CosmoLensNRE}}.

\section{Method: Posteriors from Neural Ratio Estimation}
\label{sec:nre}

Our primary goal is population-level inference through the combination of multiple individual lens likelihoods in a population: therefore, NPE is not an option.
We also seek the simplest model (i.e., the least amount of parameters and the simplest architecture): therefore, NLE is not the best option.
NRE remains the best viable approach for this task.
Next, we describe the theory and procedure for NRE in the context of cosmology, including population-level likelihood ratios and posterior sampling.
We also present a novel post hoc calibration approach to mitigate model overconfidence.

\subsection{Population-level NRE for Cosmological Parameters}
\label{sec:nre_basics}

By way of Bayes' Theorem, we write the posterior of the parameter $\eta$ as
\begin{align}
p(\eta|x) & = \frac{p(x, \eta)}{p(x)} \\
          & = r(x \mid \eta) p(\eta), 
\end{align}
\noindent where $x$ is the data, $p(x, \eta)$ is the joint distribution, $p(x)$ is the evidence; and $r(x | \eta)$ is the likelihood-to-evidence ratio (LR) and is equivalent to the ratio between the joint distribution and the marginal distribution ~\citep[][]{cranmer15}:
\begin{align}
r(x|\eta) & = \frac{p(x, \eta)}{p(x) p(\eta)}.
\end{align}
An optimal classifier may be written in terms of the joint distribution and the marginal distribution and therefore the LR:
\begin{align}
d^{*}(x, \eta) &= \frac{p(x, \eta)}{p(x, \eta) + p(x) p(\eta) } \\
               &= \frac{r(x | \eta)}{1 + r(x | \eta)},
\end{align}
where joint distribution samples have class label $y=1$ and marginal distribution samples have class label $y=0$.
Then, NRE may be constituted by writing the LR in terms of the optimal classifier~\citep{hermans20}:
\begin{align}
r(x | \eta) &= \frac{d^{*}(x, \eta)}{1 - d^{*}(x, \eta)}. 
\end{align}

In this study, the input data are the lens image $x$ and the parameters $\zeta$, $w, \om{}$.
The images are generated through the implicit likelihood of the parameters: $x \sim\ p(x | w, \om{}, \zeta, \nu)$ (see Sec.~\ref{sec:data}).
Network training marginalizes over the nuisance parameters $\nu$.
Therefore, an NRE-derived LR for a single lensing object is
\begin{align}
    r(x, \zeta | w, \om{}) &= \frac{d^{*}(x, \zeta, w, \om{})}{1 - d^{*}(x, \zeta, w, \om{})}, 
    \label{eqn:ratio}
\end{align}
where $\zeta$ is considered known (without uncertainty) for each lens, from other observations.
The joint distribution $p(x, \zeta, w, \om{})$ samples are labeled as class $y=1$ and the marginal distribution $p(x, \zeta) p(w) p(\om{})$ samples are labeled as class $y=0$.

The population-level LR for $\{x, \zeta\}$ can be obtained by combining the LR's from individual lenses under the assumption that the observations are i.i.d.: 
\begin{equation} 
    \label{eqn:joint_likelihood}
    \ln\ r(\{x, \zeta\} | w, \om{} ) = \sum_{i} \ln\  r(x_i \zeta_i| w, \om{} ).
\end{equation}

\begin{figure*}[!ht]
 \centering
    \includegraphics[width=1.12\linewidth, trim={0cm 9cm 1cm 2cm},clip]{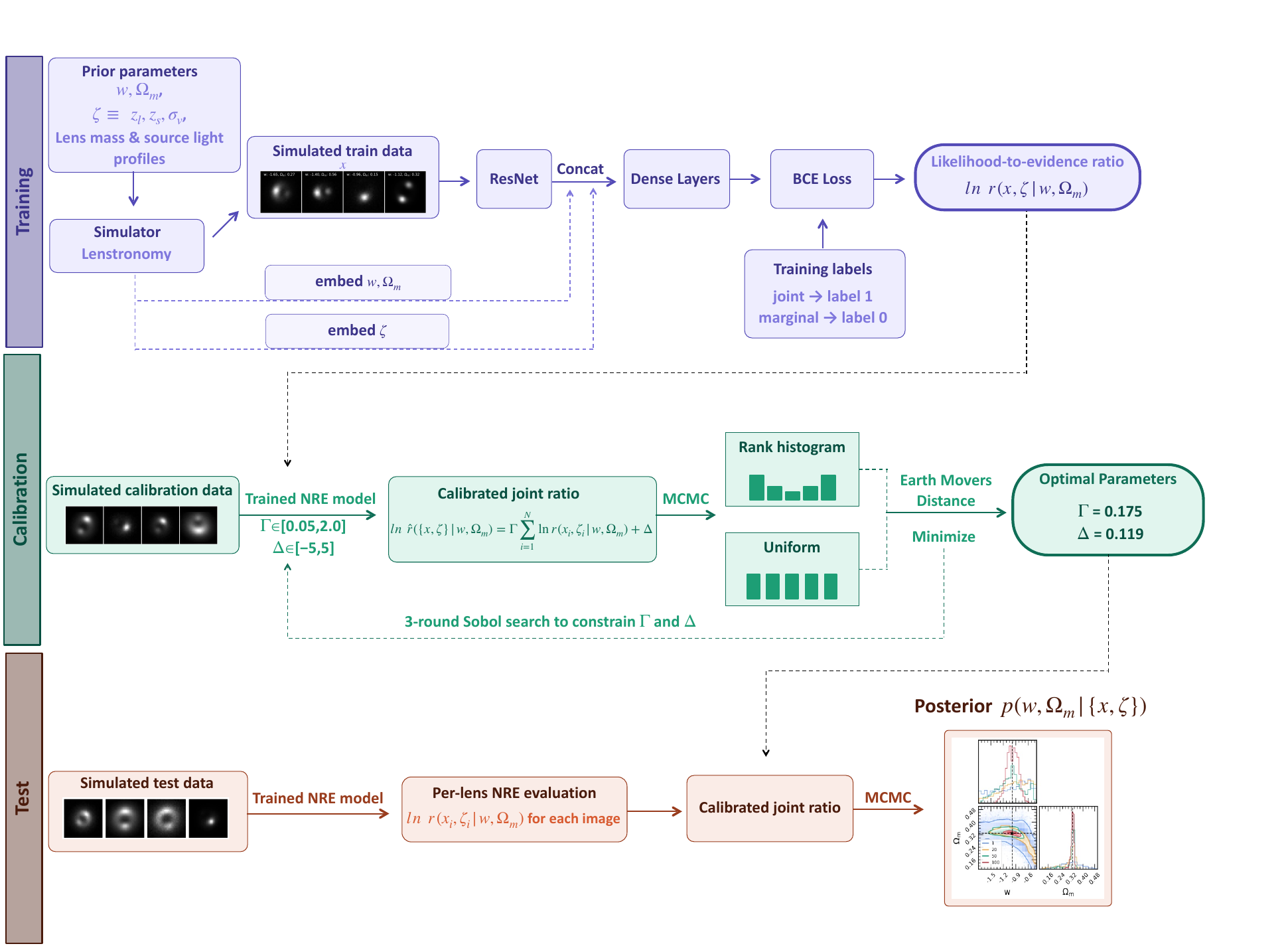}
    \caption{Workflow diagram of the NRE pipeline, including calibration.
    Top (Training): Simulated images and the corresponding cosmological parameters ($w$, $\om{}$) and astrophysics parameters ($\zeta \equiv\ \zl{}, \zs{}, \sigmav{}$) are inputs for training NRE to generate the LR $\ln r(x, \zeta \mid w, \om{}$). 
    The model architecture and training are discussed in Sec.~\ref{sec:network} and Appendix~\ref{app:network_summary}. 
    Middle (Calibration): A three-round Sobol search over scaling parameters $\Gamma$ and $\Delta$ minimizes the Earth Mover's Distance between posterior rank histograms and a uniform distribution. 
    This yields a calibrated population-level log-LR $\ln \hat{r}(\{x, \zeta\} \mid w, \om{}) = \Gamma \sum_i \ln r(x_i, \zeta_{i} \mid w, \om{}) + \Delta$. 
    The calibration method is detailed in Sec.~\ref{sec:nre_calibration}.
    Bottom (Test): The per-lens log-ratios are obtained from the trained NRE model. 
    The calibration is applied to the summed log-likelihoods and MCMC-sampled to recover the posterior $p(w, \om{} \mid \{x, \zeta\})$. The MCMC sampling is described in Sec.~\ref{sec:nre_basics}.
    }
    \label{fig:nre_pipeline}
\end{figure*}

Finally, we use MCMC sampling to infer posteriors $p(w, \om{} | \{x, \zeta\})$ from the population-level log-LR.
MCMC sampling with the Metropolis-Hastings algorithm~\citep{hastings70} was implemented with \texttt{emcee}~\citep{foreman13}.  
We initiate five walkers randomly around a starting value of $w=-1.5$ and $\om{}=0.31$. 
We implement a warm-up period with 200 steps, where the walkers converge to the target distribution. 
The position of the walkers is reset to the value that was converged on after each warm-up. 
After the warm-up, we run the sampling for 1000 steps.
The prior range for $w$ and $\om{}$ values is the same as the training prior.
These raw posteriors are not yet calibrated to account for the expected model overconfidence.
 
\subsection{Simulation-based calibration} 
\label{sec:sbc}

Simulation-based Calibration (SBC; \citealt{talts18}) methods are used to diagnose  calibration performance.
A well-calibrated posterior distribution covers the true value $x\%$ of the time within $x\%$ of its posterior probability volume. 
Additionally, the rank statistic of the prior sample relative to well-calibrated posterior samples is uniformly distributed. 
However, when the rank histogram is $\cup$-shaped, the model posterior is narrower than the true posterior and underestimates the uncertainty, indicating an over-confident model. 
Alternatively, when the histogram is $\cap$-shaped, the model posterior is broader than the true posterior, indicating an under-confident model.

SBI models are generally susceptible to overconfidence, where the neural network produces a posterior distribution that is artificially narrow and can exclude the true physical parameters.
This can lead to catastrophic model misspecification, which itself can lead to false rejection of theoretical models or the misinterpretation of systematic errors as new physics~\citep{hermans21}.
We use SBC to diagnose our models and to design a method for post hoc posterior calibration, which aims to mitigate SBI model overconfidence.

\subsection{Post hoc calibration} 
\label{sec:nre_calibration}

Algorithms like Balanced Neural Ratio Estimation (BNRE), which enforces a balancing condition on the binary neural classifier, have been shown to yield more conservative and reliable posteriors~\citep{delaunoy22}. 
There is also empirical evidence that an ensemble of models can produce more reliable posterior approximations~\citep{hermans21}. 
In our preliminary experiments, we encountered model overconfidence that could not be addressed with increased training set sizes, longer training runs, or more refined input image data.
In our attempts to use BNRE, the hyperparemter used to enforce the balance required arbitrary and non-data-driven choices -- and remained insufficient for addressing model overconfidence.
We didn't attempt ensemble modeling because it would be too computationally intensive.

Therefore, to address model overconfidence, we applied a post hoc calibration that aims to broaden posteriors across the $w$-$\om{}$ plane.
We define the calibrated LR for a population of $N$ lenses as a linearly scaled joint likelihood ratio:
\begin{equation} 
\label{eqn:calibration}
    \ln \hat{r}(\{x, \zeta\} | w, \om{}) = \Gamma \sum_{i=1}^{N} \ln r(x_i, \zeta_{i} | w, \om{}) + \Delta,
\end{equation}
\noindent where $\Gamma$ acts as a temperature-like scaling factor that adjusts the width (sharpness) of the posterior, and $\Delta$ is a global normalization offset that corrects for systematic biases in the population. 

The optimal $\Gamma$ and $\Delta$ values will yield a rank histogram close to a uniform distribution.
Scalar metrics like $\chi^2$ and the Kolmogorov-Smirnov statistic can be used to measure similarities between distributions. 
However, they estimate maximum deviation or bin-by-bin residuals and are not sufficiently sensitive to the full shape of distributions. 
Instead, we seek the posterior and the uniform distribution to have close agreement across their full support, not just at a single point or quantile.
Therefore, we use the Earth Mover's Distance ~\citep[EMD;][]{Givens1984ACO} --- aka, the Wasserstein distance --- to estimate the similarity between the posterior rank histogram and the uniform distribution. 
EMD quantifies the discrepancy between the two distributions as the minimum cost required to transform one distribution into the other, where cost is defined as the amount of probability mass moved multiplied by the distance it is transported.
A small EMD value means that the two distributions are very similar in shape and location: only a small amount of probability mass needs to be transported, and over short distances, to transform one distribution into the other.

To find optimal values of $\Gamma$ and $\Delta$, our search strategy starts with a Sobol sequence, a quasi-random sampling method designed to cover the $\Gamma$-$\Delta$ parameter space more uniformly than standard pseudo-random sampling.
We adopt broad prior ranges to ensure that the search space encompasses all physically plausible calibration states without truncation at the boundary. 
These parameters apply a single global correction to the entire prior, so the dataset only needs to be large enough to provide a reliable estimate of the average calibration needed across the full prior space. 
A much larger calibration dataset would be needed if, instead of a single global correction, one required a spatially varying correction that adapts to different Sub-Regions of the prior separately.
Ideally, one would use a calibration dataset of the same size as the test dataset (100 cosmologies), but the Sobol search is computationally expensive because each candidate $(\Gamma, \Delta)$ pair requires running MCMC chains to construct rank histograms.
We use 50 cosmologies as a practical compromise. 
To find the optimal calibration parameters, we use a held-out calibration dataset with the parameters sampled from the entire training prior of cosmological parameters.

The following algorithm describes the iterative optimization procedure: 

\renewcommand{\algorithmicrequire}{\textbf{Input:}}
\renewcommand{\algorithmicensure}{\textbf{Output:}}

\clearpage
\begin{algorithm}[H]
\begin{algorithmic}
\REQUIRE Trained NRE model (Eqn.~\ref{eqn:ratio}); calibration dataset.
\ENSURE Optimal parameters $\Gamma$ and $\Delta$.
\STATE \textbf{Objective Function Definition:}
\begin{itemize}
    \item Calibrated log-LR (Eqn.~\ref{eqn:calibration}).
    \item EMD post hoc calibration loss function $L(\Gamma, \Delta)$.
\end{itemize}
\STATE \textbf{Round 1: Global Coarse Search}
\begin{itemize}
    \item Sobol sequence: sample $\Gamma \in [0.05, 2.0]$, $\Delta \in [-5.0, 5.0]$. 
    \item For each parameter pair, sample log-LR posteriors (20 lenses per cosmology pair for 50 cosmology pairs, 4 walkers, 10 burn-in, 100 steps ).
    \item Calculate rank histograms from posterior samples.
    \item Calculate $L(\Gamma, \Delta)$ for parameter pairs.
    \item Identify top five parameter pairs for $L(\Gamma, \Delta) < 0.3$.
\end{itemize}
\STATE \textbf{Round 2: Intermediate Refinement}
\begin{itemize}
    \item Reduce Round 1 search space by a factor of 0.4.
    \item Sample posteriors for each parameter pair (50 lenses  per cosmology pair for 50 cosmology pairs, 4 walkers, 20 burn-in, 120 steps).
    \item Calculate rank histograms from posterior samples.
    \item Calculate $L(\Gamma, \Delta)$ for parameter pairs.
    \item Identify top 5 pairs for $L(\Gamma, \Delta) < 0.3$.
\end{itemize}
\STATE \textbf{Round 3: Fine Optimization}
\begin{itemize}
    \item Reduce Round 2 search space by factor of 0.4 for finer sampling.
    \item Sample posteriors (100 lenses per cosmology pair for 50 cosmology pairs, 4 walkers, 30 burn-in, 200 steps).
    \item Calculate rank histograms from posterior samples.
    \item Calculate $L(\Gamma, \Delta)$ for parameter pairs.
    \item Result: Optimal parameters at $(\Gamma, \Delta) = (0.175, 0.119$).
\end{itemize}
\end{algorithmic}
\end{algorithm}

For each Round, the sampling count regards lenses per pair of cosmological parameters.
Posterior sampling is performed using MCMC.
We apply the optimal values we found $(\Gamma, \Delta) = (0.175, 0.119)$ to the LR during parameter estimation for both the Entire Test Region and individual Sub-Regions. 

\subsection{Model Architecture and Training} 
\label{sec:network}

We use a ResNet architecture~\citep{he15}, implemented in \texttt{TensorFlow}~\citep{tensorflow2015-whitepaper}.
Details of the model architecture are shown in Table~\ref{tab:network_summary} in Appendix \ref{app:network_summary}.
The inputs are images and their cosmology and astrophysics parameters. 
Samples from the marginal distribution $p(x, \zeta)p(w)p(\om{})$ are generated by pairing each image with randomized values of $w$ and $\om{}$.
Samples from the joint distribution $p(x, \zeta, w, \om{})$ retain the true generating cosmology parameters. 
The network outputs a scalar for the logarithm of the LR as detailed in Sec.~\ref{sec:nre_basics} and shown by Eqn.~\ref{eqn:ratio}.
Samples are drawn at the batch level because randomizing at the batch level increases the diversity of the samples from the marginal distribution.

The image input is first passed through batch normalization followed by a convolutional layer with a ReLU activation, serving as the entry block of a residual network. 
The feature maps are then processed by three residual convolutional blocks with increasing channel width. 
Each block consists of two $3\times3$ convolutional layers, each followed by batch normalization and ReLU activation, and a max-pooling layer for spatial downsampling. 
Residual connections are implemented using projection shortcuts via $1\times1$ convolutions with matching stride, allowing the network to preserve low-level spatial information while enabling deeper feature extraction. 
Following the residual blocks, a separable convolutional layer is applied to increase representational capacity at reduced computational cost. 
Global average pooling is then used to collapse the spatial dimensions, yielding a fixed-length one-dimensional feature vector that summarizes the image content.

The astrophysical parameters $\zeta$ and the cosmological parameters $(w, \om{})$ are provided to the network as separate inputs. 
The astrophysical parameters are assumed to be known perfectly (i.e., no error bars) during inference and are provided to the network to reduce the degeneracy between the Einstein radius and the parameters from Eqn.~\ref{eqn:einstein_radius}. 
Each of the astrophysics and cosmology parameter vectors is independently normalized using batch normalization and passed through a dense embedding layer to project the parameters into a higher-dimensional latent feature space. 
The embedded astrophysical features and cosmological features are concatenated with the image feature vector after global average pooling. 
The concatenated feature vector is processed by a sequence of fully connected layers with batch normalization and ReLU activations. 
L2 weight regularization and dropout factor of 0.1 are applied to reduce overfitting and improve generalization. 
The final linear output layer produces a single scalar corresponding to the logarithmic density ratio.

The model is trained over 100 epochs with a batch size of 1024 using \texttt{Adam} optimizer. 
The learning rate is dynamically adjusted: when the validation loss is plateaued by a decay factor of 0.1, starting from $1e^{-2}$ to $1e^{-6}$, if the validation loss does not improve over five epochs. 
We use early stopping if the validation loss does not improve over 20 epochs. 
The model is trained on NVIDIA A100 GPU with $\sim$ 2M lens images for 80 epochs with a typical training time of $\sim$ three hours.

\section{Results: Cosmological Constraints}
\label{sec:results}

We describe the NRE model performance in producing cosmological constraints from individual lenses and populations of lenses. 
In each scenario, we compare performance without and with the post hoc calibration procedure.

\begin{figure*}[!ht]
 \centering
    \includegraphics[width=0.9\linewidth]{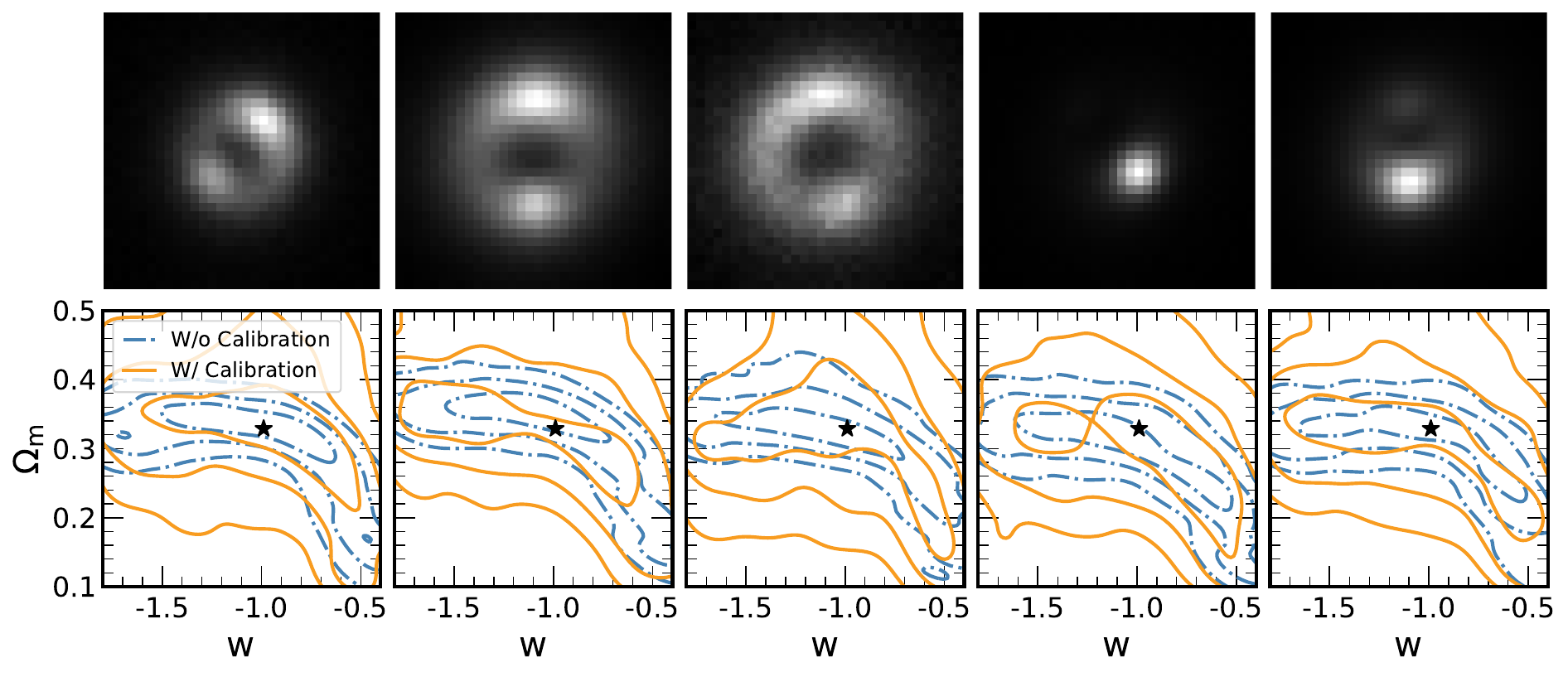}
    \caption{Examples of lensing system images and NRE-based cosmological inferences from those images. 
    All lenses have the same cosmology $w = -0.99$ and $\om{}= 0.33$, but different astrophysical parameters.
    Top: Images of lensing systems, where the lens galaxies have been removed.  
    Bottom: Joint posterior from NRE without calibration (blue, dot-dashed) and with calibration (orange, solid). The contours represent 68th, 95th, and 98th percentile. The true cosmological parameters are denoted with a star.
    \label{fig:image_posterior}
    }
\end{figure*}

\subsection{Constraints from Individual Lenses}
\label{results:individual_lens}

Each lensing system is simulated using different astrophysical parameters but the same cosmology parameters ($w = -0.99$ and $\om{}= 0.33$).
Fig.~\ref{fig:image_posterior} shows images and posteriors obtained from NRE with and without the post hoc calibration. 
With calibration, the posteriors broaden significantly for $\om{}$ and less so for $w$. 
With or without calibration, the capacity for individual lenses to constrain the cosmological parameters is relatively limited.

\begin{figure}[!ht]
\centering
\includegraphics[trim={0.34cm 0 0.5cm 0},clip, width=0.5\linewidth]{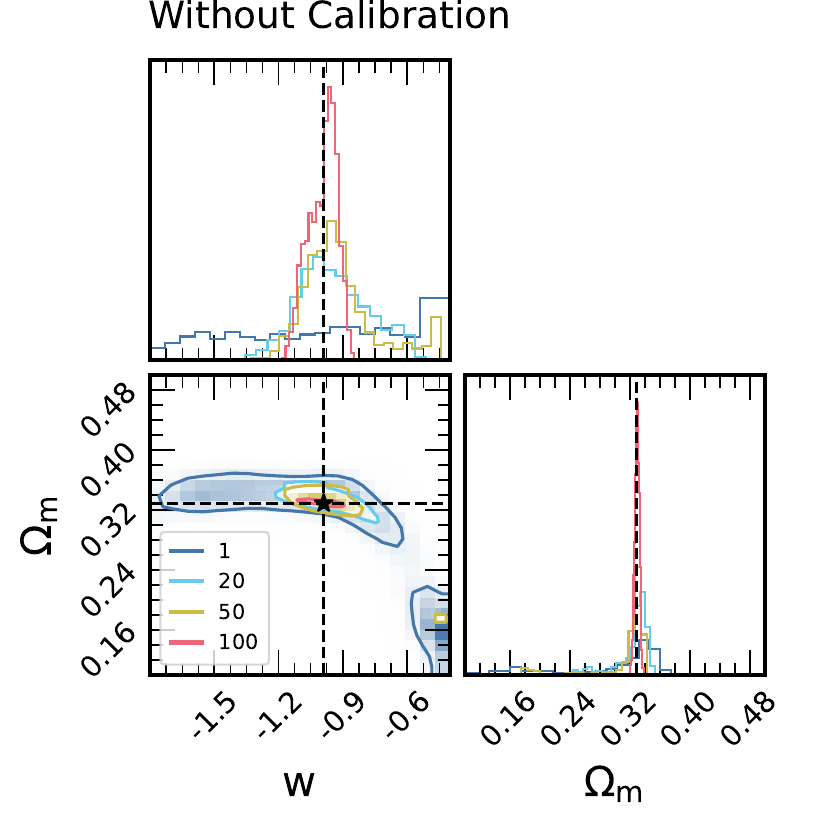}\hfill
\includegraphics[trim={0.34cm 0 0.5cm 0},clip,width=0.5\linewidth]{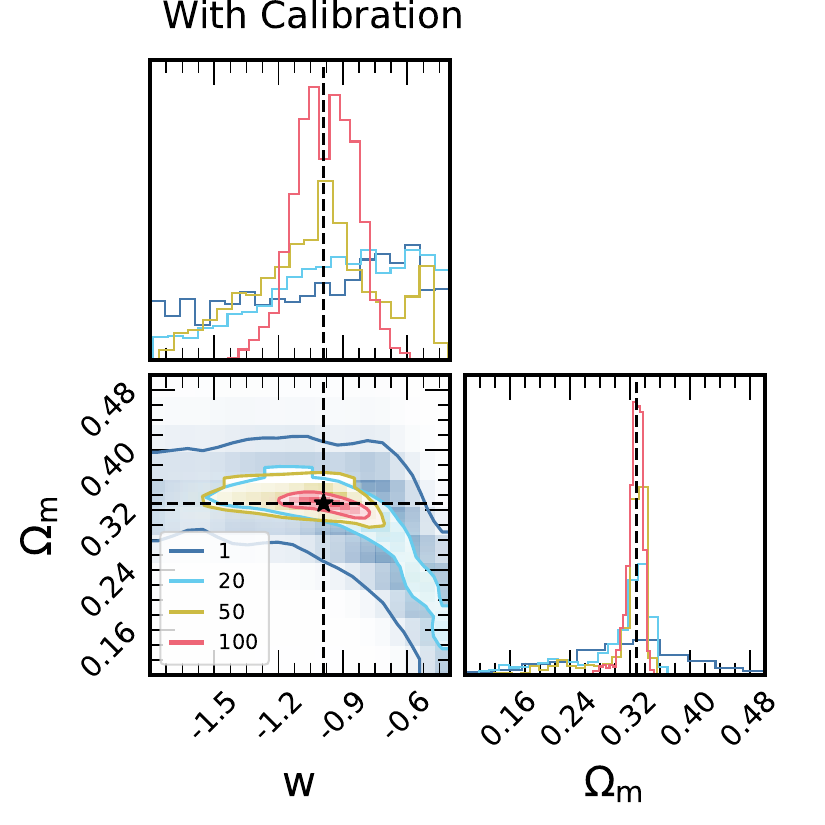}

\vspace{0.3cm}
\includegraphics[width=\linewidth]{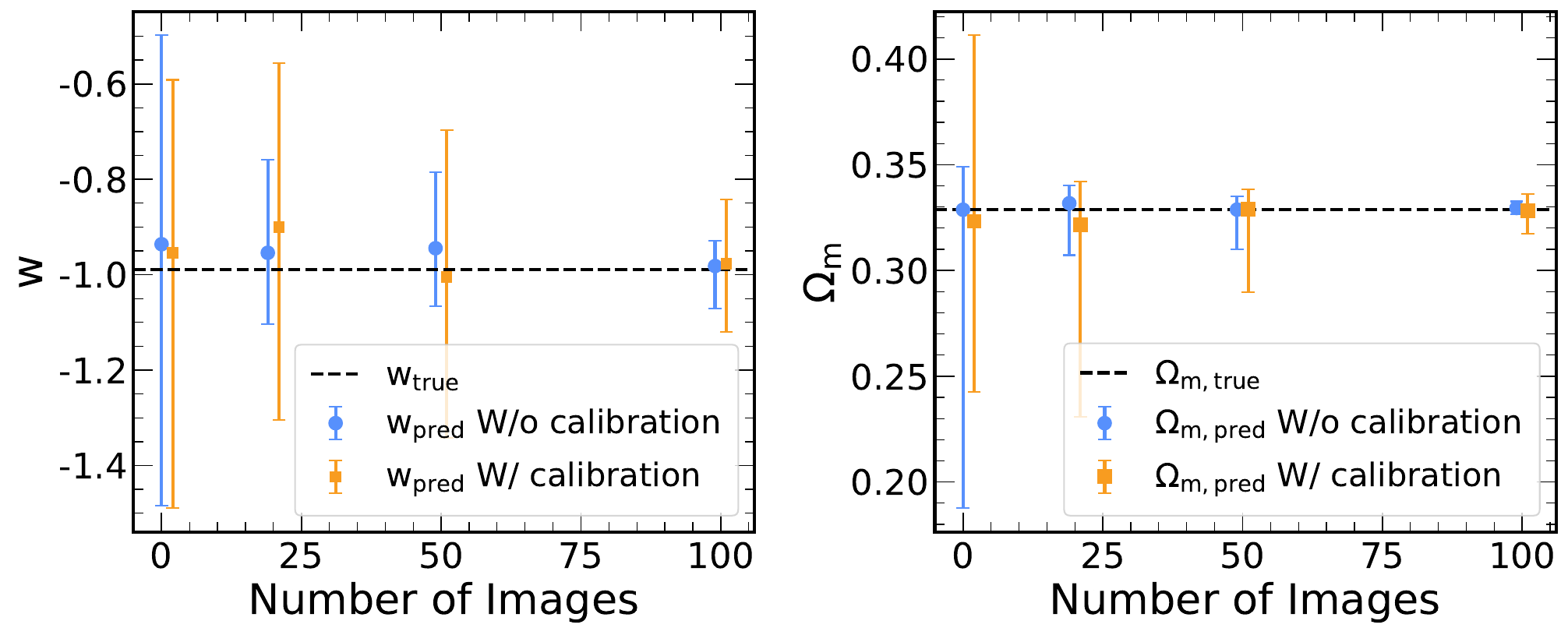}
\caption{Top: Population-level inference of cosmology parameters without calibration (left) and with calibration (right). The contours are at the 68th confidence percentile.
    Inference is performed for one (blue), 20 (cyan), 50 (yellow), and 100 (red) lenses.
    Bottom : Medians of predicted parameter posteriors of $w$ and $\om{}$ with 68th percentile intervals as a function of the number of images in the test set.
    \label{fig:single_joint_posterior}
}
\end{figure}

\subsection{Constraints from Lens Populations}
\label{sec:cosmo_constraints} 

We describe the cosmology constraints from NRE-based population-level inference  with multiple diagnostics: corner plots, parity (1-1) plots, rank histograms, posterior coverage, and the Earth Mover's Distance (EMD).
Unless otherwise stated, these analyses are performed on data from the Entire Test Region (see Fig.~\ref{fig:w_om_regions}).
The main results are discussed in this section, and additional details from each Sub-Region are discussed in Appendix~\ref{app_results}.

\subsubsection{Basic Diagnostics}
\label{sec:basic_diagnostic}

First, we study the cosmological constraints as function of increasing test set size -- one, 20, 50, and 100 lenses.
This relates most directly to the main hypothesis of this work: populations of lenses can provide constraints on cosmological parameters.
Each test set has the same true cosmological parameters ($w$ = -0.99 and $\om{} = 0.33$).
Objects in each test set are re-sampled from the astrophysical parameter prior distributions.
Several trends appear in Fig.~\ref{fig:single_joint_posterior}.
For a single lens, regardless of calibration status, NRE provides no constraint on $w$, but a small constraint on $\om{}$.
Then, as the test set size increases, the posterior widths and error bars on both cosmological parameters decrease.
For all test set sizes, the posterior widths are wider with calibration than without calibration. 

\begin{figure*}[!ht]
 \centering
    \includegraphics[width=1.0\linewidth]{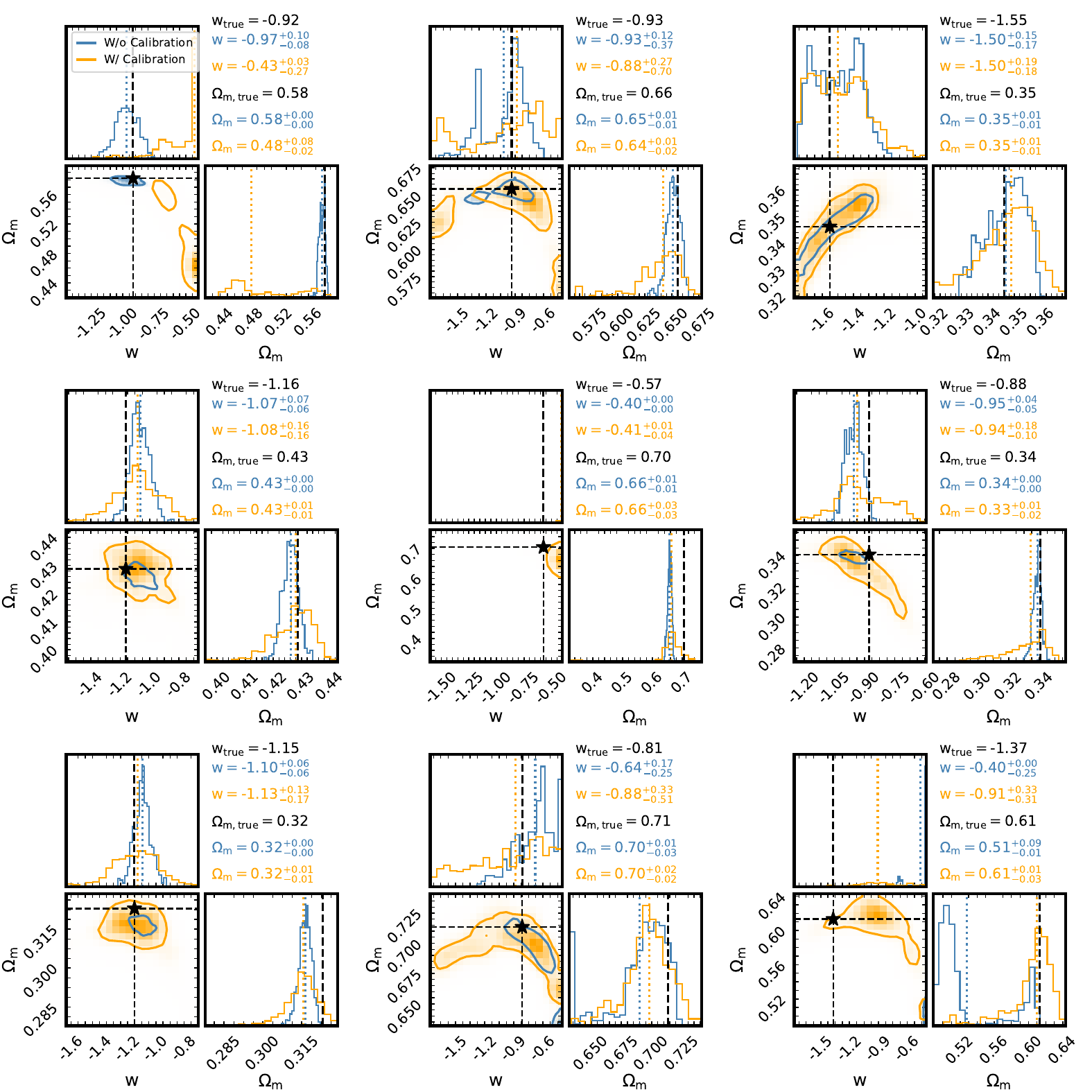}
    \caption{NRE-based posteriors -- without calibration (blue) and with calibration (orange) of nine pairs of cosmological parameters randomly selected from the Entire Test Region.
    Each example uses 100 objects with the same true cosmology parameters (black star). 
    The contours in the joint posteriors represent the 68th percentile.  
    The histograms represent the marginal posterior and show the median (dotted line) and the true value (dashed black line).
    \label{fig:joint_posterior_entireregion} 
   }
\end{figure*}

Second, we review posteriors of $w$ and $\om{}$ from a random sample of nine cosmology parameters. 
Each panel in Fig.~\ref{fig:joint_posterior_entireregion} shows the posterior obtained from population-level inference on 100 lenses -- each set has a different true cosmological parameter pair. 
The posteriors without calibration place the true cosmology outside the 68th percentile interval in five of the samples.
The posteriors with calibration are broader and place three of the samples outside the 68th percentile interval. 
Appendix~\ref{app_results} shows how the model performance varies across Sub-Regions in the cosmological parameter space: the inclusion of the true cosmology in the 68th percentile interval (without or with calibration) varies by Sub-Region.

\begin{figure*}[!ht]
 \centering
    \includegraphics[width=0.9\linewidth]{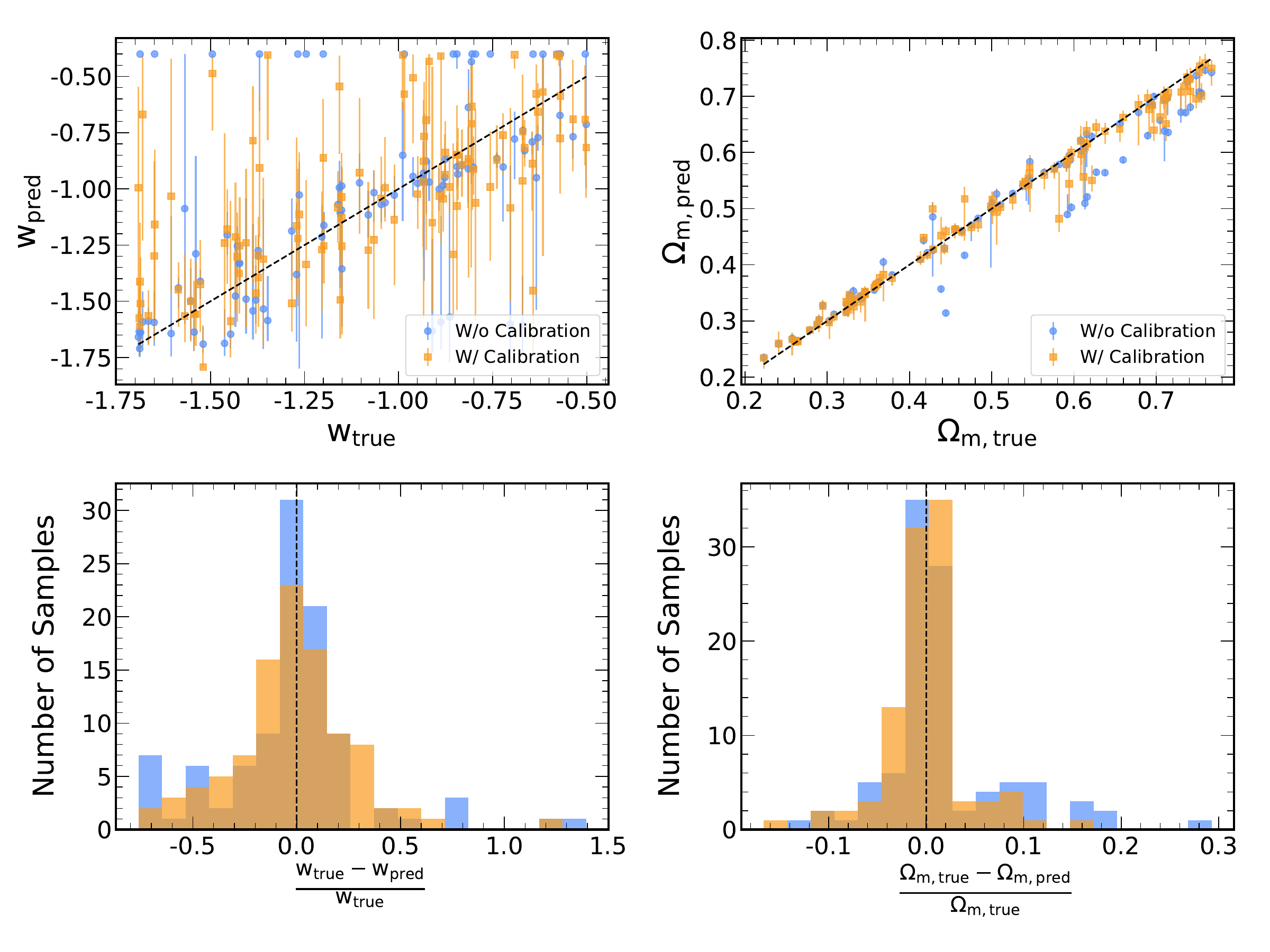}
    \caption{Top: True and predicted median values of $w$ (left) and $\om{}$ (right) from population-level posteriors for 100 cosmologies, with 100 lenses per cosmology. 
    The error bars represent the 16th and 84th percentile values.
    Bottom: The residuals of the predicted median values of the posteriors. 
    Posteriors without calibration are shown in blue, and those with calibration are shown in orange.
    \label{fig:parity_entireregion}
   }
\end{figure*}

Third, we study the model accuracy and outliers.
The top panels of Fig.~\ref{fig:parity_entireregion} present the parity of the predicted values.
Without calibration, most median values for $w$ are near parity (the one-to-one line).
Some values are far above the parity line and have the same value ($w\sim -0.4$) -- the edge of the MCMC sampling prior; most of these values have very small error bars.
With calibration, there are only a few such outliers, and they have larger error bars.
Without calibration, most median values for $\om{}$ are near parity. 
Several values lie systematically below the parity line, and most have negligible error bars. 
With calibration, fewer points for $\om{}$ lie below parity.
The bottom panels show the residuals of the median values for each parameter of the predicted posteriors with and without calibration.
The residuals corroborate the interpretation of the parity plots, and they show more succinctly the quantity and extremity of the outliers. For each of the 100 test cosmologies $i$ from the Entire Test Region, we
summarise the 1D marginal posterior of a parameter by its median $m_{i}$ and
the half-width of its 68\% credible interval,
$\sigma_{i} = (q^{84}_{i} - q^{16}_{i})/2$, where $q^{16}_{i}$ and
$q^{84}_{i}$ are the 16th and 84th percentiles of the marginal posterior for
cosmology $i$ (Fig.~\ref{fig:parity_entireregion}). We define the fractional uncertainty as $\sigma_{i}/|m_{i}|$.
Taking the median of this quantity across the 100 test cosmologies,
$\mathrm{median}(\sigma_{i}/|m_{i}|)$ yields $22.8\%$ in $w$ and $2.9\%$ in
$\om{}$. The calibrated NRE therefore constrains $\om{}$ roughly eight times more
tightly than $w$, indicating that the matter density is well determined while
the dark-energy equation of state is comparatively poorly constrained, as
expected given the weaker sensitivity of the data to $w$. We note that these values quantify precision only; the reliability of the corresponding credible intervals is established by the posterior
coverage discussed in Section~\ref{sec:posterior_coverage_calibration}.

\subsubsection{Posterior Coverage Calibration}
\label{sec:posterior_coverage_calibration}

\begin{figure*}[!ht]
 \centering
    \includegraphics[width=1\linewidth]{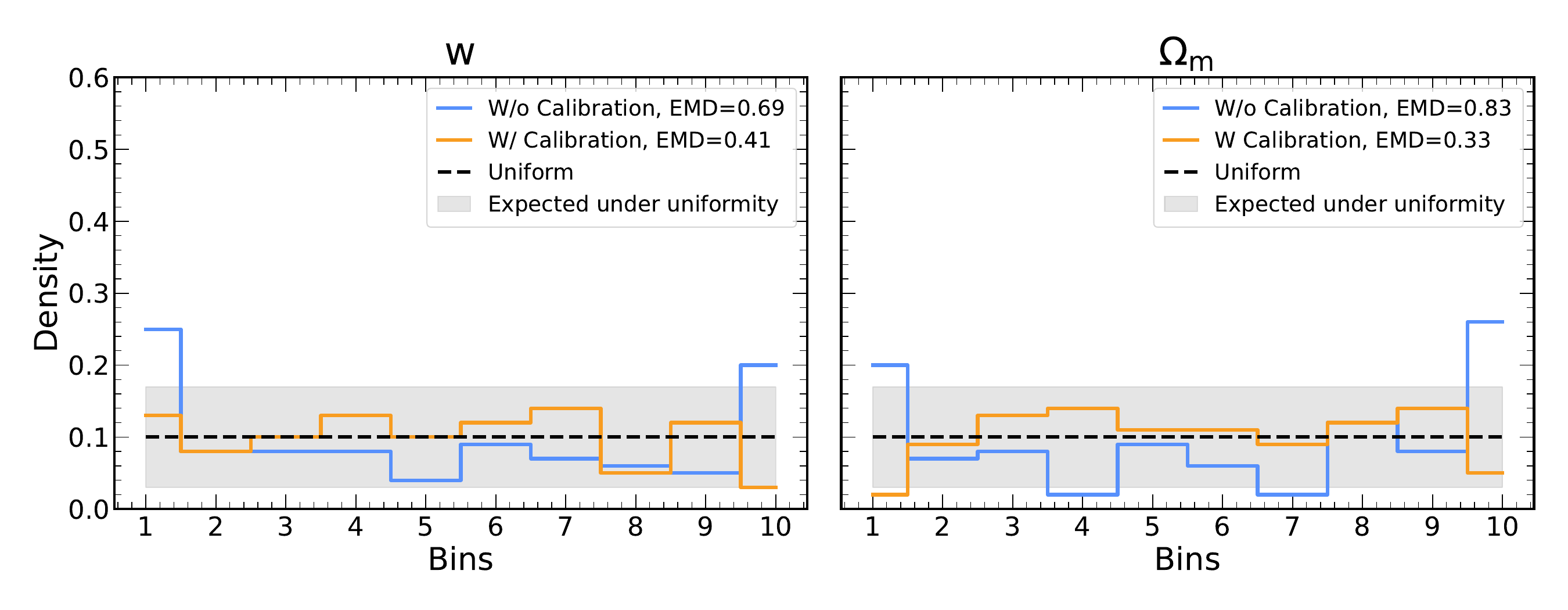}
    \caption{Rank histograms of the posteriors of $w$ (left) and $\om{}$ (right) without (blue) and with (orange) calibration.
    The uniform distribution is shown with a black dashed line.
    The 99\%-confidence interval is shown in gray.
    EMD values without and with calibration are shown in the legend.
   }
\label{fig:rank_histogram_entire_region}
\end{figure*}

\begin{figure}[!htbp]
 \centering
    \includegraphics[width=\linewidth]{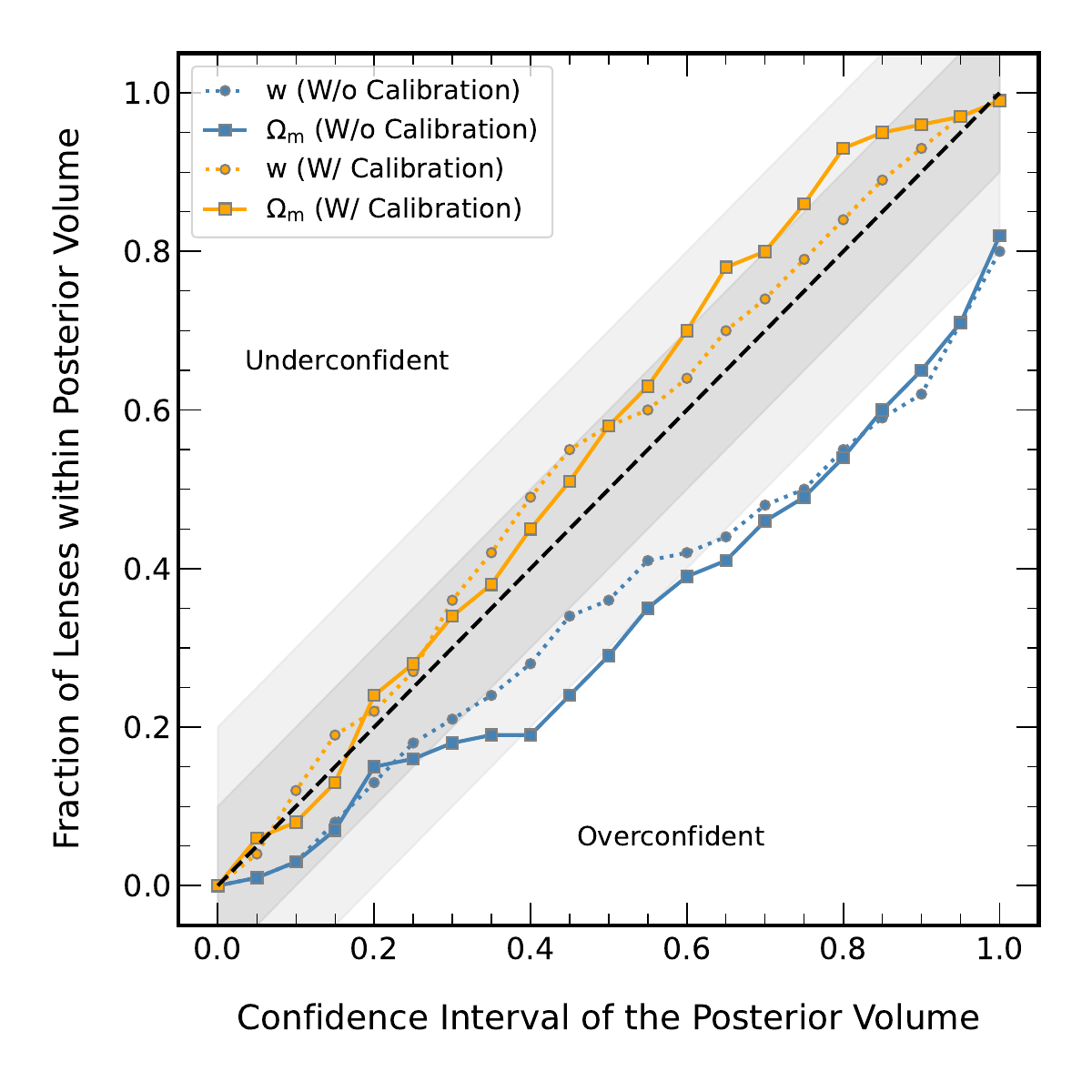}
    \caption{The posterior coverage from 100 cosmologies sampled from the Entire Test Region, with 100 lenses per cosmology. 
   Blue (orange) curves show the model performance without (with) calibration. 
   The dark- and light-gray shaded bands around the diagonal denote $10\%$ and $20\%$ thresholds, respectively. 
   \label{fig:posteror_coverage_entireregion}
   } 
\end{figure}

We study the NRE model calibration with diagnostics from SBC, which centers on the posterior coverage and which is the primary arbiter of model efficacy for this work.
First, Fig.~\ref{fig:rank_histogram_entire_region} shows the rank histograms of the posterior samples without and with calibration (blue and orange, respectively).
The histogram without calibration is $\cup$-shaped, indicating over-confident posteriors; the model with calibration is closer to uniform.
We use the EMD to examine the difference between the posterior histograms and the uniform distribution for models without and with calibration (Table~\ref{tab:emd_regions}).
With calibration, the EMD is lower, indicating that the model produces posteriors closer to the true posterior distribution.
Second, for each parameter, we plot the posterior coverage -- the fraction of the lenses whose true parameter values fall within a confidence interval as a function of the confidence interval (see Fig.~\ref{fig:posteror_coverage_entireregion}). 
The model without calibration produces over-confident posteriors, and the calibration procedure partially mitigates the overconfidence.
The calibrated curves are mostly within $10\%$ tolerance.

We also reviewed individual Sub-Regions without and with calibration.
In all Sub-Regions, the post-hoc calibration broadens the posteriors, yielding more conservative uncertainty estimates. 
In most of the Sub-Regions, there are no more major outliers, the posterior coverage is improved, and the EMD is lower with calibration (see Table~\ref{tab:emd_regions}): the model is no long catastrophically misspecified.
However, in some Sub-Regions, the EMD increases with calibration for one of the parameters (and in one of the Sub-Regions for both parameters).
A more detailed discussion of the Sub-Region results is presented in Appendix~\ref{app_results}.

\begin{table*}[ht]
\centering
\caption{Priors and Earth Mover’s Distance (EMD) for $w$ and $\om{}$ in the Entire Test Region and in each Sub-Region without (w/o) and with (w/) calibration. 
    \label{tab:emd_regions}
}
\begin{tabular}{lcccccc}
    \toprule
     &
    \multicolumn{2}{c}{Priors} &
    \multicolumn{2}{c}{$\mathrm{EMD}_{w}$} &
    \multicolumn{2}{c}{$\mathrm{EMD}_{\om{}}$} \\
    Region &
    $w$ &
    $\om{}$ &
    (w/o) &
    (w/) &
    (w/o) &
    (w/) \\
    \midrule
    Entire Region & $\mathcal{U}(-1.70,\,-0.50)$ & $\mathcal{U}(0.22,\;0.77)$ & 0.69 & 0.41 & 0.83 & 0.33 \\
    Sub-Region 1 & $\mathcal{U}(-1.70,\,-1.50)$ & $\mathcal{U}(0.22,\;0.38)$ & 1.80 & 1.13 & 1.81 & 1.07 \\
    Sub-Region 2 & $\mathcal{U}(-1.10,\,-0.90)$ & $\mathcal{U}(0.22,\;0.38)$ & 1.67 & 1.44 & 0.46 & 0.41 \\
    Sub-Region 3 & $\mathcal{U}(-0.70,\,-0.50)$ & $\mathcal{U}(0.22,\;0.38)$ & 3.19 & 1.44 & 3.19 & 1.35 \\
    Sub-Region 4 & $\mathcal{U}(-1.70,\,-1.50)$ & $\mathcal{U}(0.42,\;0.57)$ & 0.18 & 0.77 & 0.30 & 0.76 \\
    Sub-Region 5 & $\mathcal{U}(-1.10,\,-0.90)$ & $\mathcal{U}(0.42,\;0.57)$ & 0.57 & 1.00 & 1.96 & 1.08 \\
    Sub-Region 6 & $\mathcal{U}(-0.70,\,-0.50)$ & $\mathcal{U}(0.42,\;0.57)$ & 1.72 & 1.69 & 1.56 & 1.57 \\
    Sub-Region 7 & $\mathcal{U}(-1.70,\,-1.50)$ & $\mathcal{U}(0.62,\;0.77)$ & 1.71 & 1.70 & 1.44 & 1.70 \\
    Sub-Region 8 & $\mathcal{U}(-1.10,\,-0.90)$ & $\mathcal{U}(0.62,\;0.77)$ & 1.58 & 0.44 & 3.00 & 1.63 \\
    Sub-Region 9 & $\mathcal{U}(-1.70,\,-1.50)$ & $\mathcal{U}(0.62,\;0.77)$ & 0.81 & 1.31 & 1.41 & 0.41 \\
    \bottomrule
\end{tabular}
\end{table*}

\section{Discussion}
\label{sec:discussion}

We study the NRE model performance as a function of training set size, perform a comparison to analytic-based methods, and review the post hoc calibration consistency across Sub-Regions of the test parameter space.

\subsection{Effect of training set size}
\label{sec:train_size}

One possible explanation for the overconfidence observed in the uncalibrated NRE posteriors (Sec.~\ref{sec:cosmo_constraints}) is insufficient training data: a classifier trained on few examples may learn a decision boundary that does not generalize. 
We test this hypothesis by training four models on datasets of $\sim$500K, $\sim$1M, $\sim$1.5M, and $\sim$2M lens images. 
The lens, source, and cosmology parameters are drawn from the same prior distributions (Table~\ref{table:params}). 
The models are evaluated on the same held-out test dataset generated from the Entire Test Region prior range comprising 100 distinct cosmologies, with 100 lens images generated per cosmology.

\begin{figure*}[!ht]
 \centering
    \includegraphics[width=1.0\linewidth]{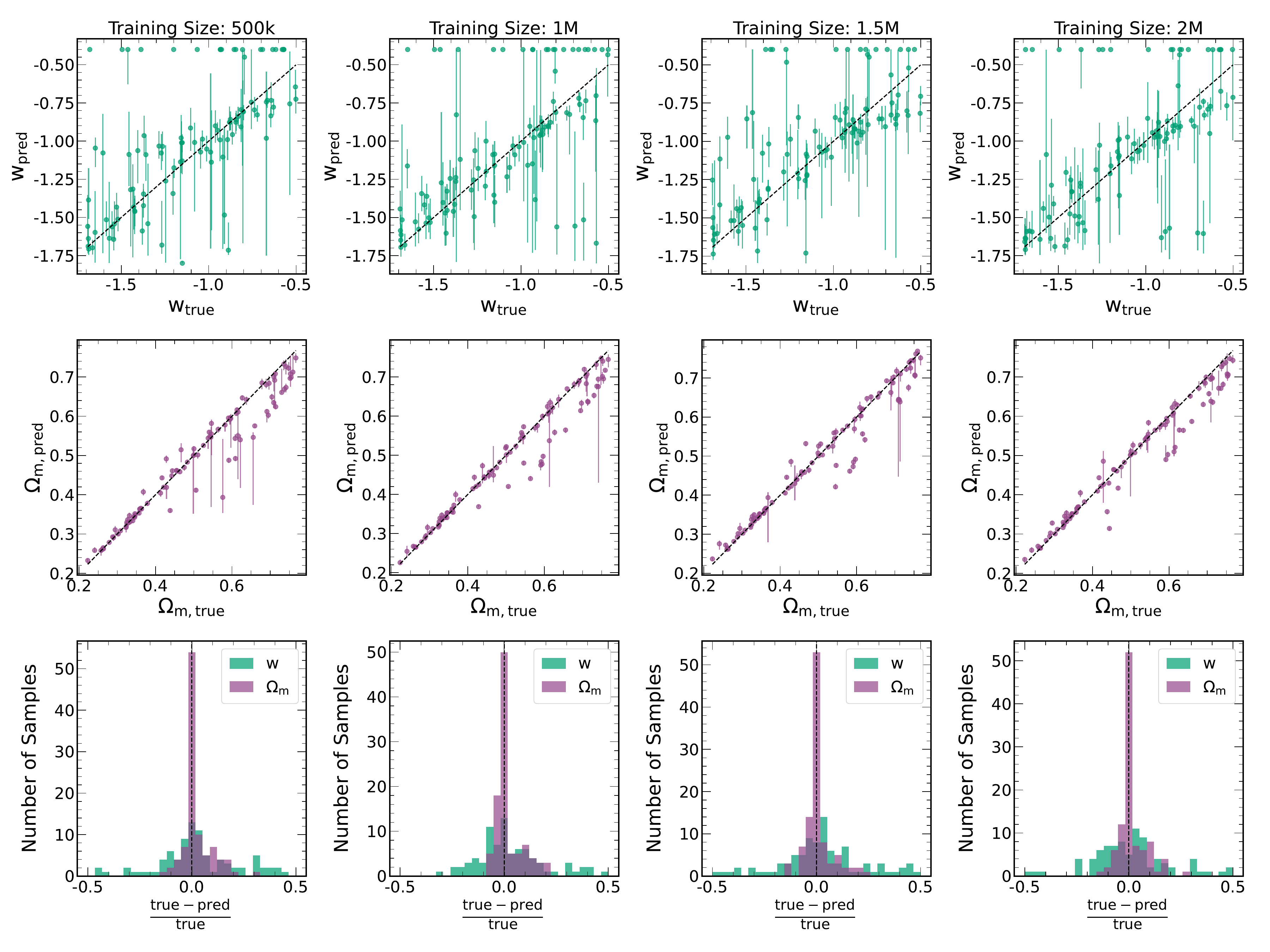}
    \caption{Model performance (before calibration) as a function of training set size (from left to right: 500K, 1M, 1.5M, and 2M). Top (green): parity (predicted vs. true) of the median value of the predicted posterior of $w$ for 100 cosmologies, each with 100 objects sampled from the Entire Test Region. The error bars indicate the 16th and 84th percentile intervals, centered on the median.
  Middle (violet): same as top row, but for $\om{}$. 
  Bottom: residuals of the median values of predicted posteriors for each cosmological parameter. 
  }
  \label{fig:compare_train_size}
\end{figure*}

Fig.~\ref{fig:compare_train_size} shows the predicted median and 68\% credible intervals for 100 cosmologies sampled from the Entire Test Region prior, along with the residuals, for each training set size. 
Increasing the training sample by a factor of four produces no detectable reduction in overconfidence, though we cannot rule out that increases of an order of magnitude or more may be required to see a significant effect. Post-hoc calibration therefore remains a practical requirement of the framework at the training set sizes explored here.

\subsection{Comparing NRE-based and analytic methods}
\label{sec:compare_analytic}

We compare our calibrated NRE likelihood with two implementations of the analytic approach -- one that we perform and one from the literature.
Generally, all methods follow the same population-level Bayesian framework where individual per-lens likelihoods are combined and the joint posterior over $w$ and $\om{}$ is sampled via MCMC (Sec.~\ref{sec:nre_basics}).

\begin{figure}[htbp]
    \centering
    \includegraphics[width=1\linewidth]{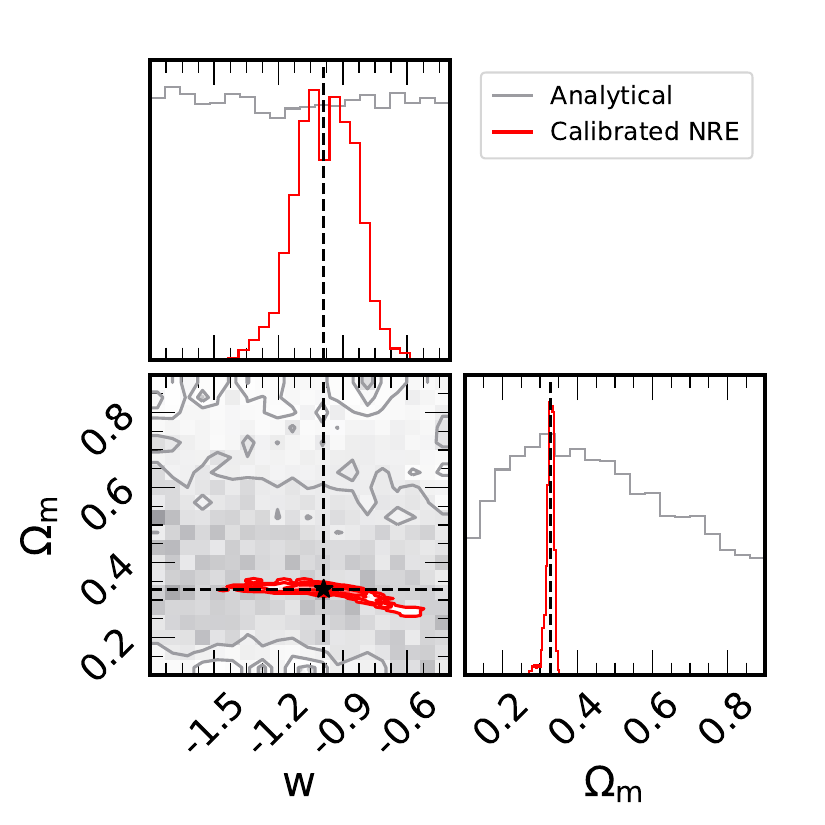}
    \caption{Comparison between the posteriors calculated from an analytic likelihood (gray) with those from an NRE-based likelihood (red). 
    True values for $w$ and $\om{}$ are marked with a dashed line.
    \label{fig:compare_analytic_nre}
    }
\end{figure}

The efficacy of traditional analytical likelihood methods is tied to the chosen model's  capacity to represent the physical complexities of the data. 
In our analytical implementation, we assume a simple SIE lens profile to estimate the Einstein radius, representing the lens image. 
Given a candidate cosmology $(w, \om{})$ and the known astrophysical parameters $\zeta_{i}$ for that lens, the predicted Einstein radius $\hat{\theta}_{E,i}$ is computed from Eqn.~\ref{eqn:einstein_radius}. 
The analytical log-likelihood for the population is then
\begin{equation}
   p(x| w,\om{})= \sum_{i=1}^{N}(\theta_{E,i} - \hat{\theta}_{E,i})^{2}.
\end{equation}
\noindent Fig.~\ref{fig:compare_analytic_nre} compares the resulting population-level posteriors. 
The calibrated NRE likelihood yields well-constrained posteriors for both $w$ and $\om{}$ (red curves).
The analytic approach fails to constrain $w$ and produces a broad constraint on $\om{}$ (gray curves). 

Alternatively, a more realistic lens profile may give a more accurate description of strong lens systems. 
In contrast to the SIE lens profile assumptions in this study, \citet{litian24} employ an analytic hierarchical Bayesian framework using a power-law mass profile and stellar luminosity to simultaneously constrain the lens population and cosmological parameters. 
They use a cosmology with true parameters  $w = -1.0$ and $\om{} = 0.3$.
From 10,000 mock observations, they recover $w = -0.98 \pm 0.11$ and $\om{} = 0.300 \pm 0.015$. 
Applying our work's calibrated NRE approach to 100 simulated lens images with true cosmology  $w = -0.99$ and $\om{} = 0.33$, we obtain $w = -1.01 \pm 0.39$ and $\om{} = 0.39 \pm 0.10$. 
We evaluate and compare the reliability of these predictions using the Z-score, 
which measures the distance between the predicted mean ($\hat{\theta_{\mu}}$) and the ground truth ($\theta$) in units of the predicted standard deviation ($\hat{\theta_{\sigma}}$):
$Z = \frac{\hat{\theta_{\mu}} - \theta}{\hat{\theta_{\sigma}}}$.
This is also considered as the number of standard deviations that the predicted mean lies from the true value. 
A Z-score close to 0 indicates high agreement, and values within $\pm$1 are considered good.
The analytical approach by \citet{litian24} yields $Z_{w} = 0.18$ and $Z_{\om{}} = 0.00$, and our NRE-based inference yields $Z_{w} = -0.05$ and $Z_{\om{}} = 0.60$. 
All scores fall within $\pm 1$, both methodologies are considered statistically reliable. 
The NRE model uses a sample size 100 times smaller than that required by \citet{litian24}. 

\subsection{Sub-Region calibration consistency}
\label{sec:calibration_discussion}

Our calibration procedure aims to minimize the EMD between posterior rank distributions and a uniform distribution by re-scaling the joint LR across the $w$-$\om{}$ parameter space. The re-scaling is determined by Eqn.~\ref{eqn:calibration}, where the parameters $\Gamma$ and $\Delta$ are optimized by a held-out calibration data set across the parameter space.
The resulting calibration produces posteriors across the Entire Test Region that are reduced from overconfident to moderately underconfident (see Fig.~\ref{fig:posteror_coverage_entireregion}). 
However, models in individual test Sub-Regions are affected differently: for example, while most models become either well-calibrated or underconfident for both cosmological parameters, some models are more overconfident in one parameter and slightly underconfident in another. 
Consider Sub-Region 5, where the coverage curve after calibration is S-shaped in $\om{}$ indicating that the posterior is overconfident in the centre and underconfident in the tails.
Similarly, in Sub-Regions 7 and 8, models remain significantly overconfident in $w$ and slightly underconfident in $\om{}$, respectively, but no longer catastrophically misspecified. 

More detailed analysis of the model behavior in all Sub-Regions, with the corresponding, figures can be found in Appendix~\ref{app_results}. Future work will focus on uncovering the causes and mitigating the variable efficacy of calibration in all Sub-Regions.

\section{Conclusion}
\label{sec:conclusion}

In this work, we presented the first NRE-based population-level inference of $w$ and $\om{}$ from galaxy-galaxy strong lensing images. 
The main conclusions in this study are:

\begin{itemize}
    \item 
    NRE (without or with a calibration procedure) produces very limited constraints based on individual lensing systems (Fig.'s~\ref{fig:image_posterior} and~\ref{fig:single_joint_posterior}). This is true especially for $w$. $\om{}$ can still be somewhat constrained from a single lens.

    \item 
    NRE (without or with a calibration procedure) produces tighter constraints when invoked for population-level inference (Fig.~\ref{fig:single_joint_posterior}): the size of the posterior contours decreases as the number of objects in the tested population increases.
     
    \item 
    NRE without calibration produces overconfident posteriors for individual and population-level inference. 
    We used a post hoc calibration procedure that aims to reduce the confidence (increase the posterior width). The calibration performance depends on the region of parameter space where the inference is taking place. 
    This calibration mitigates overconfidence throughout most of the parameter space that we tested (Fig.'s~\ref{fig:image_posterior},~\ref{fig:single_joint_posterior},~\ref{fig:joint_posterior_entireregion},~\ref{fig:parity_entireregion},~\ref{fig:rank_histogram_entire_region}, and~\ref{fig:posteror_coverage_entireregion}).
    
    \item 
    There does not appear to be a change in performance of the NRE approach when training sizes are increased from 500k to 2M, which is less than order of magnitude  (Fig.~\ref{fig:compare_train_size}).
    
    \item 
    The calibrated NRE model has a Z-score comparable to analytic likelihood on simulated data from \cite{litian24}; see Sec.~\ref{sec:compare_analytic}. 
    
    \item 
    We have not yet uncovered the cause of variable efficacy of the calibration from Sub-Region to Sub-Region or what allows the calibration to succeed for the Entire Test Region, while not consistently successfully across the Sub-Regions. This is the next major step for future studies.
\end{itemize}

Recall the context, assumptions, and caveats for this proof-of-concept study.
First, we performed experiments on data simulated to mimic ideal lensing systems: the idealization pertains to redshift and morphological configurations and to DES observational noise, primarily regarding pixel scale and exposure time. 
We also assume perfect lens light subtraction and perfect deblending. 
Additionally, the present analysis is limited to $g$-band imaging.
Incorporating multi-band photometry in future work would provide richer color information, reduce degeneracies between lens and source parameters, potentially allowing for stronger cosmological constraints. 

Second, we treat the astrophysical parameters $\zl{}$, $\zs{}$, and $\sigmav{}$ as perfectly measured, ignoring observational uncertainties. 
In practice, these quantities carry non-negligible measurement errors and their observation typically requires significant experimental effort beyond photometric survey imaging. 
Incorporating these uncertainties in future studies would enhance the realism of cosmological constraint estimates derived from NRE-based population-level inference.
 
Third, the present study aims to derive constraints over a large, uniformly sampled parameter space (Entire Test Region); some Sub-Regions may require denser sampling for the algorithm to learn accurately and consistently across that large space.
Similarly, this underdense sampling may be the cause of the non-uniformity of the calibration in specific parameter Sub-Regions. 
Furthermore, the MCMC search strategy for the calibration is computationally expensive and sensitive to the initial conditions leading to possible convergence on local optima.
These computational inefficiencies may be mitigated by a secondary neural network that uses the data to predict the scaling factors for the calibration. 
Alternatively, replacing the current staged search with a more sample-efficient approach, like Bayesian Optimization, may improve convergence efficiency.

Finally, bridging the gap between simulations and real observations is an active area of research in the field. 
Promising approaches include domain adaptation techniques that align the feature distributions of simulated and observed data ~\citep[e.g.,][]{ciprijanovic23, pandya25, treyer26}, and nuisance-randomized training strategies that render the inference robust to nuisance variables ~\citep[e.g.,][]{puli21, gandrakota24}. 

In this study, we have taken initial steps toward demonstrating the potential of NRE for population-level inference in a cosmological context. 
The next, primary quandary to address is the inconsistency of the model calibration and performance across the $w$-$\om{}$ parameter space. 
This would begin with an investigation of objects in the Sub-Regions where calibration was less successful. 
After addressing and mitigating inconsistencies, assumptions, and caveats, future work may be able to use NRE to derive state-of-the-art constraints from strong lens populations observed with Rubin, Euclid and Roman.

\section*{Acknowledgments}

\textit{Funding:} This work was produced by Fermi Forward Discovery Group, LLC under Contract No. 89243024CSC000002 with the U.S. Department of Energy, Office of Science, Office of High Energy Physics. Publisher acknowledges the U.S. Government license to provide public access under the DOE Public Access Plan DOE Public Access Plan
(\url{http://energy.gov/downloads/doe-public-access-plan}).

The work of S.J. and B.N. (while at Fermilab) was supported by the DOE Grant No. 0000258217 (FNAL 21-25).\\

\noindent \textit{Author Contributions:} 
\begin{description}
    \item[Jarugula] Methodology, Formal analysis, Software, Validation, Data curation, Writing - Original draft.
    \item[Nord] Conceptualization, Methodology, Writing - Review \& Editing (including AI-removal editing), Supervision, Project administration, Funding acquisition. 
    Nord did not use AI in the writing or editing of this manuscript.
    \item[\'Ciprijanovi\'c] Methodology, Writing - Review \& Editing, Supervision, Project administration.
    \item[Trivedi] Methodology, Writing - Review \& Editing.
\end{description}

\noindent \textit{AI Disclosure:} 
The generative AI tool Claude Sonnet 4.6 was used in the preparation of this manuscript. 
The Abstract, Introduction, Sec.s 6.2 and 6.3, and the Conclusion were drafted by the authors and polished with the help of AI; they were then rewritten by humans, attempting to remove the influence of AI. 
Claude was used to generate the algorithm table in Sec. 4.2 in LaTeX format. 
Claude was used to edit the code for visualization (layout and formatting of Fig.'s 5, 7, and 11) and the code for the calibration.
It was also used to assist in identifying statistical metrics for comparing posterior distributions with literature values; this led to the use of the Z-score.
The interpretation of all calculations (including the Z-score) was performed by the authors.
Finally, authors who did not use AI performed a final pass through the full manuscript to rewrite or significantly edit the sections where AI was used.

\software{\texttt{astropy}~\citep{2013A&A...558A..33A,2018AJ....156..123A},  \texttt{deeplenstronomy}~\citep{morgan21}, \texttt{lenstronomy}~\citep{birre15, birrer18}.}

\clearpage
\appendix

\section{Network architecture}
\label{app:network_summary}

\begin{table}[h!]
\centering
\begin{tabular}{lll}
\hline
Layer Type & Output Shape & No. of Parameters \\\hline

Input Images $x$ 
& (Batch, 32, 32, 1) & 0 \\

Conv2D ($3\times3$, stride 2, 8 filters) 
& (Batch, 16, 16, 8) 
& $3\!\times\!3\!\times\!1\!\times\!8 + 8 = 80$ \\

\multicolumn{3}{c}{\textbf{Residual Block 1 (16 filters)}} \\

Conv2D ($3\times3$) 
& (Batch, 16, 16, 16) 
& $3\!\times\!3\!\times\!8\!\times\!16 + 16 = 1{,}168$ \\

Conv2D ($3\times3$) 
& (Batch, 16, 16, 16) 
& $3\!\times\!3\!\times\!16\!\times\!16 + 16 = 2{,}320$ \\

MaxPooling (stride 2) 
& (Batch, 8, 8, 16) & 0 \\

Residual Projection Conv2D ($1\times1$, stride 2) 
& (Batch, 8, 8, 16) 
& $1\!\times\!1\!\times\!8\!\times\!16 + 16 = 144$ \\

Add 
& (Batch, 8, 8, 16) & 0 $\leftarrow$ \\

\multicolumn{3}{c}{\textbf{Residual Block 2 (32 filters)}} \\

Conv2D ($3\times3$) 
& (Batch, 8, 8, 32) 
& $3\!\times\!3\!\times\!16\!\times\!32 + 32 = 4{,}640$ \\

Conv2D ($3\times3$) 
& (Batch, 8, 8, 32) 
& $3\!\times\!3\!\times\!32\!\times\!32 + 32 = 9{,}248$ \\

MaxPooling (stride 2) 
& (Batch, 4, 4, 32) & 0 \\

Residual Projection Conv2D ($1\times1$, stride 2) 
& (Batch, 4, 4, 32) 
& $1\!\times\!1\!\times\!16\!\times\!32 + 32 = 544$ \\

Add 
& (Batch, 4, 4, 32) & 0 $\leftarrow$ \\

\multicolumn{3}{c}{\textbf{Residual Block 3 (45 filters)}} \\

Conv2D ($3\times3$) 
& (Batch, 4, 4, 45) 
& $3\!\times\!3\!\times\!32\!\times\!45 + 45 = 13{,}005$ \\

Conv2D ($3\times3$) 
& (Batch, 4, 4, 45) 
& $3\!\times\!3\!\times\!45\!\times\!45 + 45 = 18{,}270$ \\

MaxPooling (stride 2) 
& (Batch, 2, 2, 45) & 0 \\

Residual Projection Conv2D ($1\times1$, stride 2) 
& (Batch, 2, 2, 45) 
& $1\!\times\!1\!\times\!32\!\times\!45 + 45 = 1{,}485$ \\

Add 
& (Batch, 2, 2, 45) & 0 $\leftarrow$ \\

SeparableConv2D ($3\times3$, 64 filters) 
& (Batch, 2, 2, 64) 
& $3\!\times\!3\!\times\!45 + 45\!\times\!64 + 64 = 3{,}349$ \\

Global Average Pooling (Embedding A)
& (Batch, 64) & 0 \\

\hline
Input Astrophysics Params $(\zl{},\zs{},\sigmav{})$ 
& (Batch, 3) & 0 \\

Dense Embedding  (Embedding B)
& (Batch, 64) 
& $3\!\times\!64 + 64 = 256$ \\
\hline
Input Cosmology Params ($w$,$\om{}$) 
& (Batch, 2) & 0 \\

Dense Embedding (Embedding C)
& (Batch, 64) 
& $2\!\times\!64 + 64 = 192$ \\
\hline
Concatenate Embeddings A, B, C
& (Batch, 192) & 0 \\

Dense 
& (Batch, 192) 
& $192\!\times\!192 + 192 = 37{,}056$ \\

Dropout 
& (Batch, 192) & 0 \\

Dense 
& (Batch, 64) 
& $192\!\times\!64 + 64 = 12{,}352$ \\

Dropout 
& (Batch, 64) & 0 \\

Dense (ln $r$) 
& (Batch, 1) 
& $64\!\times\!1 + 1 = 65$ \\

\hline
\end{tabular}

\caption{Summary of the NRE network architecture with output shapes and parameter counts for $32\times32\times1$ input images. 
Residual skip connections are indicated by the arrows and implemented through projection shortcuts and \textit{Add} layers. 
Batch normalization and activation functions are applied after each convolutional and dense layer.}
\label{tab:network_summary}
\end{table}

\section{Model performance in test Sub-Regions}
\label{app_results}

We report on NRE model performance in each of the test Sub-Regions.
These figures may be compared with those in Sec.~\ref{sec:results}, where we analyze the Entire Test Region.
The post hoc calibration was trained using data from the Entire Test Region of the test data and applied to each Sub-Region individually. 
This analysis provides information regarding which Sub-Regions could be better calibrated with the post hoc procedure.
Investigations into the causes of the more and less successful calibrations (by Sub-Region) will likely aid in determining how to improve the procedure for the Entire Test Region.

\subsection{Sub-Region 1}
\label{app:region1}

\begin{figure*}[!ht]
 \centering
    \includegraphics[width=1.0\linewidth]{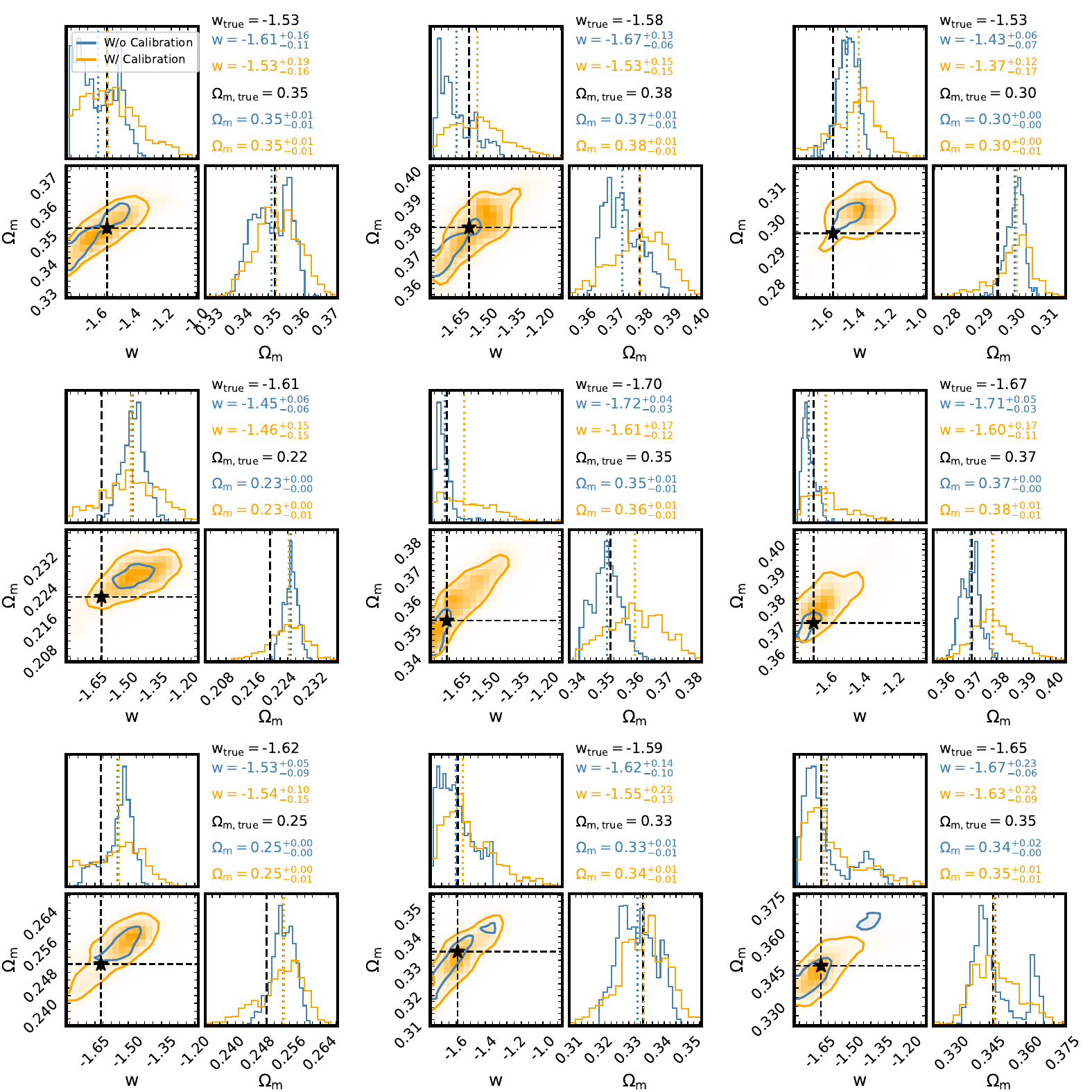}
    \caption{Same as Fig.~\ref{fig:joint_posterior_entireregion}, but for Sub-Region 1.}   \label{fig:joint_posterior_region1}
\end{figure*}

\begin{figure*}[!ht]
 \centering
    \includegraphics[width=0.6\linewidth]{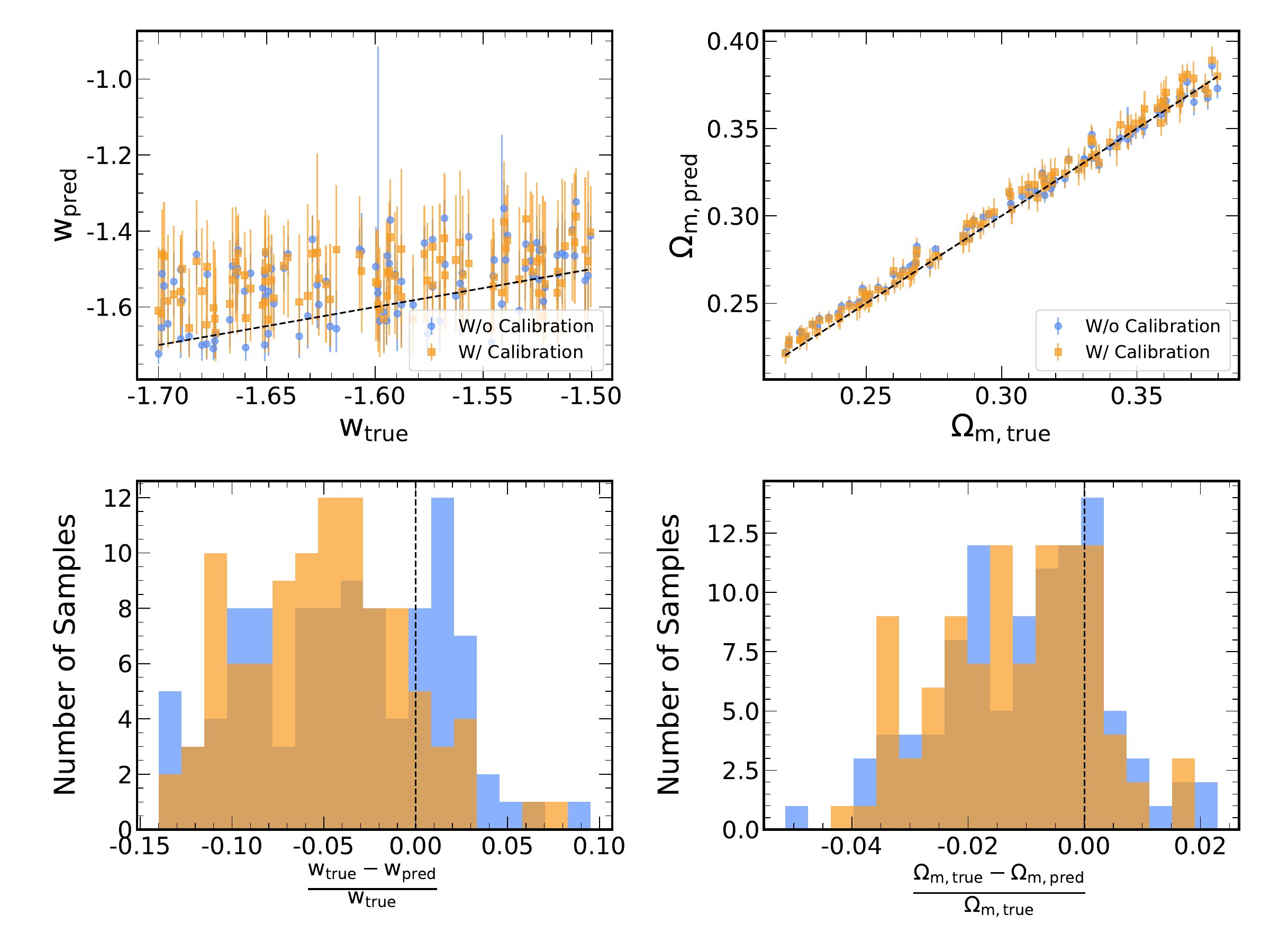}
    \caption{Same as Fig.~\ref{fig:parity_entireregion}, but for Sub-Region 1.}
   \label{fig:parity_region1}
\end{figure*}

\begin{figure*}[!ht]
 \centering
    \includegraphics[width=0.6\linewidth]{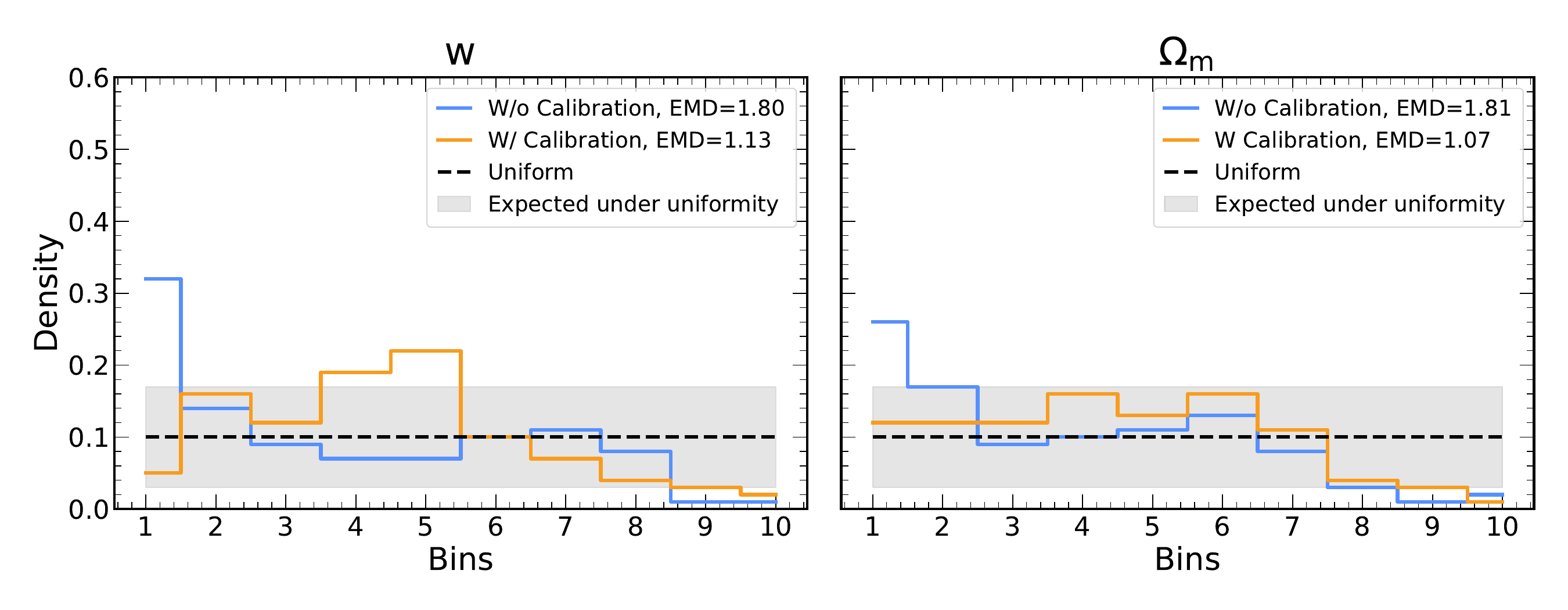}
    \caption{Same as Fig.~\ref{fig:rank_histogram_entire_region}, but for Sub-Region 1.}
   \label{fig:rank_region1}
\end{figure*}

\begin{figure*}[!ht]
 \centering
    \includegraphics[width=0.4\linewidth]{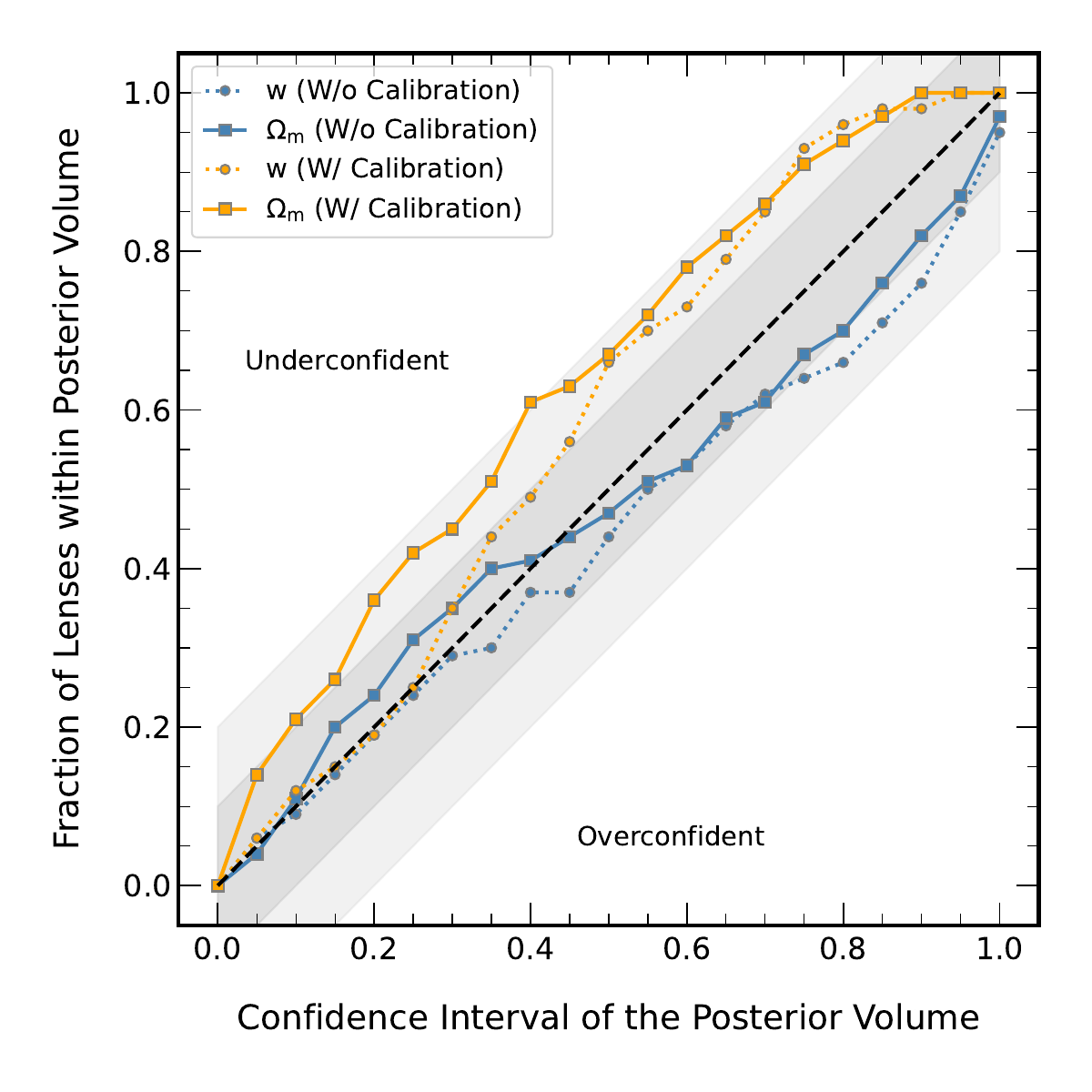}
    \caption{Same as Fig.~\ref{fig:posteror_coverage_entireregion}, but for Sub-Region 1.}
   \label{fig:posterior_coverage_region1}
\end{figure*}

\begin{description}
    \item[Posterior Contours (Fig. \ref{fig:joint_posterior_region1})] 
    Posteriors are larger with calibration (orange) than without calibration (blue); all posteriors with calibration contain the true points, but one posterior without calibration excludes the true value.
    \item[Parity and Residuals (Fig. \ref{fig:parity_region1})] 
    Median $w$ values are much higher and median $\om{}$ values are slightly higher than truth both with and without calibration
    \item[Rank Histogram (Fig. \ref{fig:rank_region1})] Histograms for both $w$ and $\om{}$ are left-skewed without calibration, and less so with calibration. 
    The EMD is lowered but nonzero with calibration.  
    \item[Posterior Coverage (Fig. \ref{fig:posterior_coverage_region1})] For both parameters, the model is partially overconfident without calibration; it is fully underconfident with calibration.
\end{description}

For $w$, with calibration, the model is made less biased and underconfident. 
For $\om{}$, with calibration, the model bias is left unchanged, and it is made underconfident.

\clearpage

\subsection{Sub-Region 2}
\label{app:region2}

\begin{figure*}[!ht]
 \centering
    \includegraphics[width=1.0\linewidth]{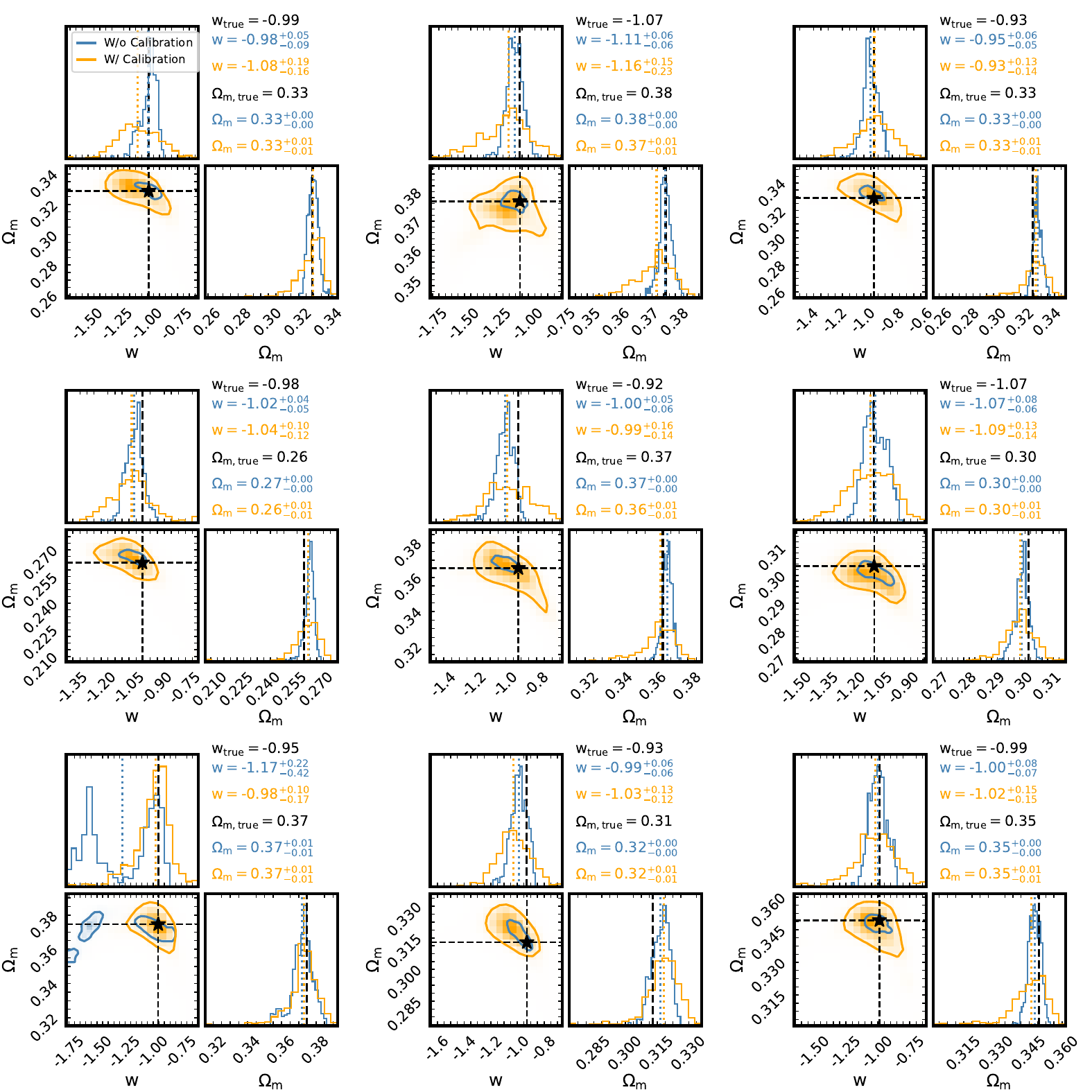}
    \caption{Same as Fig.~\ref{fig:joint_posterior_entireregion}, but for Sub-Region 2.}
   \label{fig:joint_posterior_region2}
\end{figure*}

\begin{figure*}[!ht]
 \centering
    \includegraphics[width=0.6\linewidth]{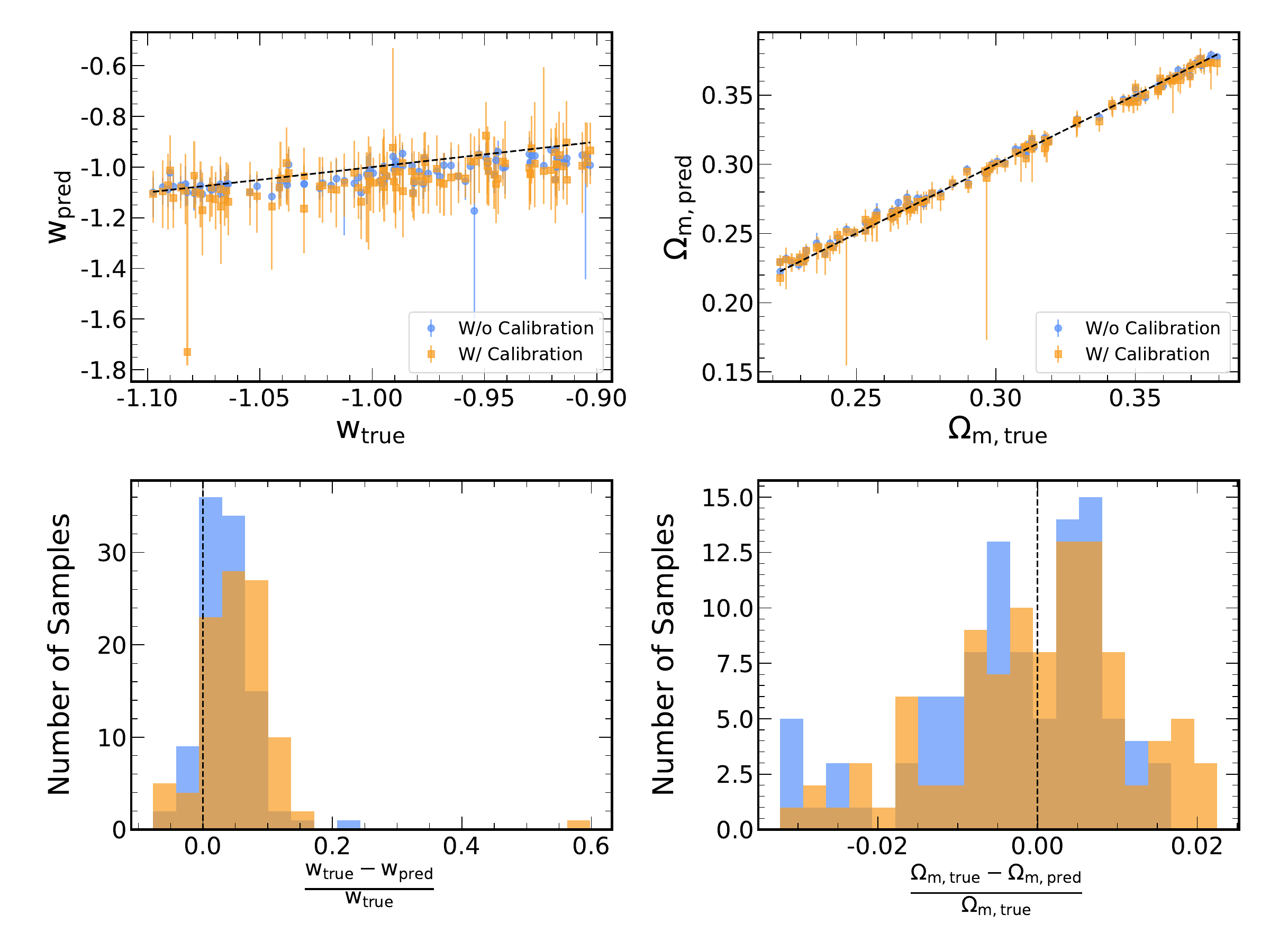}
    \caption{Same as Fig.~\ref{fig:parity_entireregion}, but for Sub-Region 2.}
   \label{fig:parity_region2}
\end{figure*}

\begin{figure*}[!ht]
 \centering
    \includegraphics[width=0.6\linewidth]{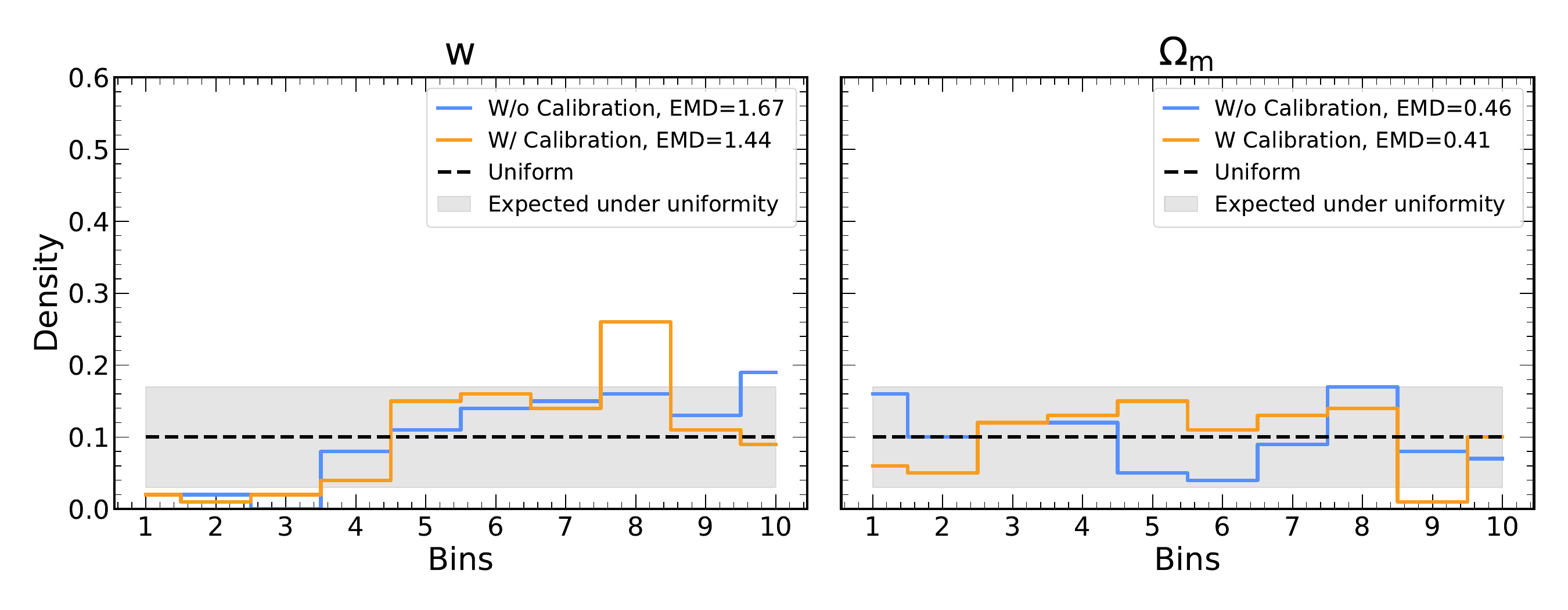}
    \caption{Same as Fig.~\ref{fig:rank_histogram_entire_region}, but for Sub-Region 2.}
   \label{fig:rank_region2}
\end{figure*}

\begin{figure*}[!ht]
 \centering
    \includegraphics[width=0.4\linewidth]{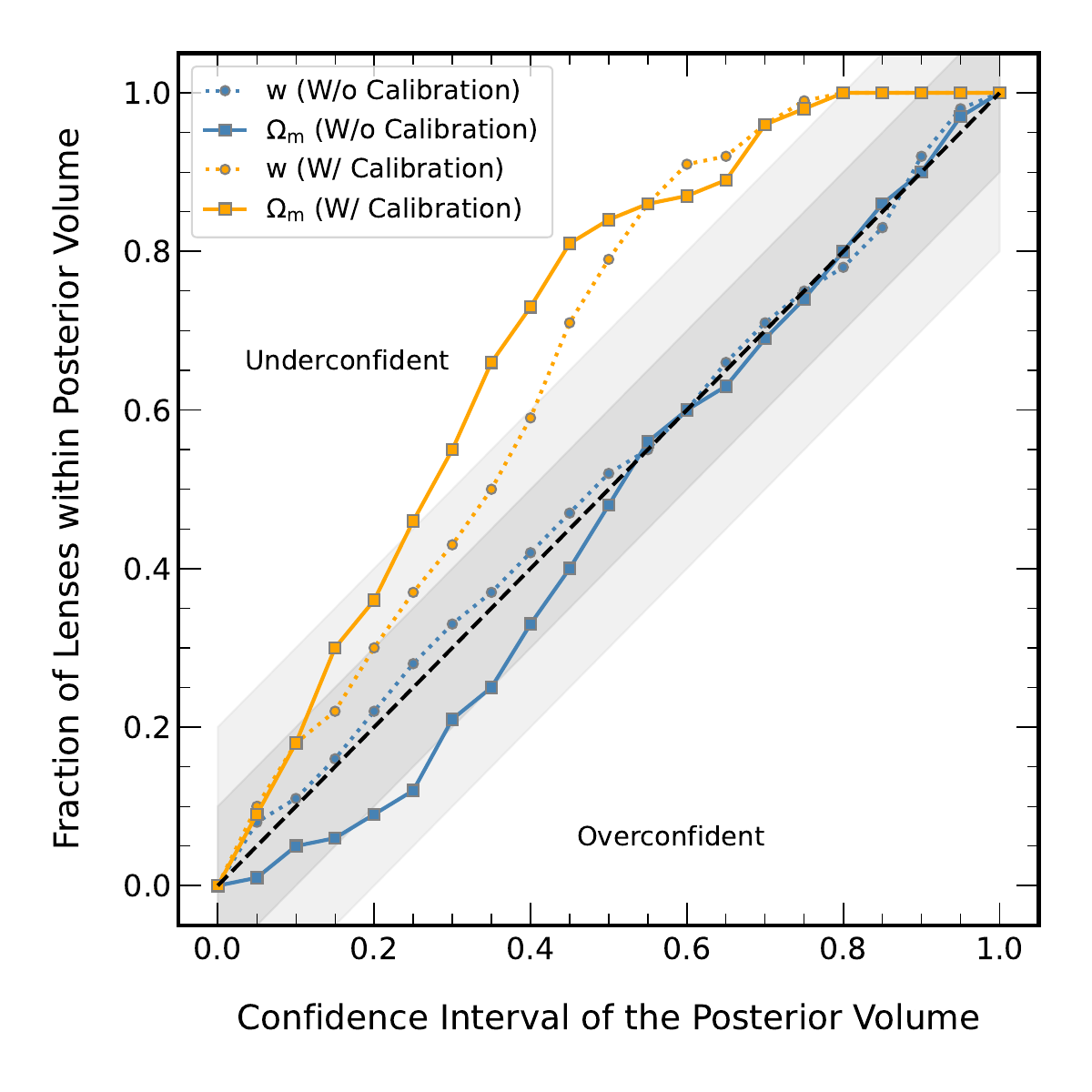}
    \caption{Same as Fig.~\ref{fig:posteror_coverage_entireregion}, but for Sub-Region 2.}
   \label{fig:posterior_coverage_region2}
\end{figure*}

\begin{description}
    \item[Posterior Contours (Fig. \ref{fig:joint_posterior_region2})] The contours are larger with calibration than without calibration; contours in both scenarios contain the true values. 
    \item[Parity and Residuals (Fig. \ref{fig:parity_region2})] With and without calibration, the median $w$ values are lower than truth, and the median $\om{}$ values are near truth.
    \item[Rank Histogram (Fig. \ref{fig:rank_region2})] With and without calibration, the $w$ ranks are right-skewed; $\om{}$ is consistent with uniform. 
    The EMD for $w$ is lower with calibration;  $\om{}$ same with and without calibration.
    \item[Posterior Coverage (Fig. \ref{fig:posterior_coverage_region2})] With calibration, the model becomes highly underconfident for both parameters.
\end{description}

The model accuracy is unchanged with calibration. 
For $w$, the model confidence is lessened from almost perfect to significantly underconfident; for $\om{}$, the model confidence is lessened from partially overconfident to significantly underconfident.

\clearpage

\subsection{Sub-Region 3} 
\label{app:region3}

\begin{figure*}[!ht]
 \centering
    \includegraphics[width=1.0\linewidth]{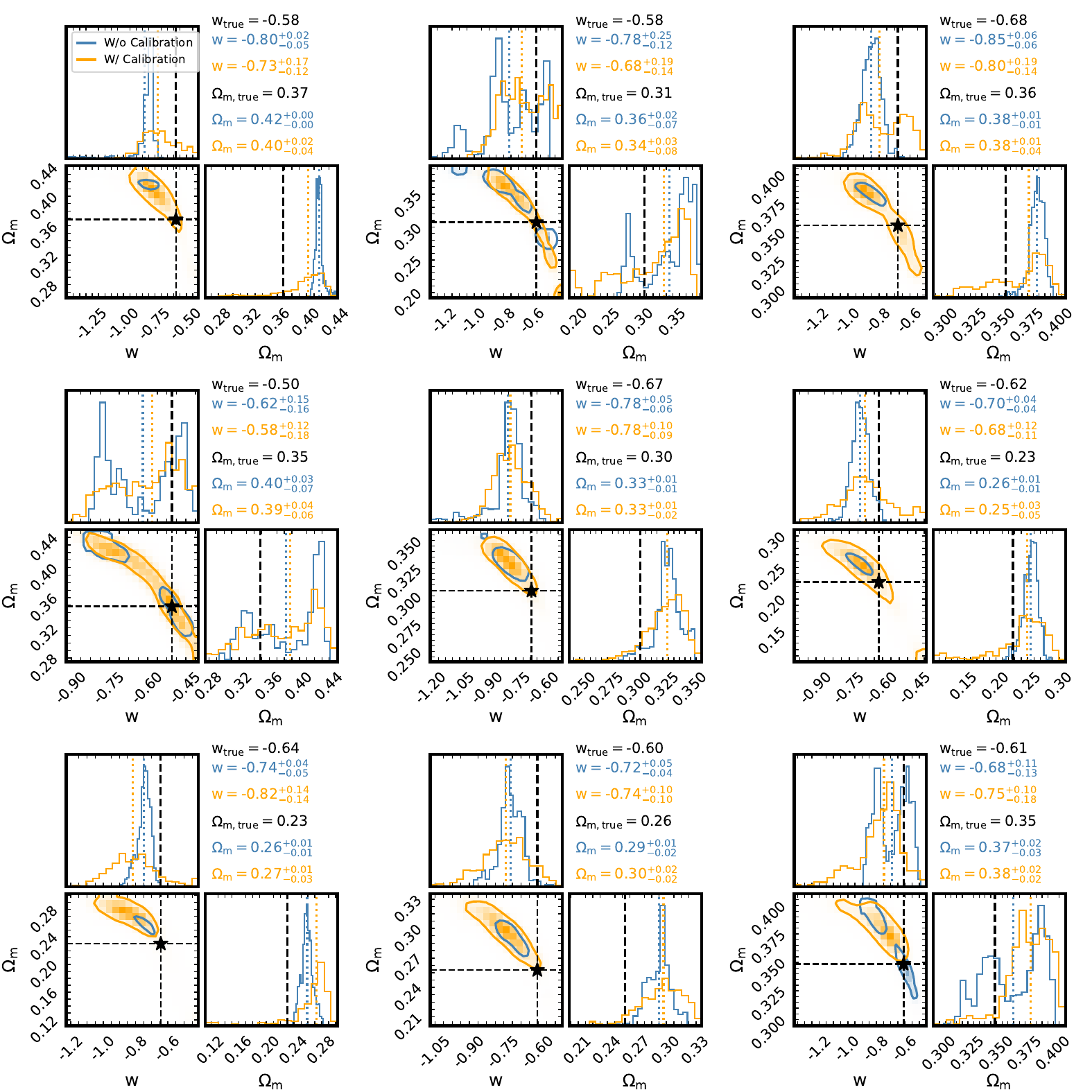}
    \caption{Same as Fig.~\ref{fig:joint_posterior_entireregion}, but for Sub-Region 3.}
   \label{fig:joint_posterior_region3}
\end{figure*}

\begin{figure*}[!ht]
 \centering
    \includegraphics[width=0.6\linewidth]{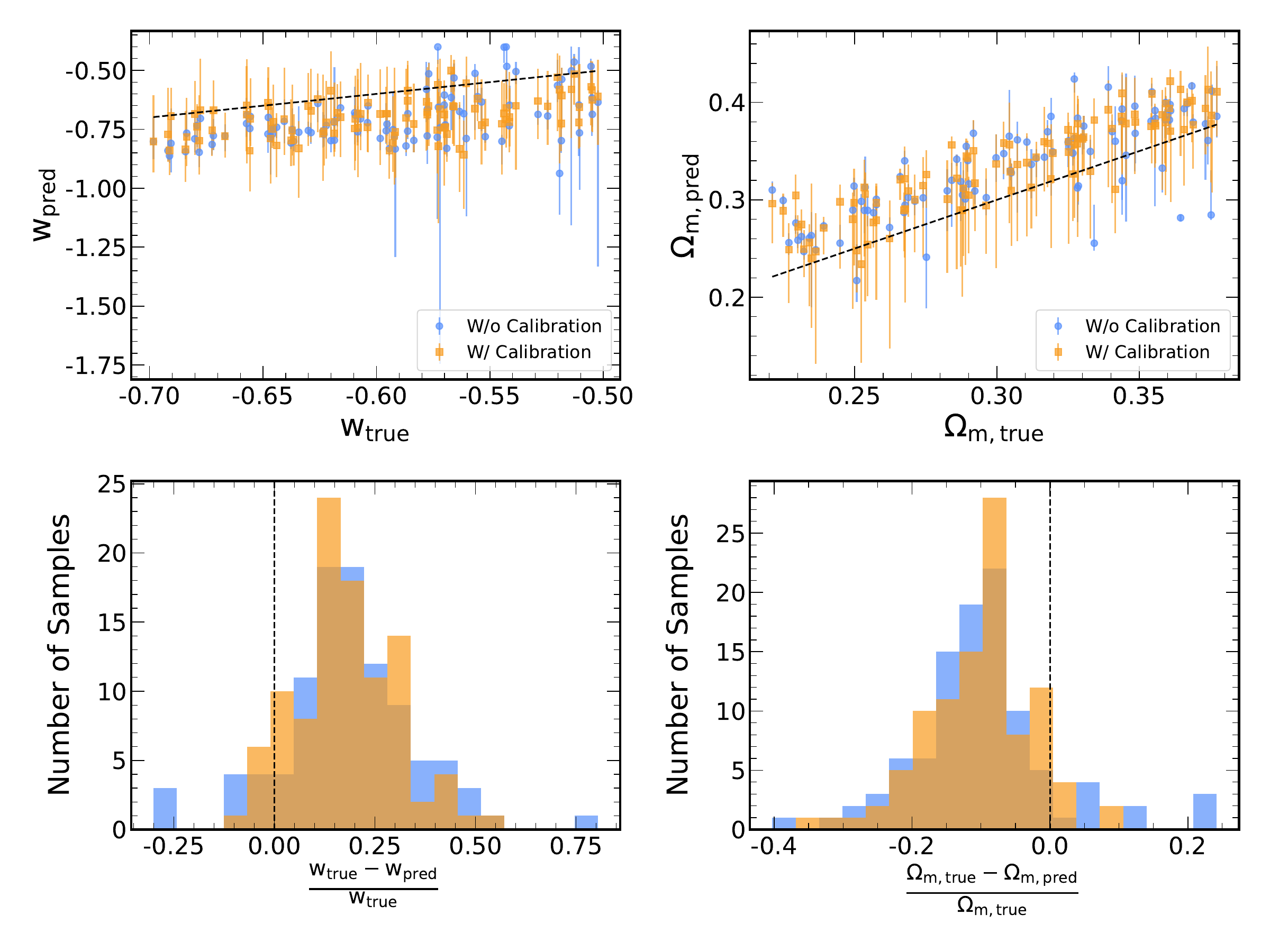}
    \caption{Same as Fig.~\ref{fig:parity_entireregion}, but for Sub-Region 3.}
   \label{fig:parity_region3}
\end{figure*}

\begin{figure*}[!ht]
 \centering
    \includegraphics[width=0.6\linewidth]{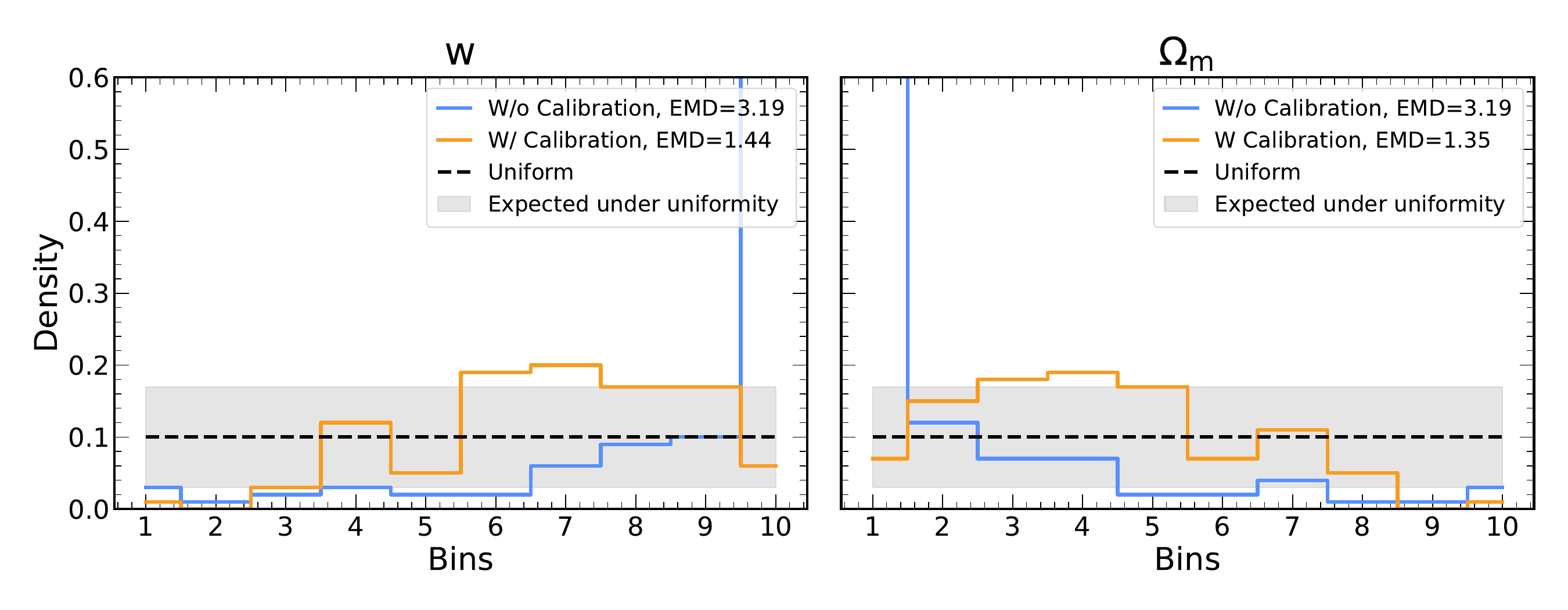}
    \caption{Same as Fig.~\ref{fig:rank_histogram_entire_region}, but for Sub-Region 3.}
   \label{fig:rank_region3}
\end{figure*}

\begin{figure*}[!ht]
 \centering
    \includegraphics[width=0.4\linewidth]{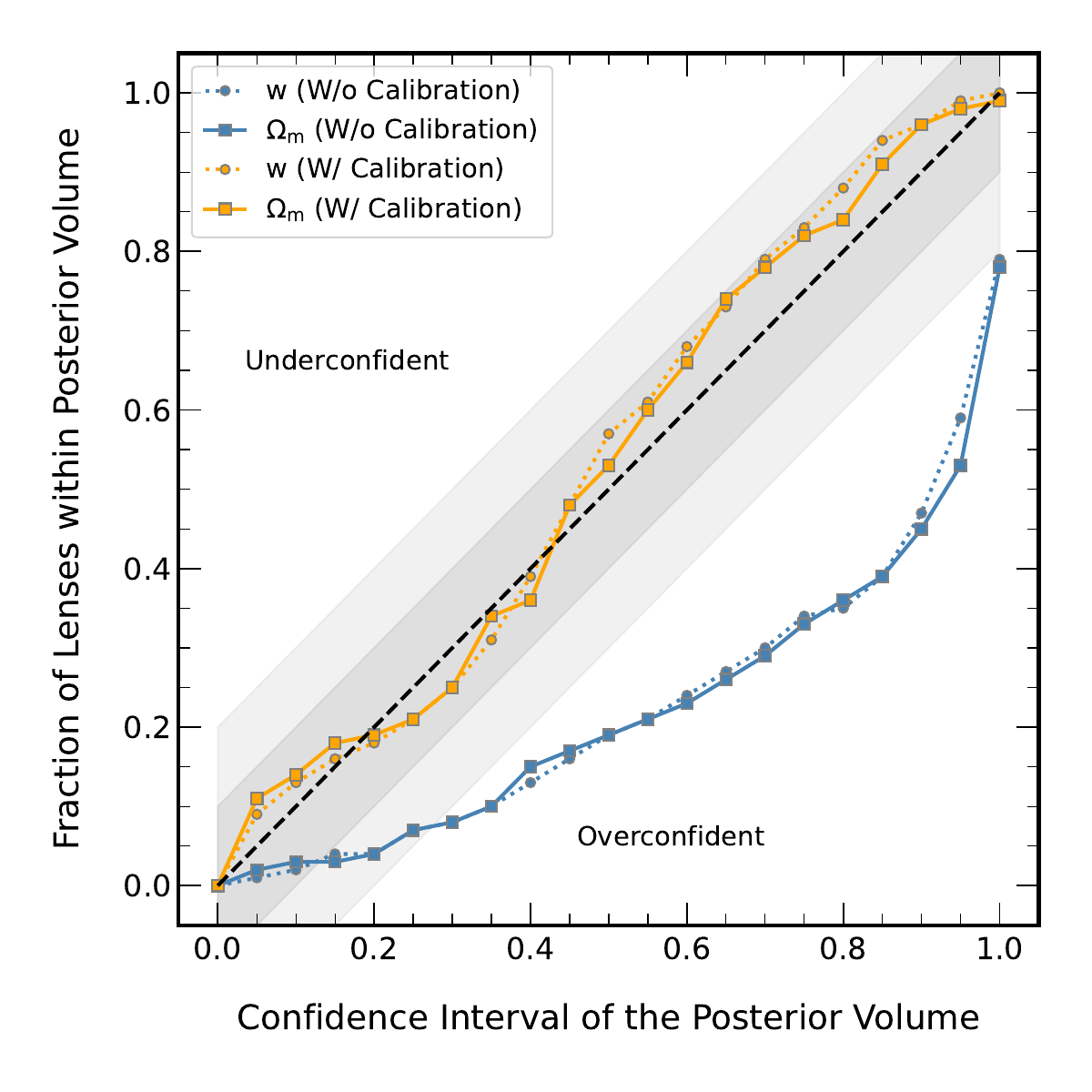}
    \caption{Same as Fig.~\ref{fig:posteror_coverage_entireregion}, but for Sub-Region 3.}
   \label{fig:posterior_coverage_region3}
\end{figure*}

\begin{description}
    \item[Posterior Contours (Fig. \ref{fig:joint_posterior_region3})] Without calibration, most contours exclude the true value.
    With calibration, seven of the nine contours include the true value.
    \item[Parity and Residuals (Fig. \ref{fig:parity_region3})] There is almost no change in bias with calibration: there are fewer outliers for both parameters.
    \item[Rank Histogram (Fig. \ref{fig:rank_region3})] The ranks for $w$ are highly right-skewed without calibration and less skewed with calibration. The ranks for $\om{}$ are highly left-skewed without calibration and less skewed with calibration. With calibration, the EMD is reduced by almost 50\% for both. 
    \item[Posterior Coverage (Fig. \ref{fig:posterior_coverage_region3})] Without calibration, the model is highly overconfident for both parameters. With calibration, the model is nearer to perfect confidence.
\end{description}

For both parameters, the model is biased with and without calibration, but it is nearly perfectly confident with calibration.

\clearpage

\subsection{Sub-Region 4}
\label{app:region4}

\begin{figure*}[!ht]
 \centering
    \includegraphics[width=1.0\linewidth]{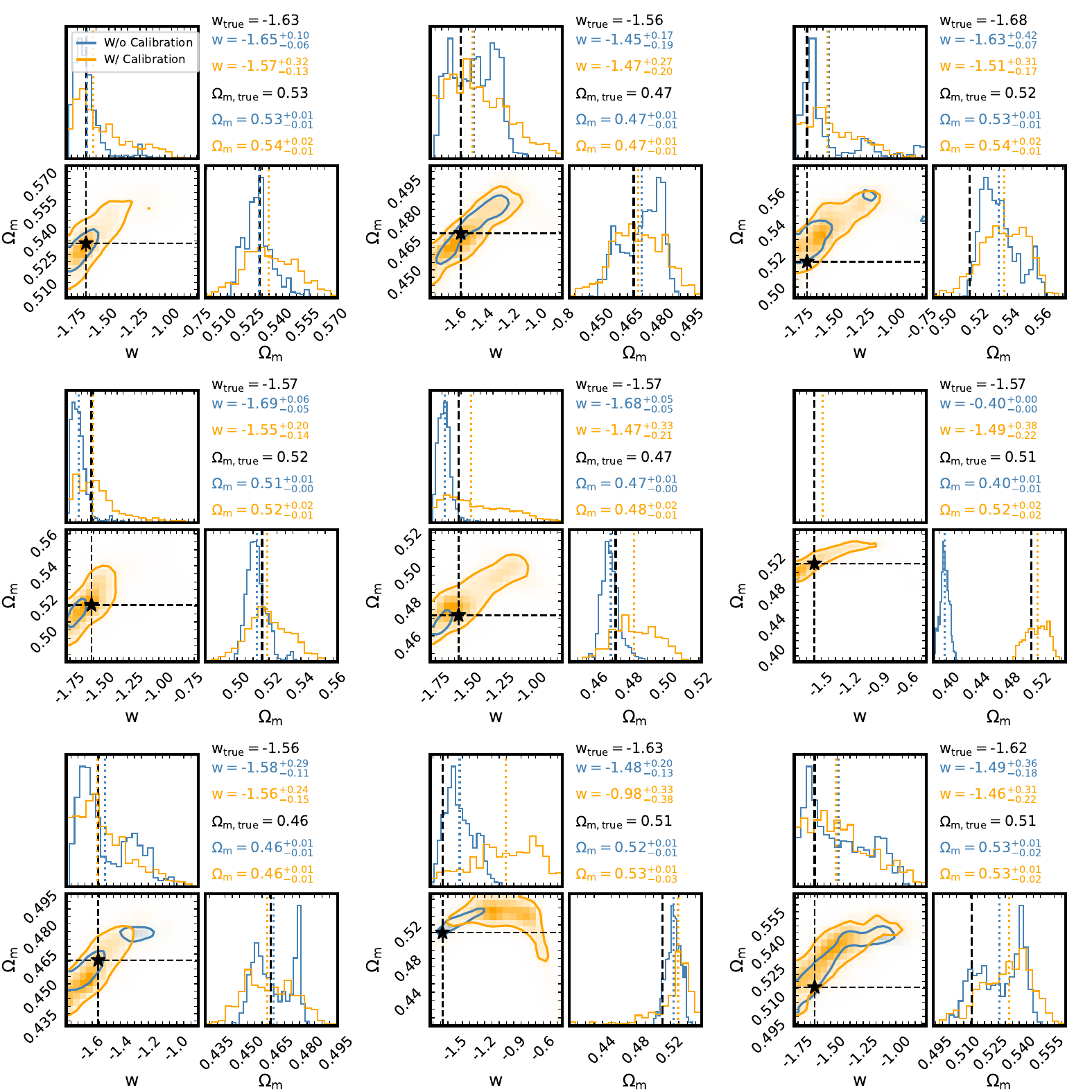}
    \caption{Same as Fig.~\ref{fig:joint_posterior_entireregion}, but for Sub-Region 4.}
   \label{fig:joint_posterior_region4}
\end{figure*}

\begin{figure*}[!ht]
 \centering
    \includegraphics[width=0.6\linewidth]{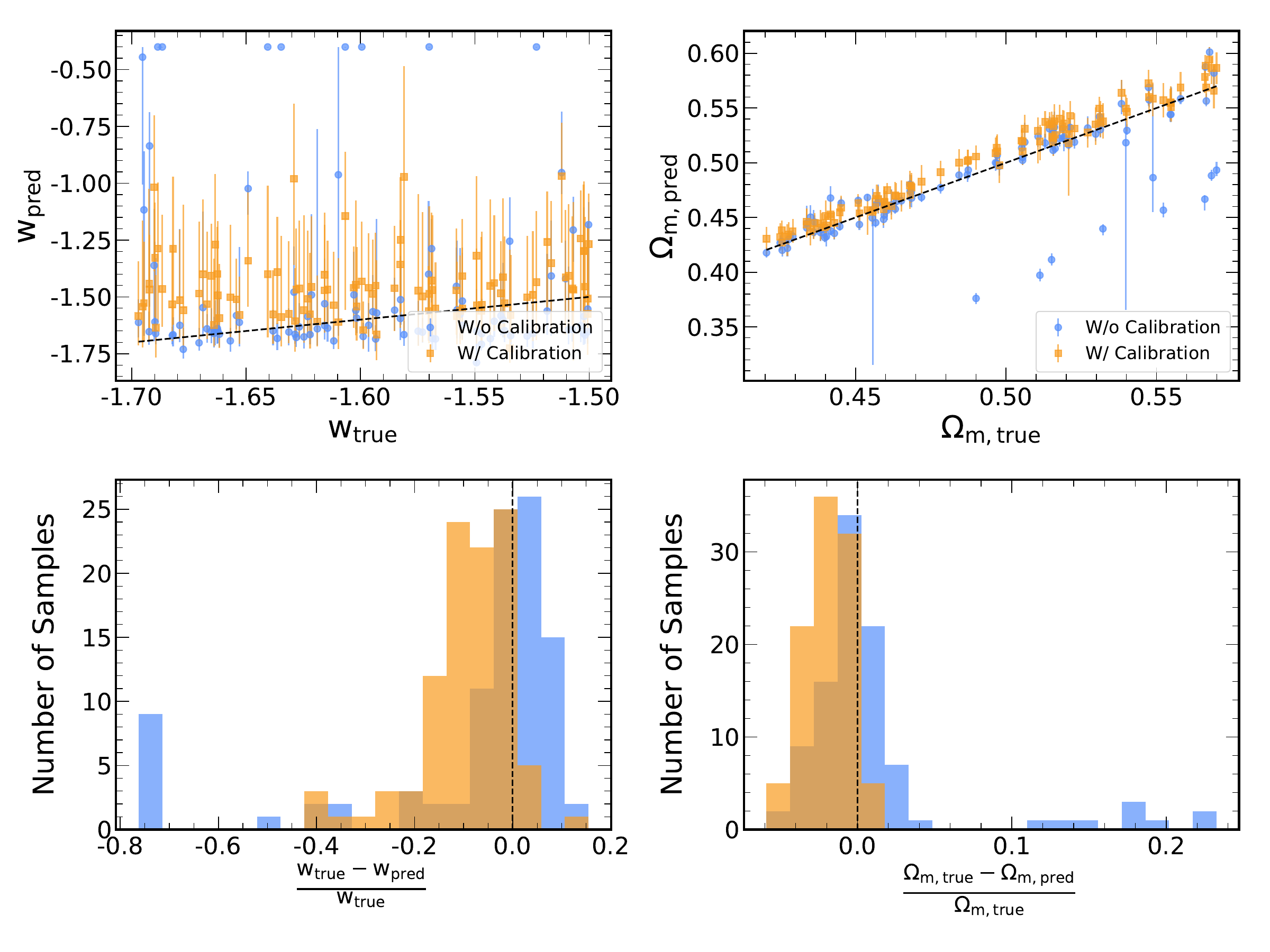}
    \caption{Same as Fig.~\ref{fig:parity_entireregion}, but for Sub-Region 4.}
   \label{fig:parity_region4}
\end{figure*}

\begin{figure*}[!ht]
 \centering
    \includegraphics[width=0.6\linewidth]{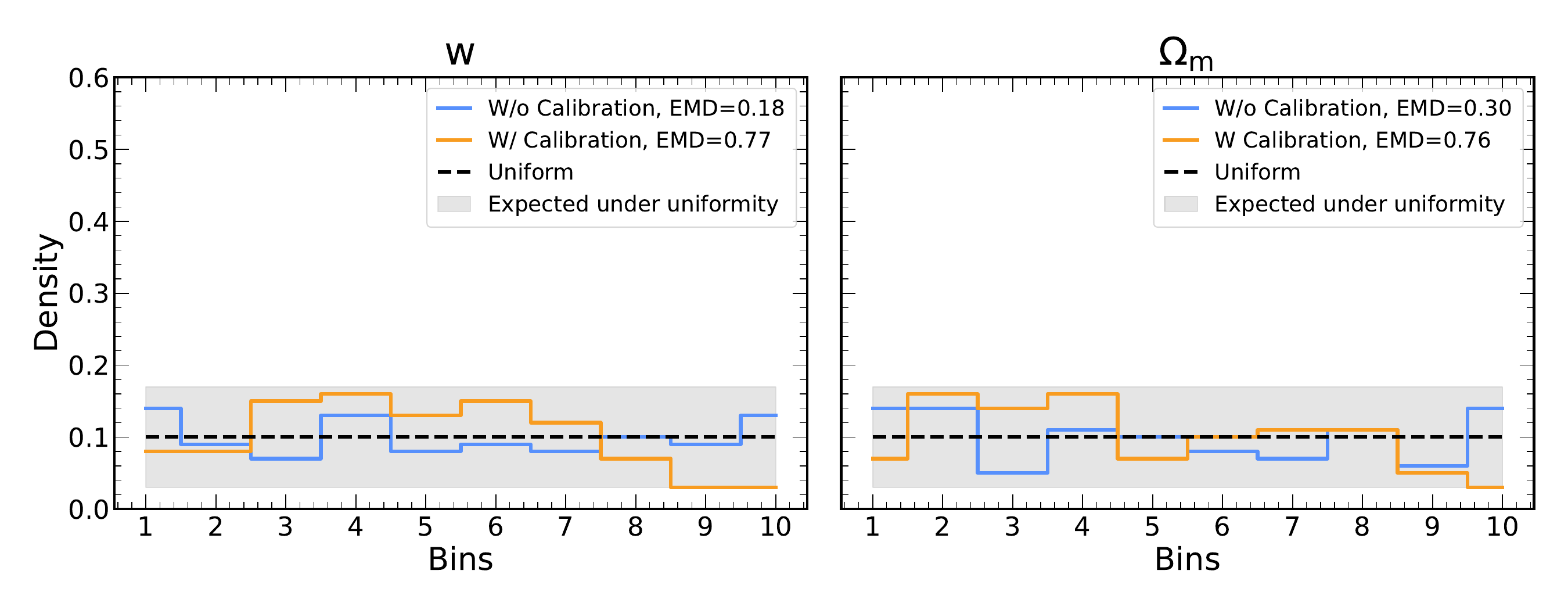}
    \caption{Same as Fig.~\ref{fig:rank_histogram_entire_region}, but for Sub-Region 4.}
   \label{fig:ranks_region4}
\end{figure*}

\begin{figure*}[!ht]
 \centering
    \includegraphics[width=0.4\linewidth]{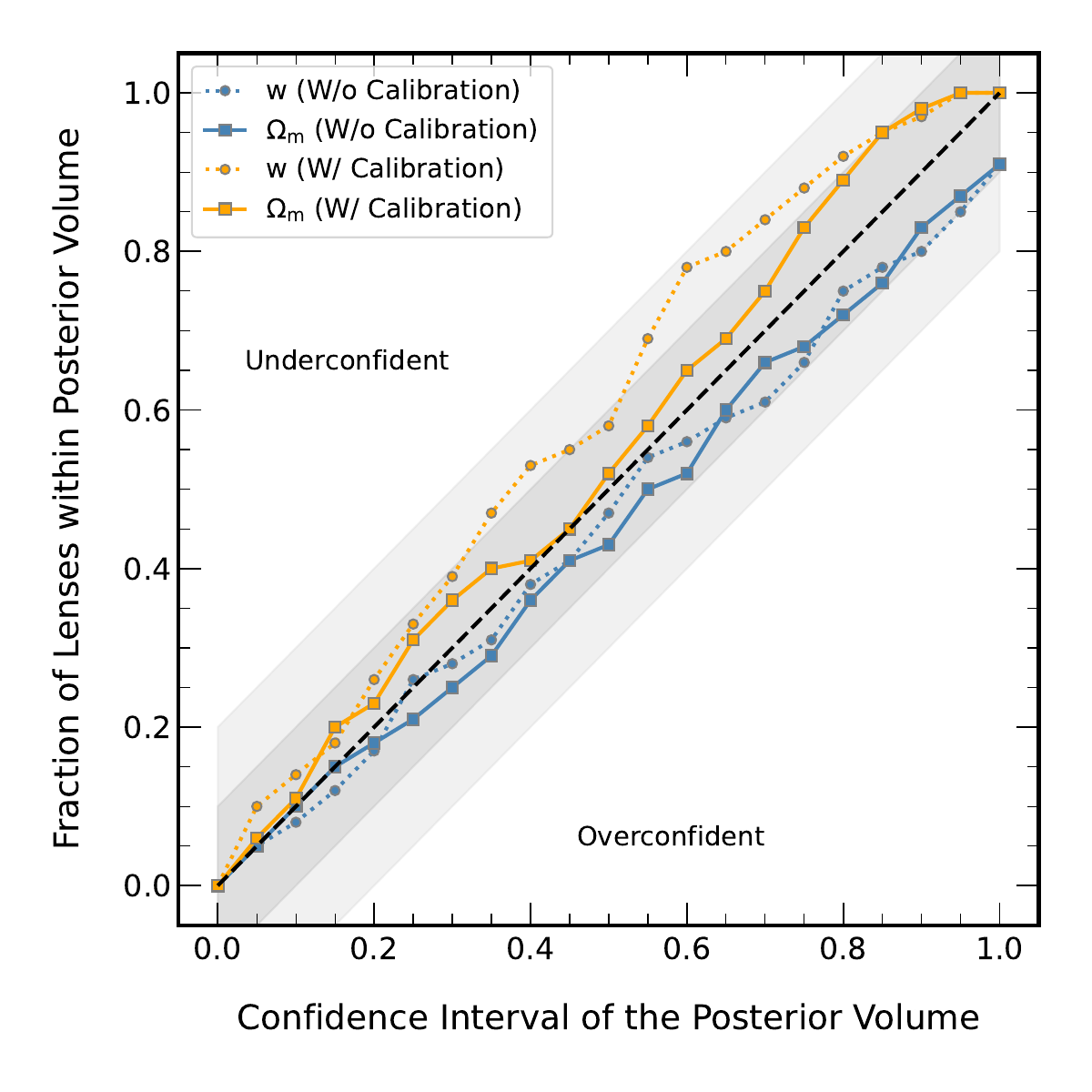}
    \caption{Same as Fig.~\ref{fig:posteror_coverage_entireregion}, but for Sub-Region 4.}
   \label{fig:posterior_coverage_region4}
\end{figure*}

\begin{description}
    \item[Posterior Contours (Fig. \ref{fig:joint_posterior_region4})] Contours without calibration include the true value in five out of nine examples. Contours with calibration include the true value in eight of out nine examples. Those models disagree in one example.
    \item[Parity and Residuals (Fig. \ref{fig:parity_region4})] Without calibration, the model is less biased, but has a large number of outliers for both parameters. With calibration, for both parameters, the outliers are lessened significantly, but there is a slight bias.
    \item[Rank Histogram (Fig. \ref{fig:ranks_region4})] For both parameters, without calibration, the ranks are nearly uniform. With calibration, there is slight left-skew for both parameters. The EMD is more than doubled with calibration.
    \item[Posterior Coverage (Fig. \ref{fig:posterior_coverage_region4})] Without calibraiton, the model is slightly overconfident for both parameters. With calibraiton, the model is moderately underconfident.
\end{description}

With calibration, for both parameters, the model is slightly reduced in accuracy, but is made underconfident.

\clearpage

\subsection{Sub-Region 5}
\label{app:region5}

\begin{figure*}[!ht]
 \centering
    \includegraphics[width=1.0\linewidth]{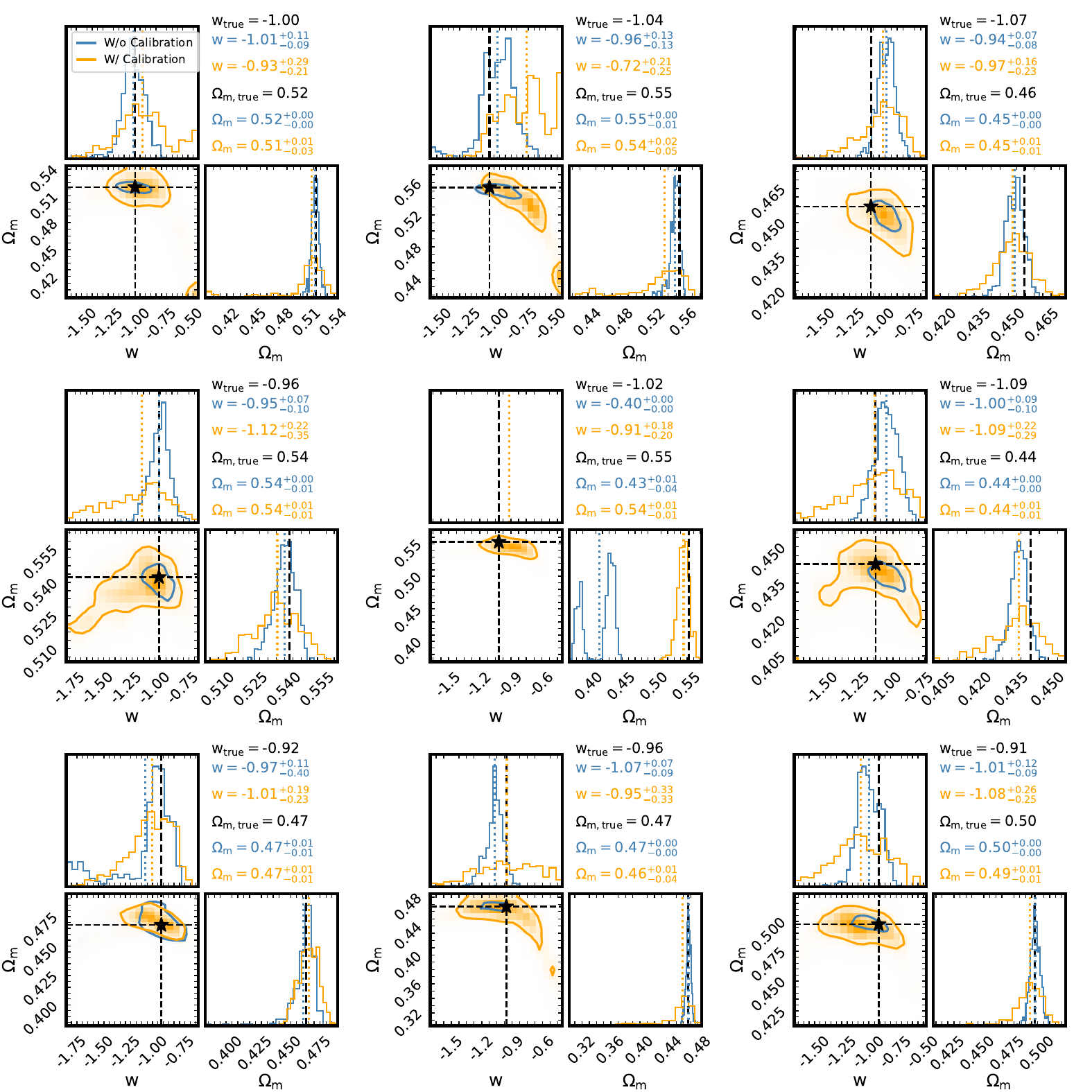}
    \caption{Same as Fig.~\ref{fig:joint_posterior_entireregion}, but for Sub-Region 5.}
   \label{fig:joint_posterior_region5}
\end{figure*}

\begin{figure*}[!ht]
 \centering
    \includegraphics[width=0.6\linewidth]{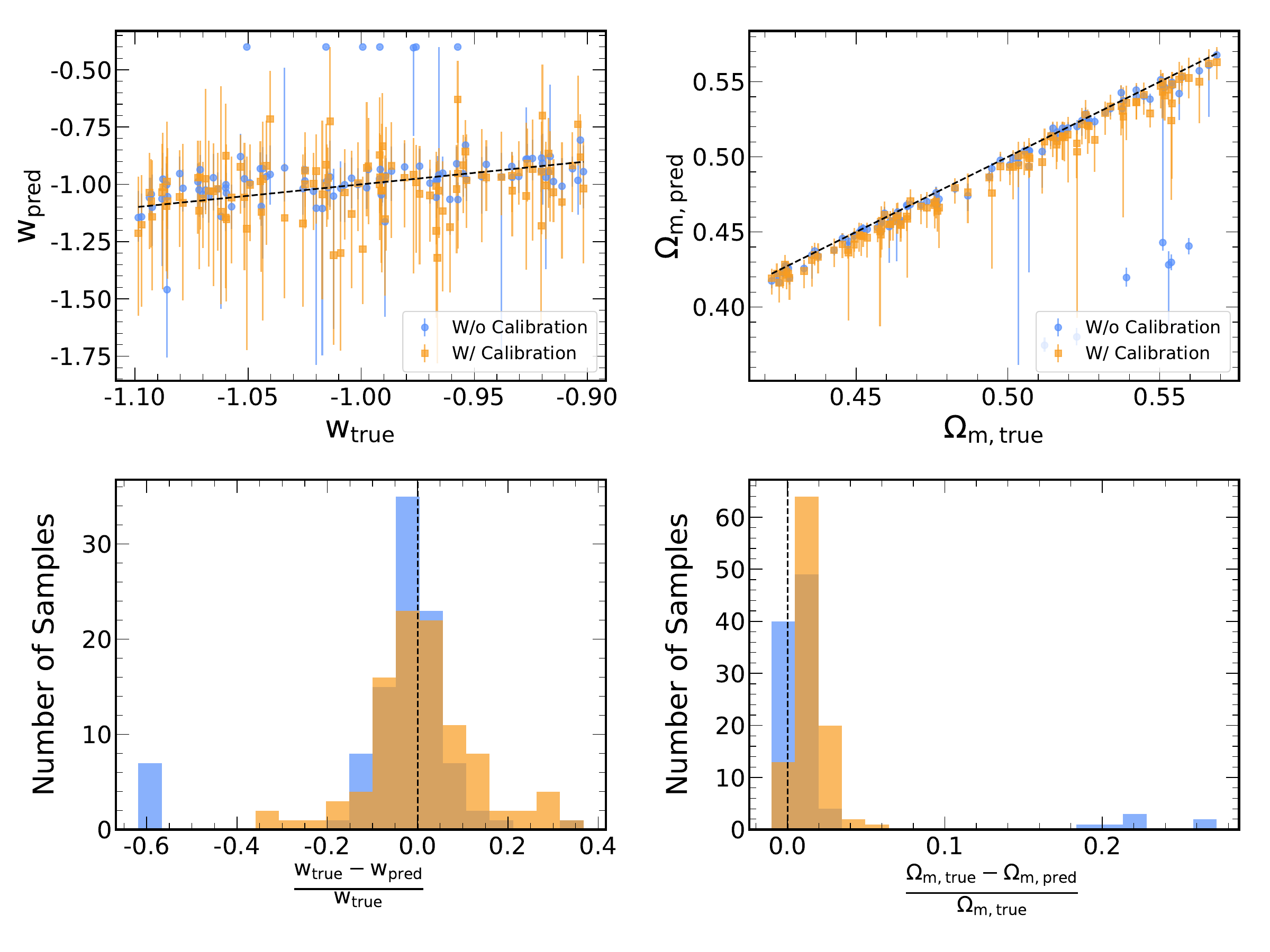}
    \caption{Same as Fig.~\ref{fig:parity_entireregion}, but for Sub-Region 5.}
   \label{fig:parity_region5}
\end{figure*}

\begin{figure*}[!ht]
 \centering
    \includegraphics[width=0.6\linewidth]{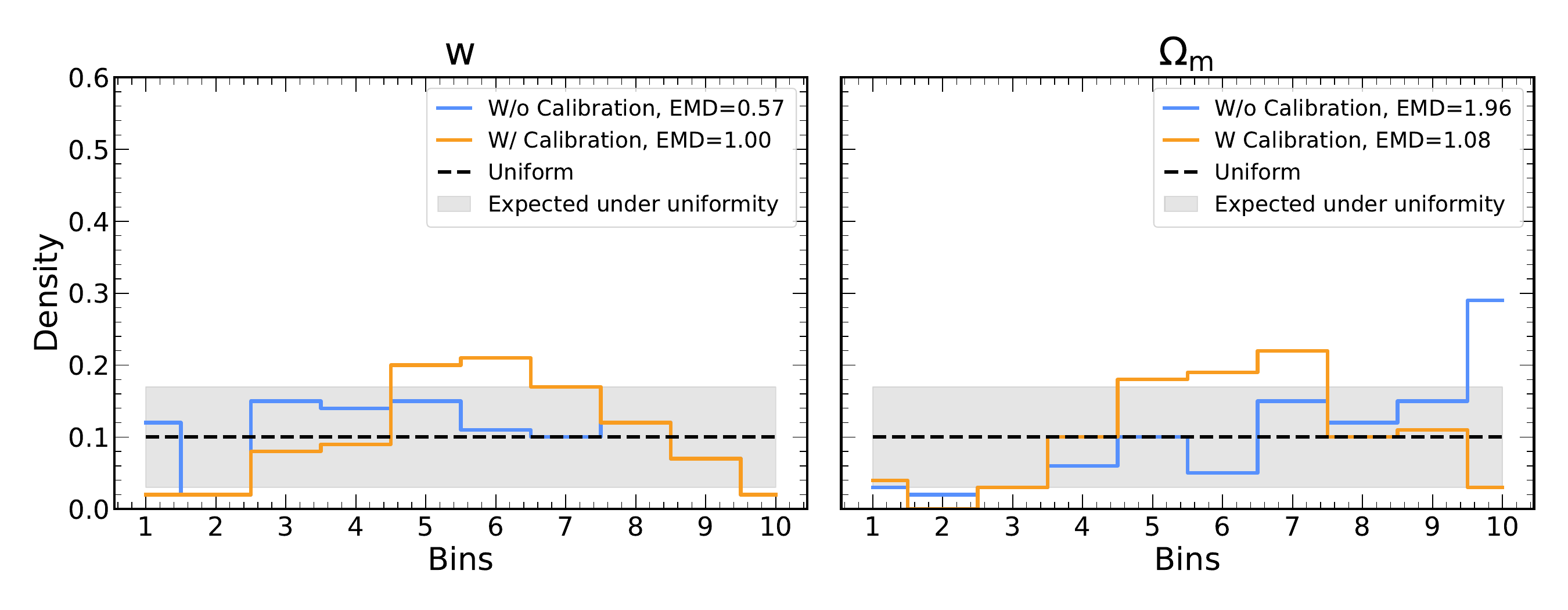}
    \caption{Same as Fig.~\ref{fig:rank_histogram_entire_region}, but for Sub-Region 5.}
   \label{fig:ranks_region5}
\end{figure*}

\begin{figure*}[!ht]
 \centering
    \includegraphics[width=0.4\linewidth]{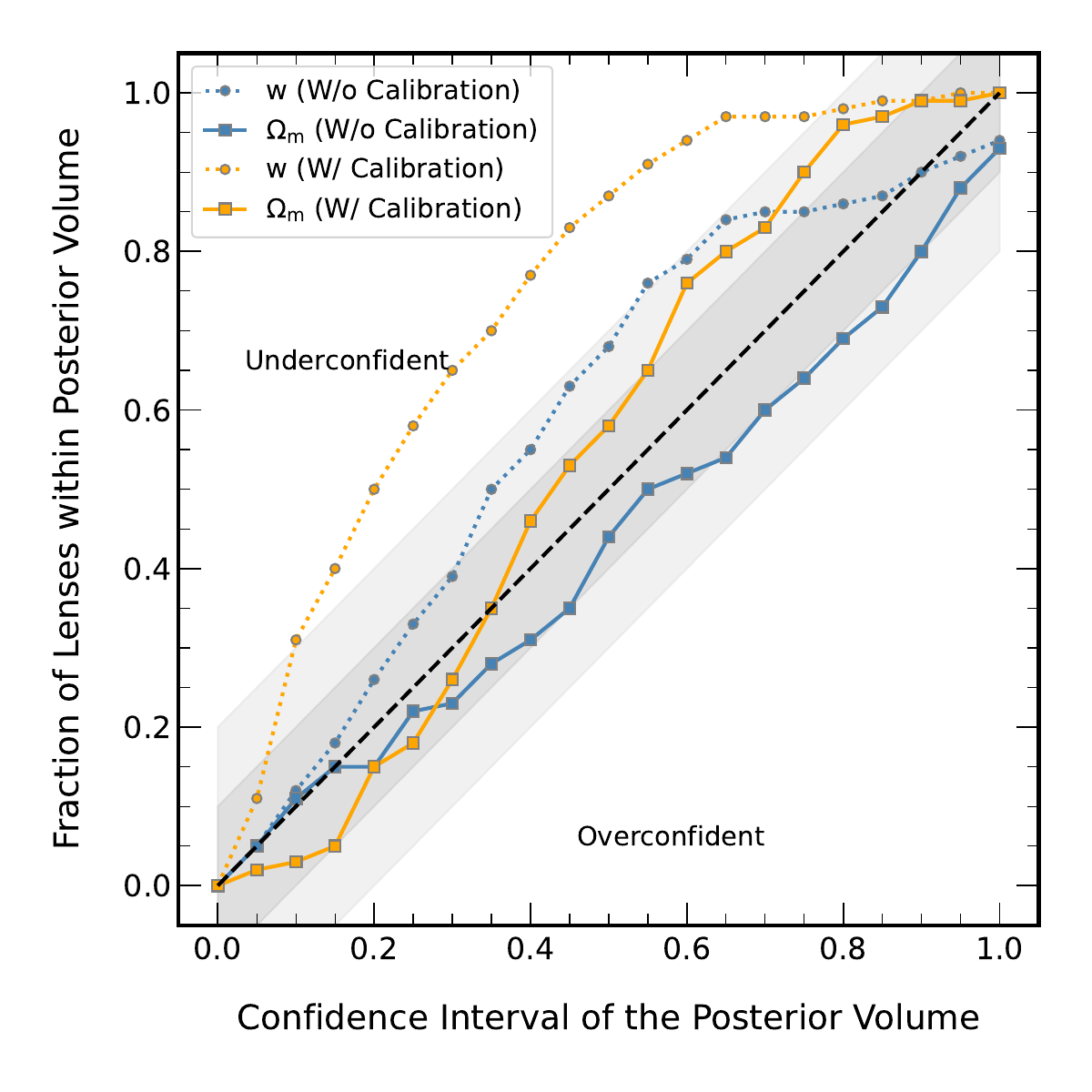}
    \caption{Same as Fig.~\ref{fig:posteror_coverage_entireregion}, but for Sub-Region 5.}
   \label{fig:posterior_coverage_region5}
\end{figure*}

\begin{description}
    \item[Posterior Contours (Fig. \ref{fig:joint_posterior_region5})] Without calibration, the contours exclude one true value out of the nine examples. 
    With calibration the contours are larger and don't exclude any true values. 
    \item[Parity and Residuals (Fig. \ref{fig:parity_region5})] Without calibration, the model is unbiased for both parameters, but has a large number of outliers.
    With calibration, the model has no outliers, and it is unibased for $w$, while having a slight bias for $\om{}$.
    \item[Rank Histogram (Fig. \ref{fig:ranks_region5})] Without calibration, $w$ ranks are slightly left-skewed, and with calibration, they are $\cap$-shaped. 
    Without calibration, the $\om{}$ ranks are highly right-skewed, and with calibration, they are $\cap$-shaped and slightly right-skewed. 
    The EMD for $w$ is nearly doubled with calibration, and it is nearly halved for $\om{}$.
    \item[Posterior Coverage (Fig. \ref{fig:posterior_coverage_region5})] Without calibration, the model is moderately undeconfident for $w$ and moderately overconfident for $\om{}$. 
    With calibration, the model is highly underconfident for $w$ and partially S-shaped for $\om{}$ indicating that the posterior is overconfident in the centre and underconfident in the tails.
\end{description}

For both parameters, with calibration, the model's accuracy is mostly unchanged, but it is made underconfident (to different degrees).

\clearpage

\subsection{Sub-Region 6}
\label{app:region6}

\begin{figure*}[!ht]
 \centering
    \includegraphics[width=1.0\linewidth]{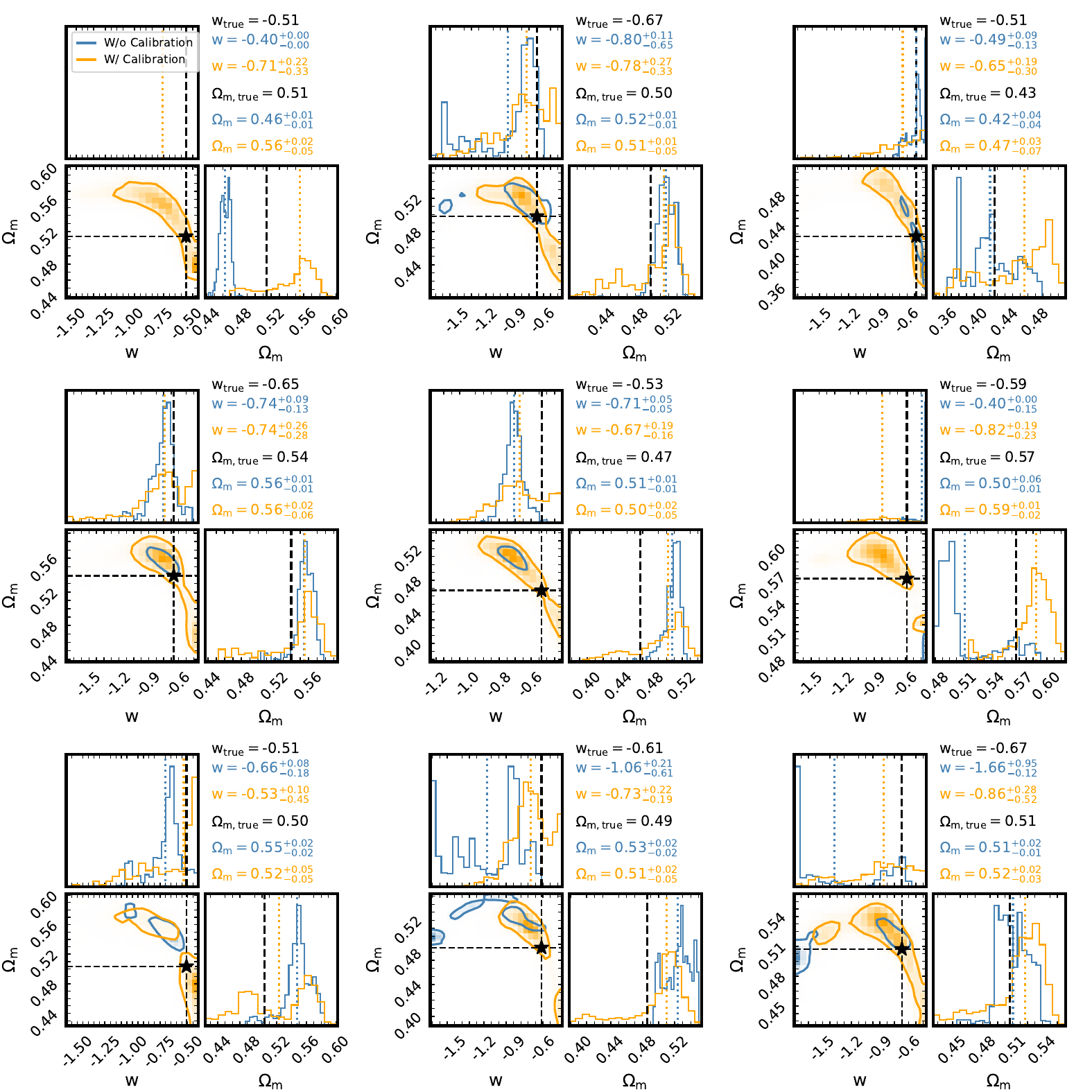}
    \caption{Same as Fig.~\ref{fig:joint_posterior_entireregion}, but for Sub-Region 6.}
   \label{fig:joint_posterior_region6}
\end{figure*}

\begin{figure*}[!ht]
 \centering
    \includegraphics[width=0.6\linewidth]{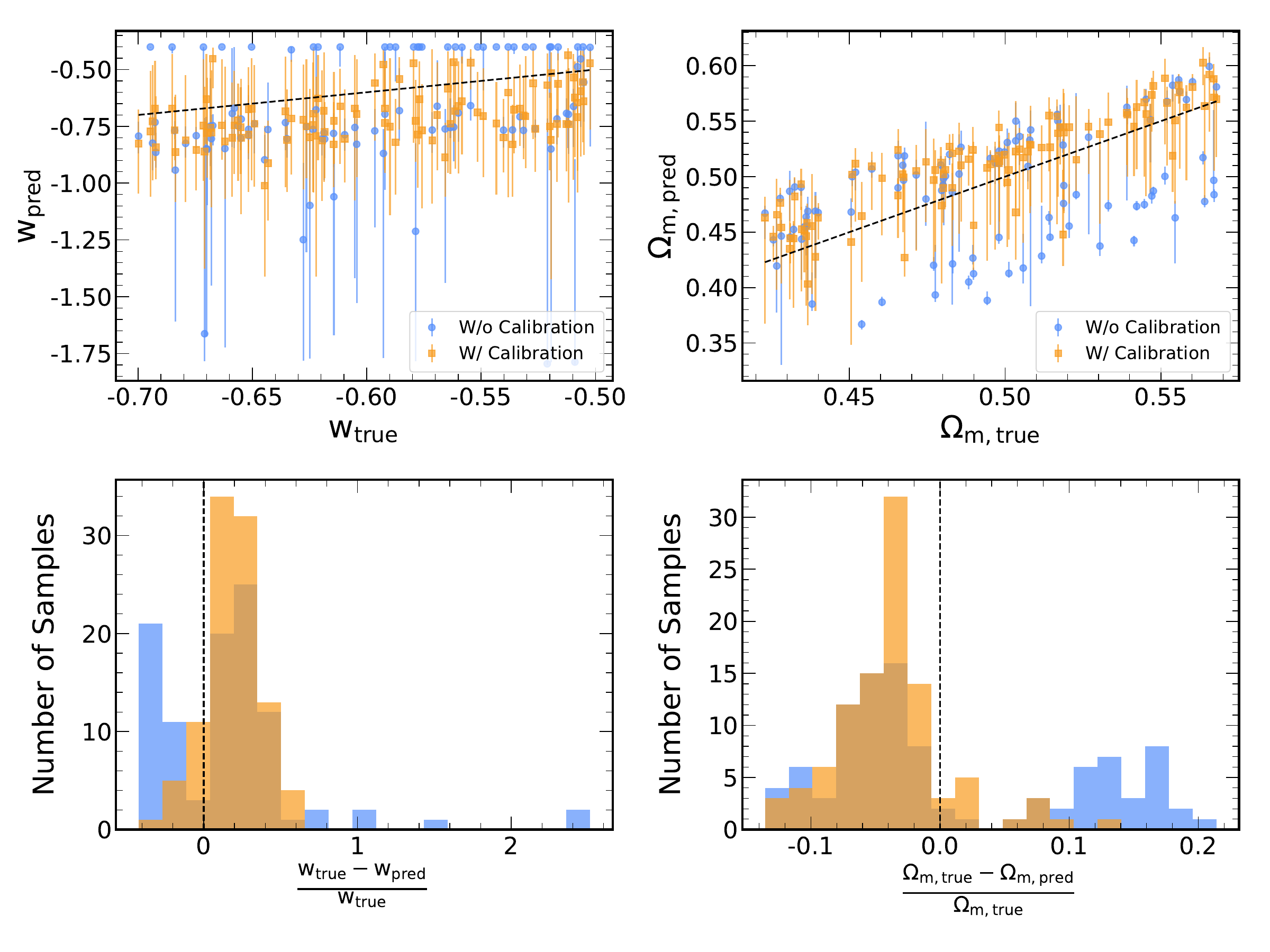}
    \caption{Same as Fig.~\ref{fig:parity_entireregion}, but for Sub-Region 6.}
    \label{fig:parity_region6}
\end{figure*}

\begin{figure*}[!ht]
 \centering
    \includegraphics[width=0.6\linewidth]{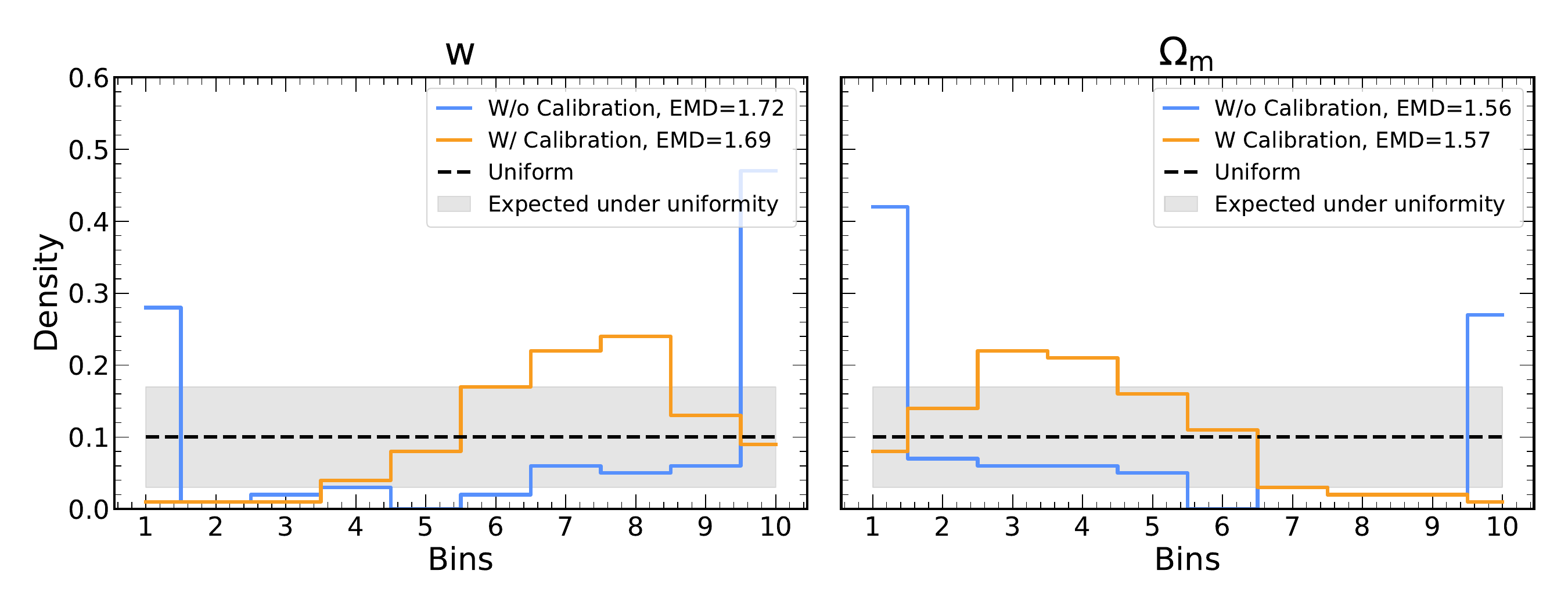}
    \caption{Same as Fig.~\ref{fig:rank_histogram_entire_region}, but for Sub-Region 6.}
   \label{fig:ranks_region6}
\end{figure*}

\begin{figure*}[!ht]
 \centering
    \includegraphics[width=0.4\linewidth]{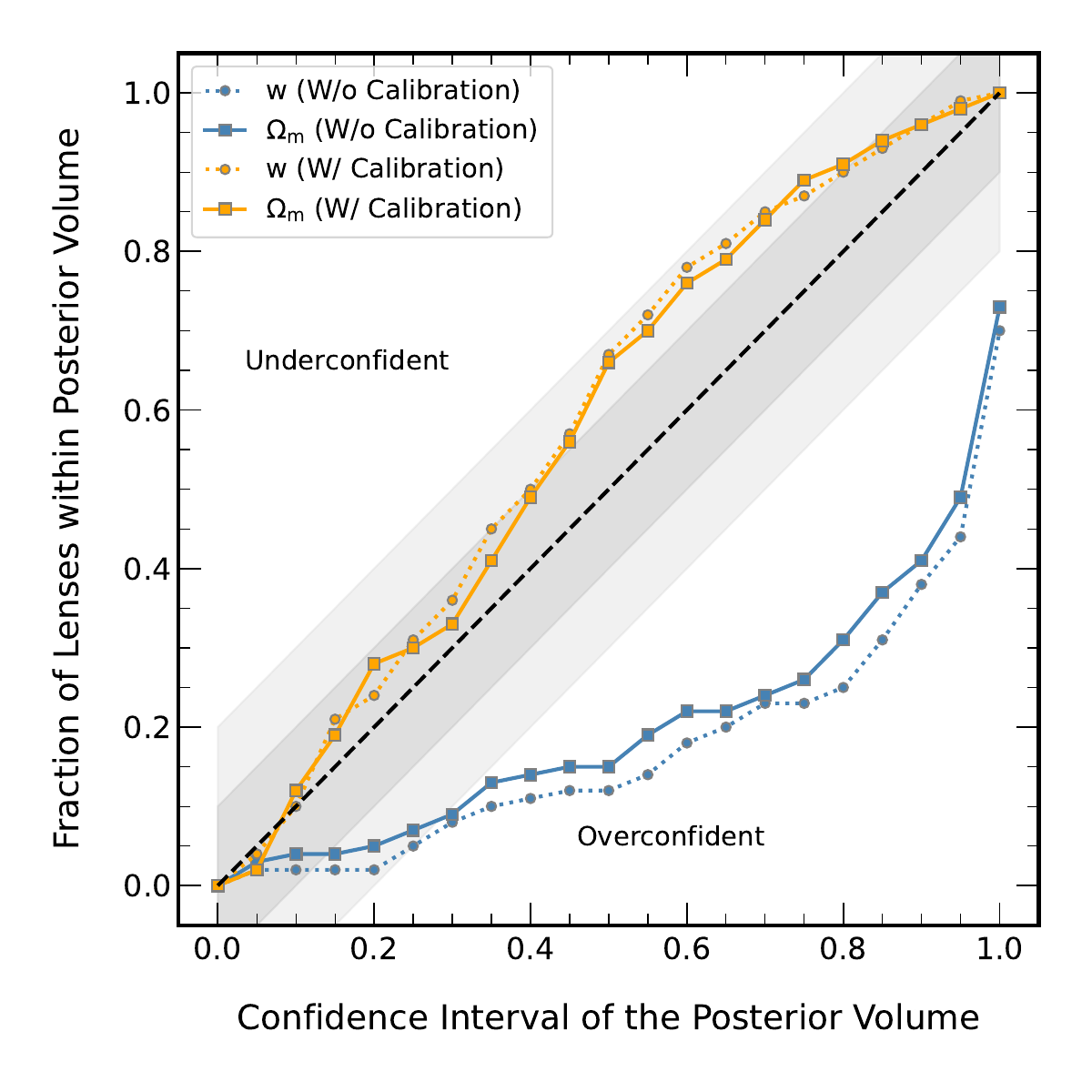}
    \caption{Same as Fig.~\ref{fig:posteror_coverage_entireregion}, but for Sub-Region 6.}
   \label{fig:posterior_coverage_region6}
\end{figure*}

\begin{description}
    \item[Posterior Contours (Fig. \ref{fig:joint_posterior_region6})] Without calibration, the model excludes the true value in the majority of examples. With calibration, the model contours are larger and include all true values.
    \item[Parity and Residuals (Fig. \ref{fig:parity_region6})] For both parameters, without calibration, the model has many outliers. For $\om{}$, the model is nearly bimodal, and thus highly biased.  
    With calibration, the model remains equivalently biased, but has many fewer outliers.
    \item[Rank Histogram (Fig. \ref{fig:ranks_region6})] Without calibration, the ranks are $\cup$-shaped for both parameters. 
    With calibration, the ranks become $\cap$-shaped -- but also slightly right-skewed for $w$ and slightly left-skewed for $\om{}$. 
    With calibration, the EMD is very slightly reduced for both parameters.
    \item[Posterior Coverage (Fig. \ref{fig:posterior_coverage_region6})] For both parameters, without calibration, the models are equally and highly overconfident. 
    With calibration, the models are equally and moderately underconfident.
\end{description}

With calibration, the model is mostly unchanged with respect to accuracy, but it is made more conservative in terms of confidence.

\clearpage

\subsection{Sub-Region 7}
\label{app:region7}

\begin{figure*}[!ht]
 \centering
    \includegraphics[width=1.0\linewidth]{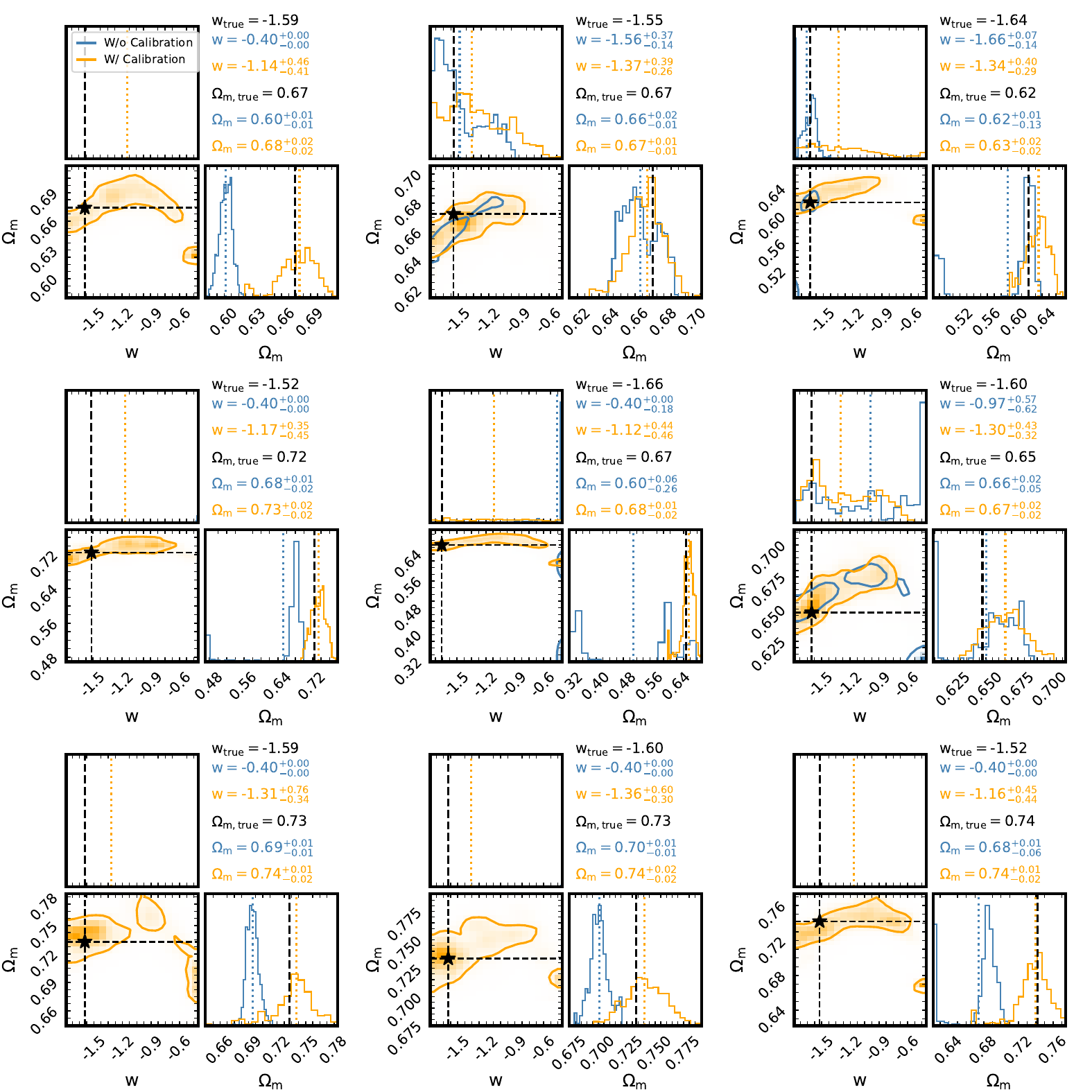}
    \caption{Same as Fig.~\ref{fig:joint_posterior_entireregion}, but for Sub-Region 7.}
   \label{fig:joint_posterior_region7}
\end{figure*}

\begin{figure*}[!ht]
 \centering
    \includegraphics[width=0.6\linewidth]{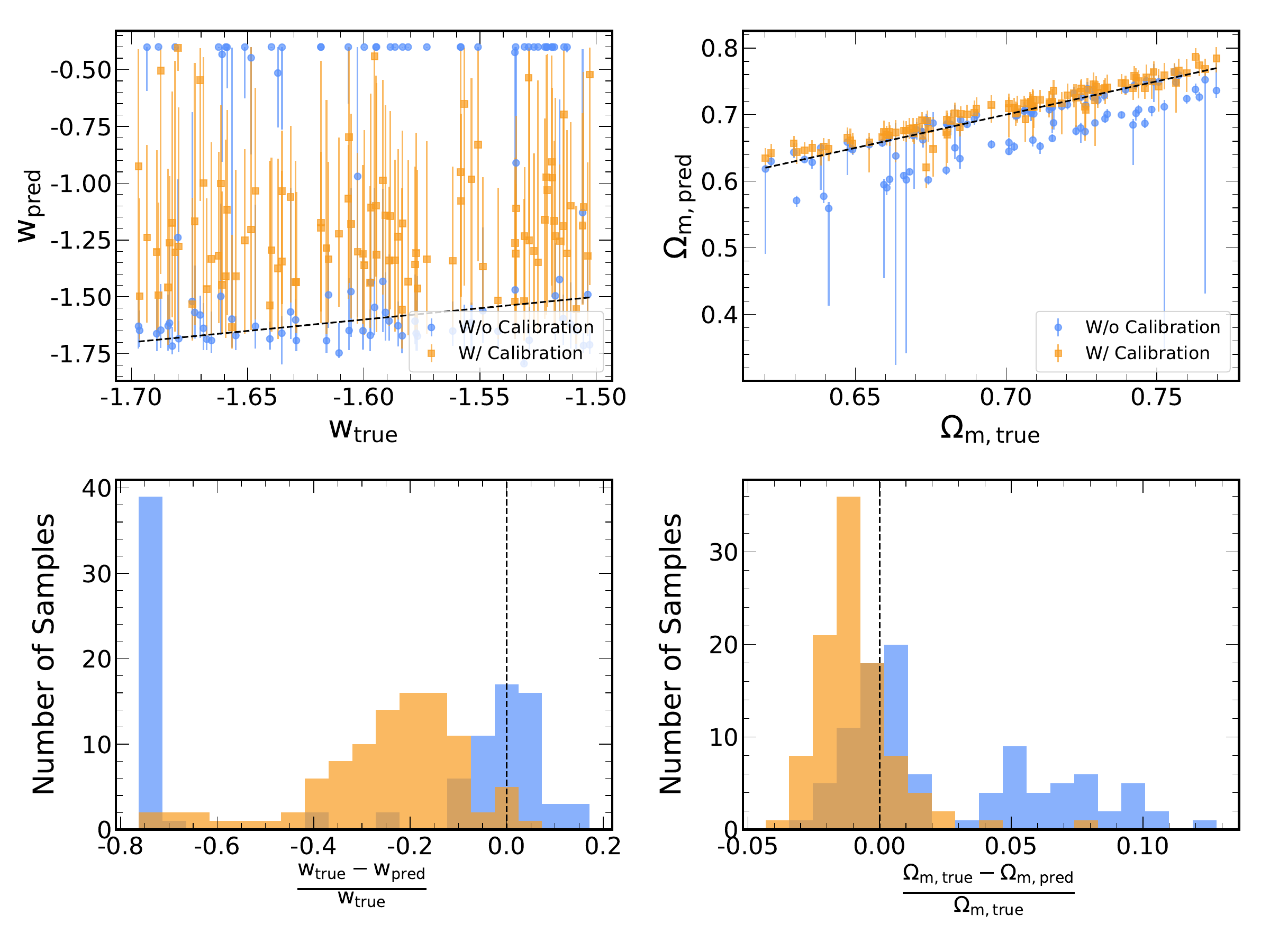}
    \caption{Same as Fig.~\ref{fig:parity_entireregion}, but for Sub-Region 7.}
   \label{fig:parity_region7}
\end{figure*}

\begin{figure*}[!ht]
 \centering
    \includegraphics[width=0.6\linewidth]{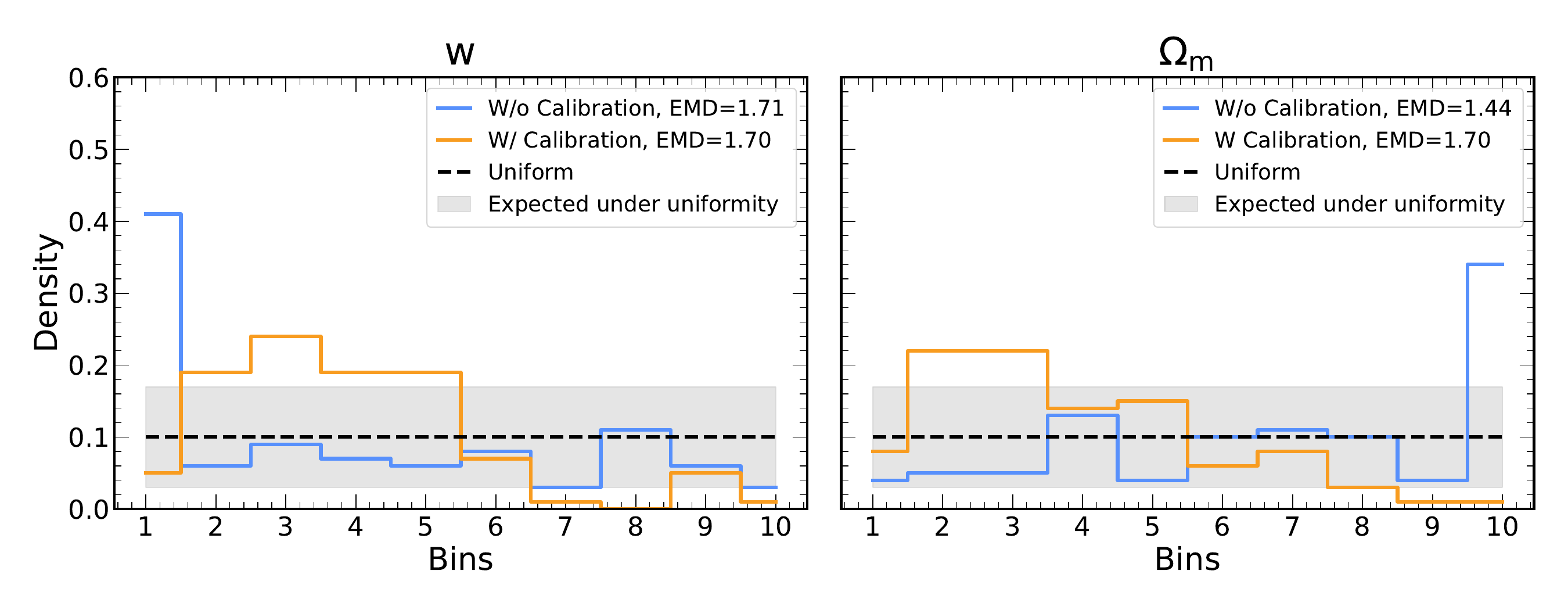}
    \caption{Same as Fig.~\ref{fig:rank_histogram_entire_region}, but for Sub-Region 7.}
   \label{fig:ranks_region7}
\end{figure*}

\begin{figure*}[!ht]
 \centering
    \includegraphics[width=0.4\linewidth]{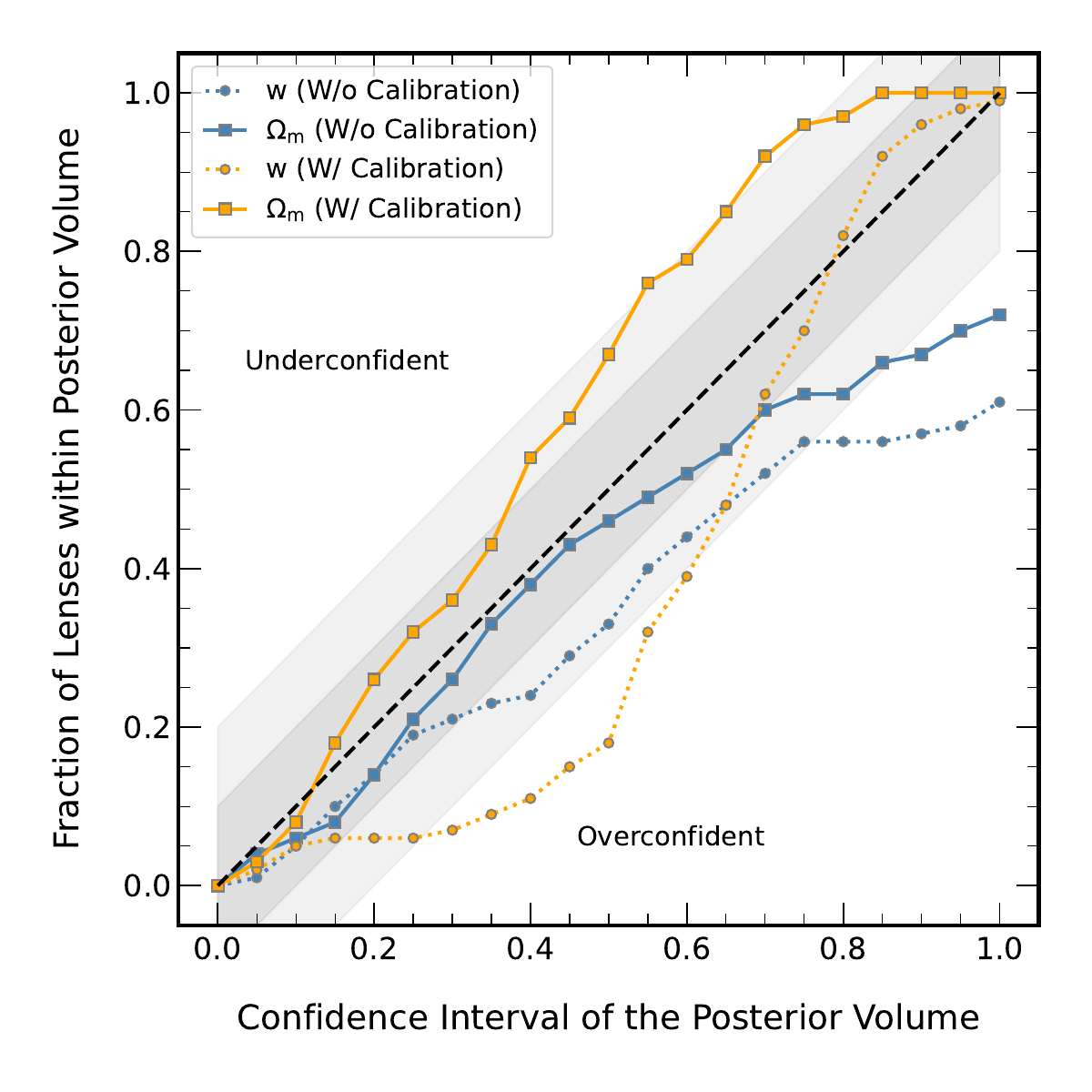}
    \caption{Same as Fig.~\ref{fig:posteror_coverage_entireregion}, but for Sub-Region 7.}
   \label{fig:posterior_coverage_region7}
\end{figure*}

\begin{description}
    \item[Posterior Contours (Fig. \ref{fig:joint_posterior_region7})] Without calibration, the model contours exclude true values for almost all the examples, but include true values in all examples with calibration.
    \item[Parity and Residuals (Fig. \ref{fig:parity_region7})] For $w$, without calibration, the model is unbiased, but has many outliers. 
    For $\om{}$, the model is nearly bimodal, and thus highly biased. 
    With calibration, for $w$, the model is more biased, but with fewer outliers from the main distribution. For $\om{}$, the model is less biased.
    \item[Rank Histogram (Fig. \ref{fig:ranks_region7})] Without calibration, the ranks are highly left-skewed for $w$ and highly right-skewed for $\om{}$. 
    With calibration, the ranks are moderately left-skewed for both parameters.  EMD is unchanged for $w$ with calibration and is increased for $\om{}$ with calibration.
    \item[Posterior Coverage (Fig. \ref{fig:posterior_coverage_region7})] Without calibration, the models are catastrophically misspecified for both parameters. 
    With calibration, the model is mostly overconfident for $w$ and moderately underconfident for $\om{}$.
\end{description}

For both parameters, the model is improved with calibration, but remains overconfident for $w$.

\clearpage

\subsection{Sub-Region 8}
\label{app:region8}

\begin{figure*}[!ht]
 \centering
    \includegraphics[width=1.0\linewidth]{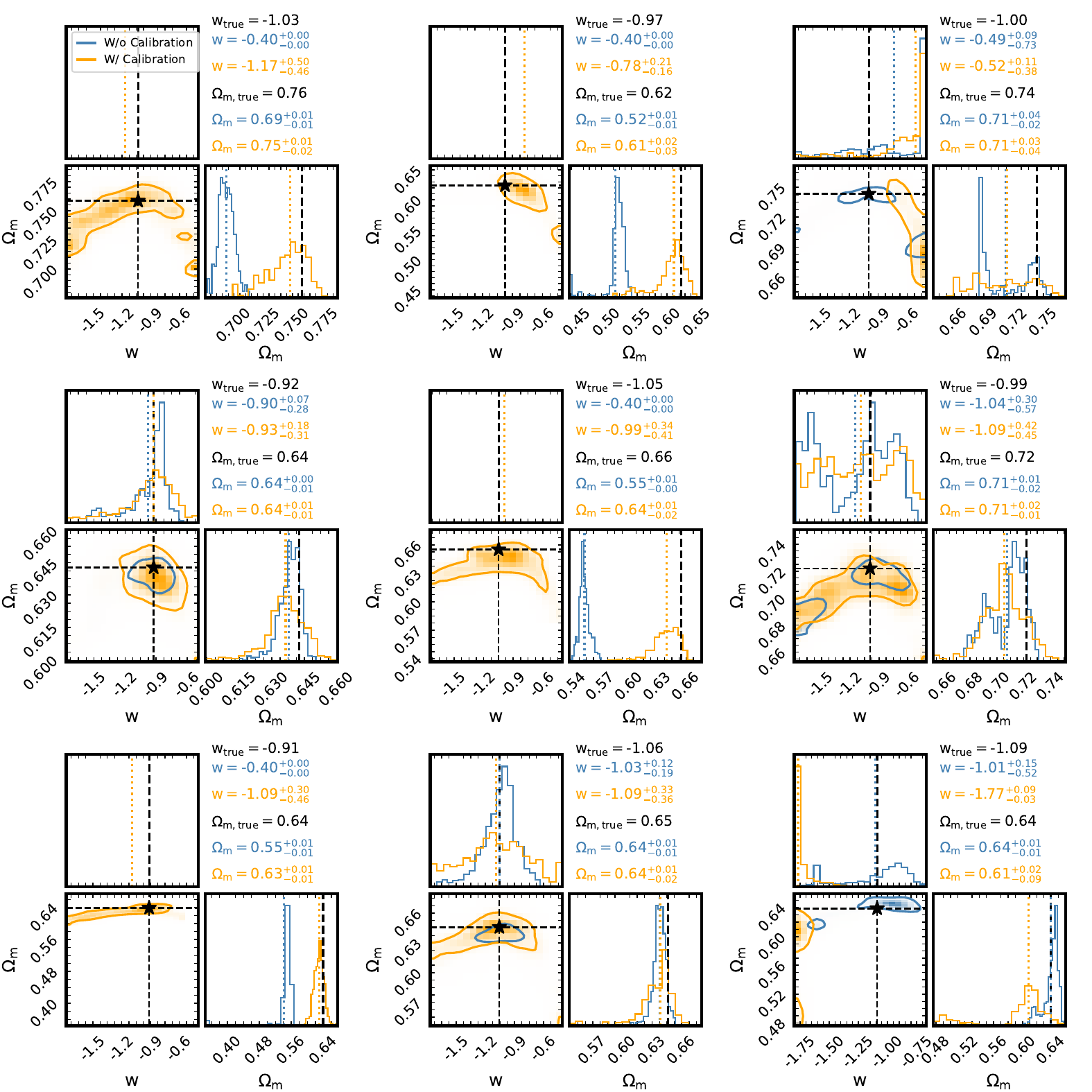}
    \caption{Same as Fig.~\ref{fig:joint_posterior_entireregion}, but for Sub-Region 8.}
   \label{fig:joint_posterior_region8}
\end{figure*}

\begin{figure*}[!ht]
 \centering
    \includegraphics[width=0.6\linewidth]{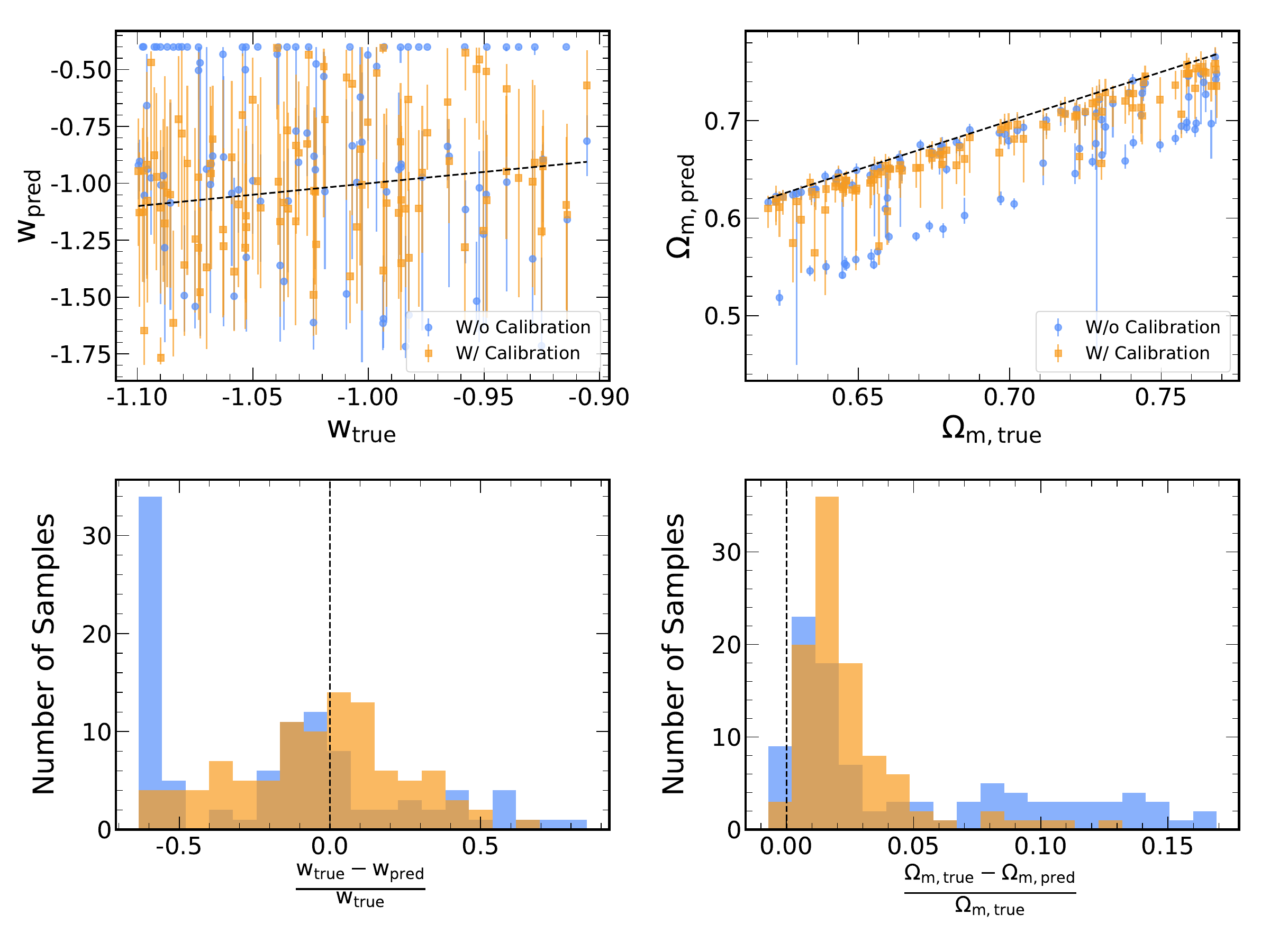}
    \caption{Same as Fig.~\ref{fig:parity_entireregion}, but for Sub-Region 8.}
   \label{fig:parity_region8}
\end{figure*}

\begin{figure*}[!ht]
 \centering
    \includegraphics[width=0.6\linewidth]{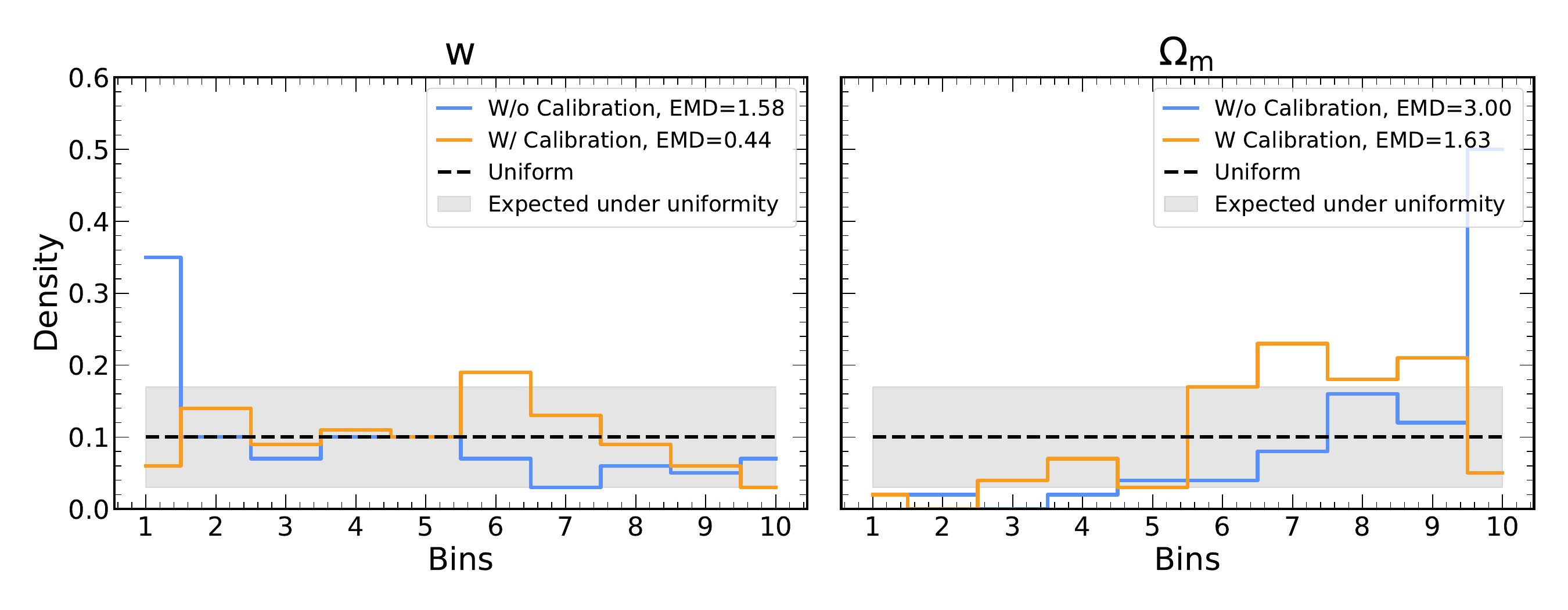}
    \caption{Same as Fig.~\ref{fig:rank_histogram_entire_region}, but for Sub-Region 8.}
   \label{fig:ranks_region8}
\end{figure*}

\begin{figure*}[!ht]
 \centering
    \includegraphics[width=0.4\linewidth]{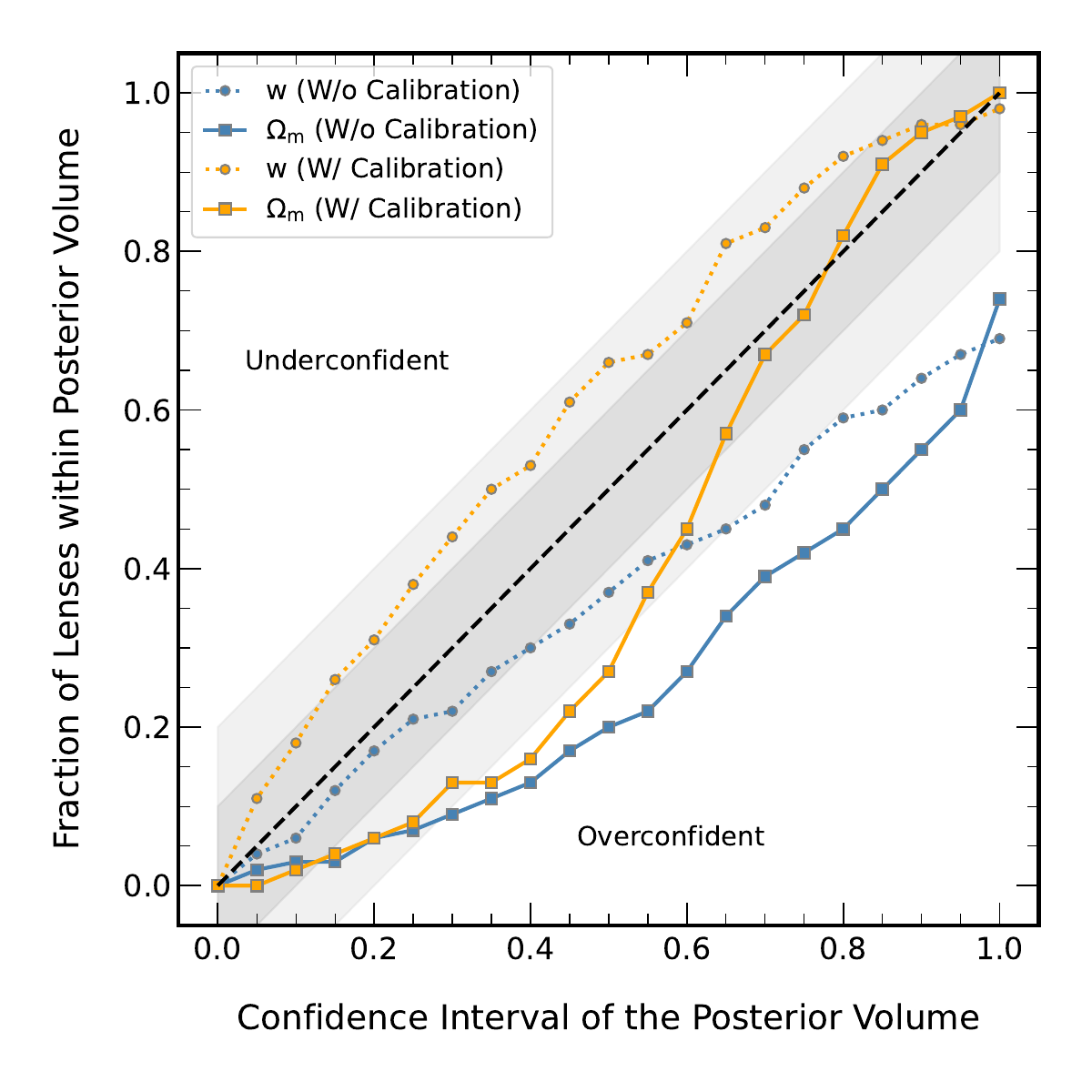}
    \caption{Same as Fig.~\ref{fig:posteror_coverage_entireregion}, but for Sub-Region 8.}
   \label{fig:posterior_coverage_region8}
\end{figure*}

\begin{description}
    \item[Posterior Contours (Fig. \ref{fig:joint_posterior_region8})] Without calibration, the model contours exclude the true value for nearly half the examples. With calibration, the model contours exclude true values in two of the examples.
    \item[Parity and Residuals (Fig. \ref{fig:parity_region8})] Without calbiration, the parity is bimodal in both parameters. With calibration, the model parity is unimodal and unbiased for $w$, and unimodal and slightly biased for $\om{}$.
    \item[Rank Histogram (Fig. \ref{fig:ranks_region8})] Without calibration, the $w$ ranks are highly left-skewed and the $\om{}$ ranks are highly right skewed. With calibration, this skewness persists, but is reduced. With calibration, the EMD for $w$ is reduced by over 70\%, the EMD for $\om{}$ is reduced by over 40\%.
    \item[Posterior Coverage (Fig. \ref{fig:posterior_coverage_region8})] Without calibration, the models are catastrophically misspecified for both parameters. With calibration, the model is moderately underconfident for $w$ and mostly overconfident $\om{}$ -- but no long catastrophically misspecified.
\end{description}

Calibration improves the model behavior for both parameters, but remains overconfident for $\om{}$.

\clearpage

\subsection{Sub-Region 9}
\label{app:region9}

\begin{figure*}[!ht]
 \centering
    \includegraphics[width=1.0\linewidth]{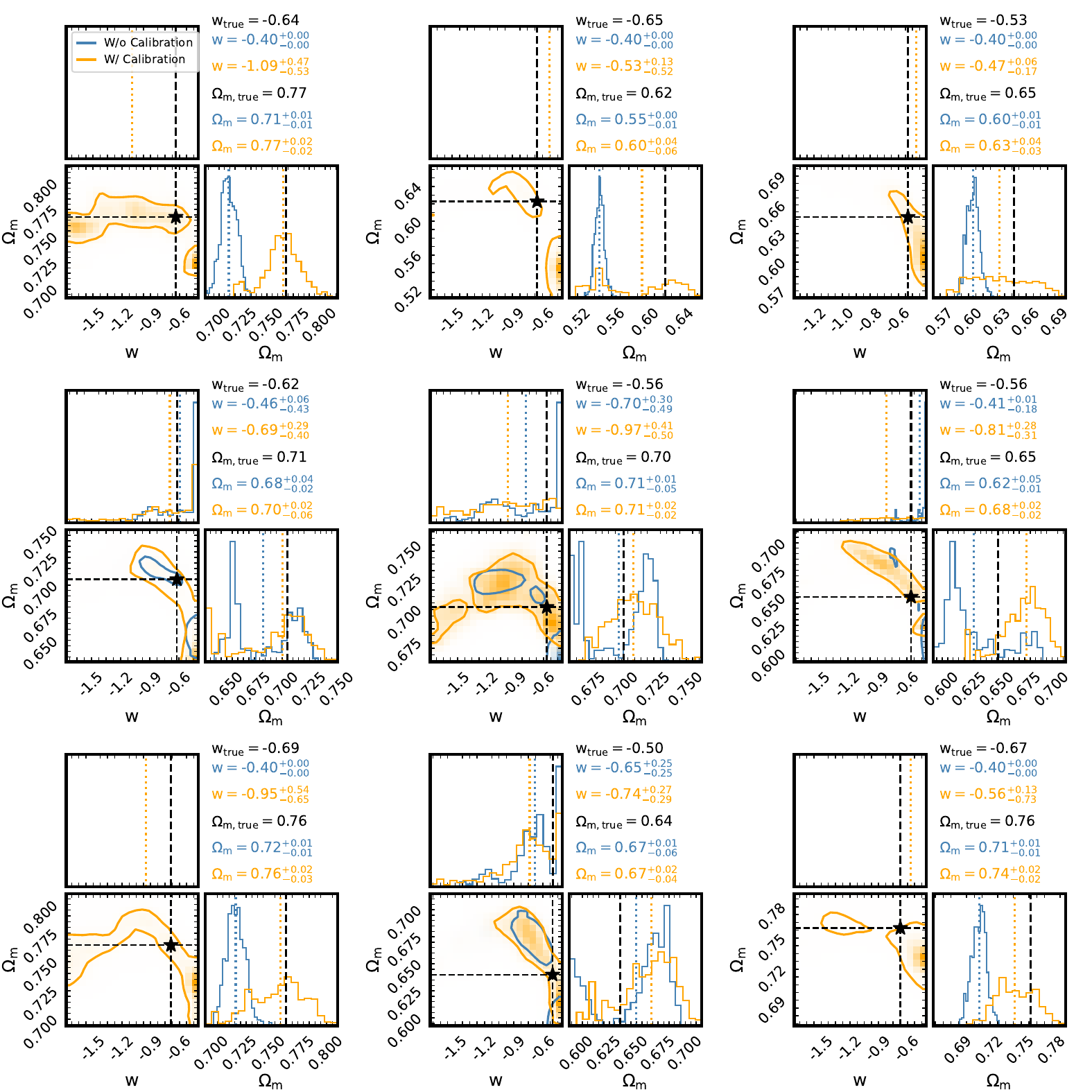}
    \caption{Same as Fig.~\ref{fig:joint_posterior_entireregion}, but for Sub-Region 9.}
   \label{fig:joint_posterior_region9}
\end{figure*}

\begin{figure*}[!ht]
 \centering
    \includegraphics[width=0.6\linewidth]{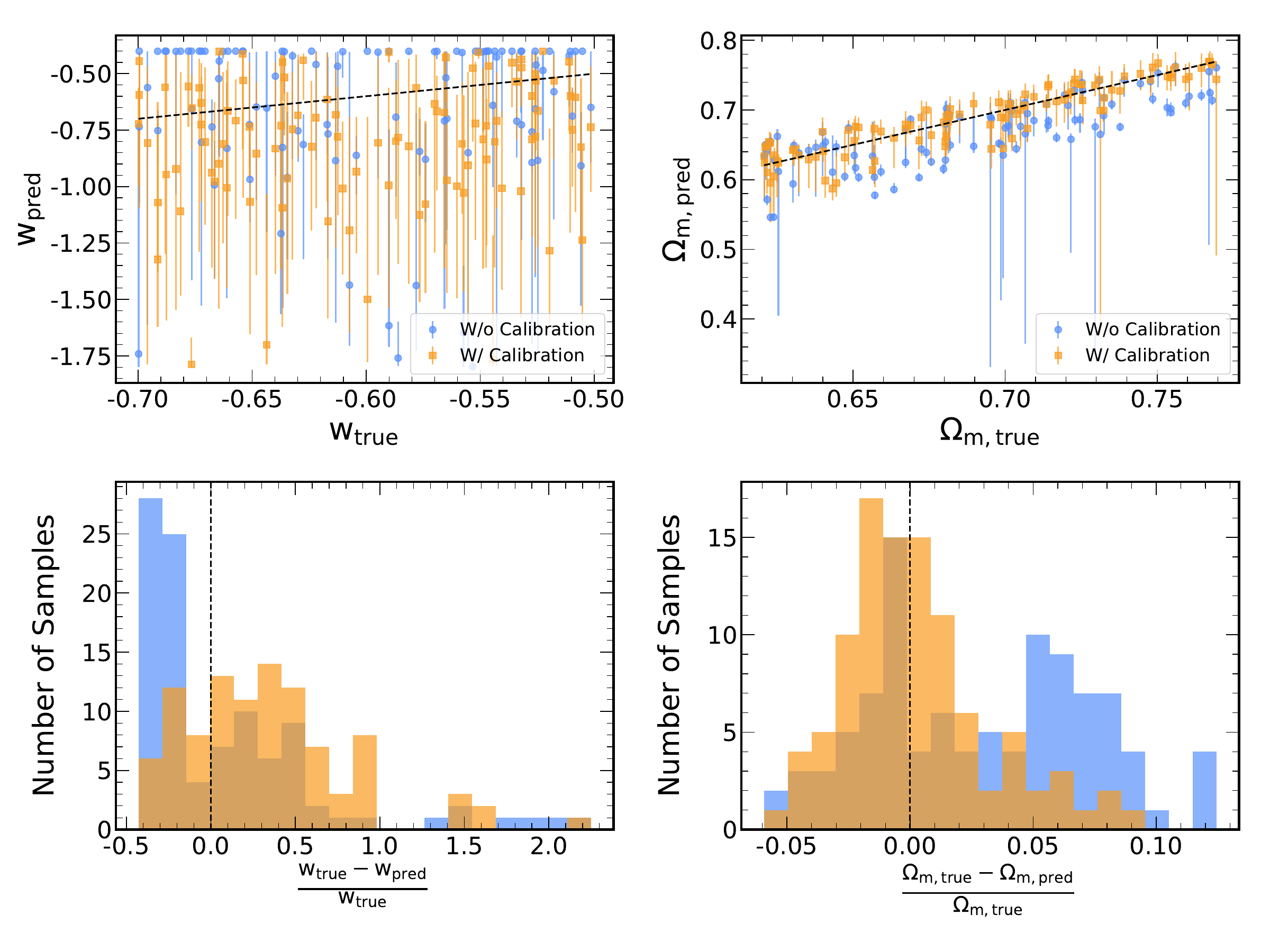}
    \caption{Same as Fig.~\ref{fig:parity_entireregion}, but for Sub-Region 9.}
   \label{fig:parity_region9}
\end{figure*}

\begin{figure*}[!ht]
 \centering
    \includegraphics[width=0.6\linewidth]{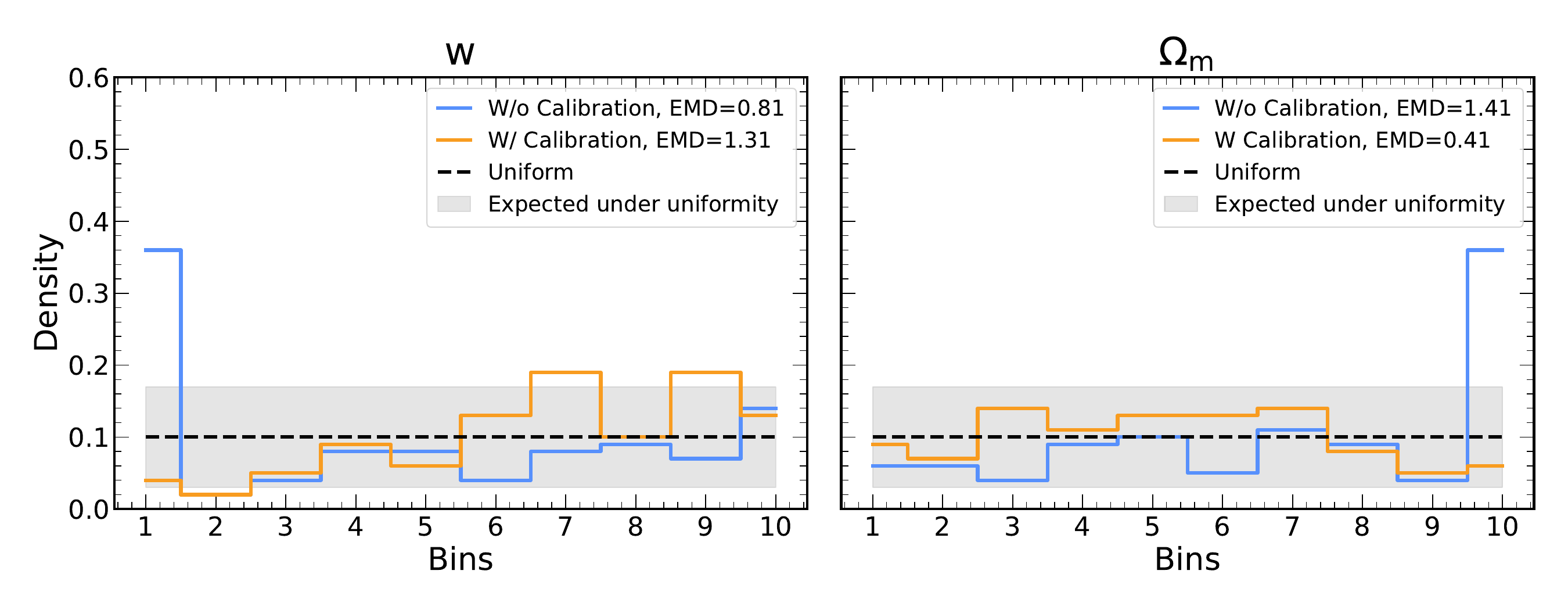}
    \caption{Same as Fig.~\ref{fig:rank_histogram_entire_region}, but for Sub-Region 9.}
   \label{fig:ranks_region9}
\end{figure*}

\begin{figure*}[!ht]
 \centering
    \includegraphics[width=0.4\linewidth]{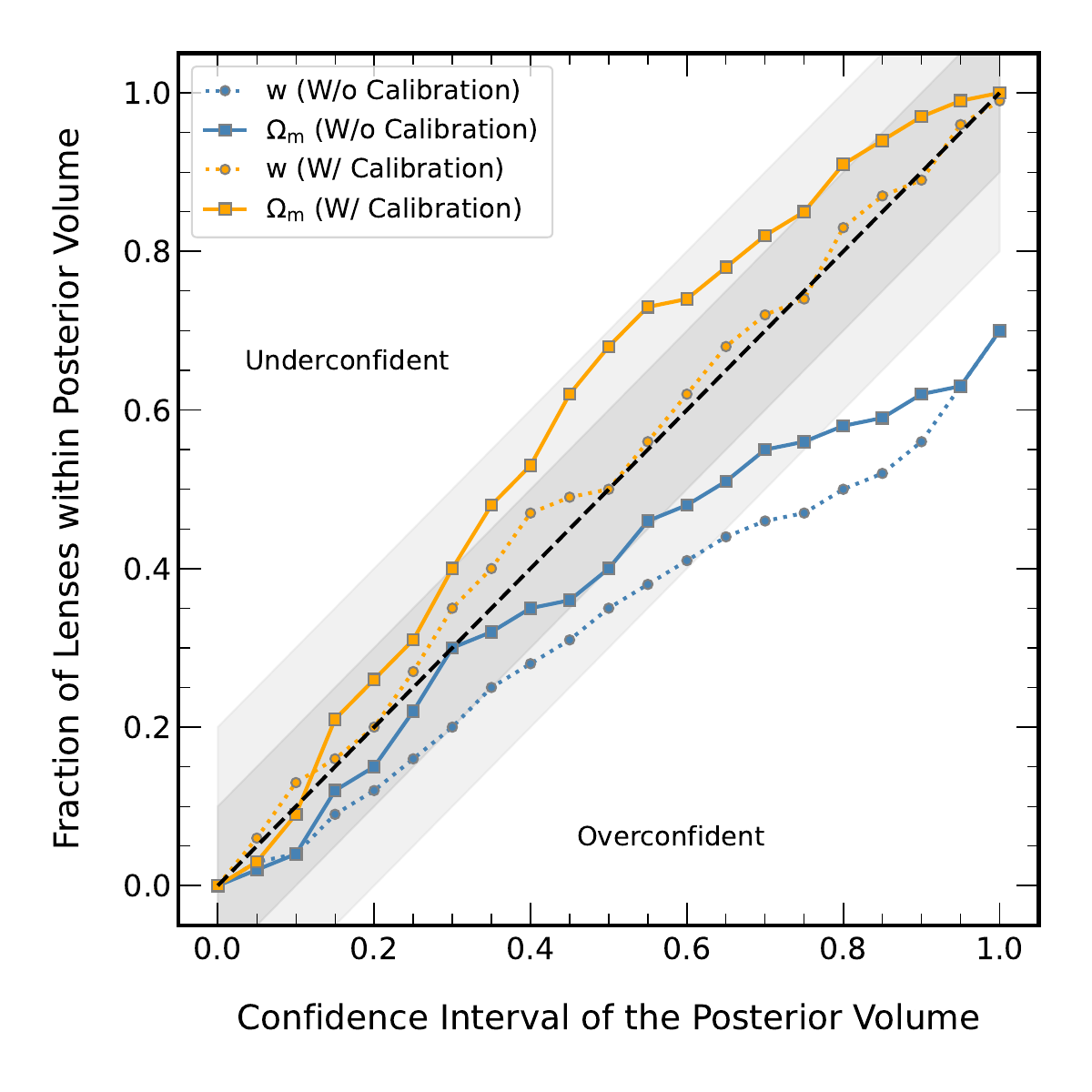}
    \caption{Same as Fig.~\ref{fig:posteror_coverage_entireregion}, but for Sub-Region 9.}
   \label{fig:posterior_coverage_region9}
\end{figure*}

\begin{description}
    \item[Posterior Contours (Fig. \ref{fig:joint_posterior_region9})] Without calibration, the model excludes the true value in all nine examples. With calibration, the model includes the true value in all nine examples.
    \item[Parity and Residuals (Fig. \ref{fig:parity_region9})] Without calibration, the model parity is bimodal and has outliers for both parameters. With calibration, the modal nearly unimodal and slightly biased for $w$ and unimodal for $\om{}$.
    \item[Rank Histogram (Fig. \ref{fig:ranks_region9})]Without calibration, the ranks for $w$ are highly left-skewed, and for $\om{}$ highly right-skewed. With calibration, the $w$ ranks are moderately right-skewed, and the $\om{}$ ranks are nearly uniform. With calibration, the EMD is increased significantly for $w$ and decreased by over 50\% for $\om{}$. 
    \item[Posterior Coverage (Fig. \ref{fig:posterior_coverage_region9})] Without calibration, the model is overconfident and misspecified for both parameters. With calibration, the model is nearly perfectly confident for $w$ and moderately underconfident for $\om{}$.
\end{description}

The model is made underconfident and well-specifed with calibration, with slight bias.

\clearpage

\bibliography{bibliography}{}
\bibliographystyle{aasjournal}


\end{document}